\documentclass[5p]{elsarticle}

\usepackage{hyperref,amsmath}
\usepackage[dvipsnames]{xcolor} 
\usepackage{graphicx}
\usepackage{multicol}
\usepackage{comment}
\usepackage{amsmath}
\usepackage{siunitx}

\begin{document}

\begin{frontmatter}

\title{High-power ultrafast radially and azimuthally polarized accelerating Airy beams and their particle-like lattice topologies}

\author[mymainaddress]{Justas Ber\v{s}kys \corref{mycorrespondingauthor}}
\cortext[mycorrespondingauthor]{Corresponding author}
\ead{justas.berskys@ftmc.lt}

\author[mymainaddress]{Paulius \v{S}levas}

\author[mymainaddress]{Sergej Orlov \corref{mycontactauthor}}
\cortext[mycontactauthor]{Contact author}
\ead{sergejus.orlovas@ftmc.lt}

\address[mymainaddress]{Center for Physical Sciences and Technology, Coherent Optics Laboratory, Sauletekio Ave. 3, Vilnius, Lithuania}

\begin{abstract}

Accelerating Airy beams, known for their non-diffracting nature, self-healing properties, and curved propagation trajectories, are solutions to the paraxial wave equation. In this work, we theoretically and experimentally investigate nonuniform (radially and azimuthally) polarized vector Airy beams. We provide an analytical representation of their spatial spectra and examine their electromagnetic field distributions in space. To validate the theoretical model, we have used a nanograting inscribed inside a glass volume by a femtosecond laser to create geometrical phase elements, suitable for the realization of high-power ultrafast radially and azimuthally polarized vector Airy beams. We have conducted experiments that confirm the successful generation of high-power vector beams and have performed Stokes parameter measurements of these beams. Additionally, we explore both theoretically and experimentally their topology, identifying particle-like formations such as skyrmionic and antiskyrmionc accelerating lattices within the high-power ultrafast electric fields and their Stokes parameters. 

\end{abstract}


\end{frontmatter}

\section{Introduction}
Electromagnetic waves are described by four main parameters: wavelength, phase, polarization, and amplitude. Today, each of these parameters can be varied independently and accurately. The development of tools such as lasers, various optical elements, and more recent meta-elements \cite{dorrah2022tunable, shimotsuma2003self} has led to the development of areas such as metrology, manufacturing, quantum technologies, and many others. 
The optical fields that can be obtained by only manipulating the parameters mentioned above span from optical tweezers \cite{yang2021optical}, optical polarization Möbius strips \cite{bauer2015observation}, non-diffracting beams \cite{durnin1987diffraction, mcgloin2005bessel}, and singular beams \cite{gbur2016singular, dennis2002polarization} to topological particles of light \cite{sugic2021particle, shen2024optical} and complex field structures for cryptography and data transmission.

The family of non-diffracting beams has a particular interest, the light field that propagates does not perceive any diffraction, like the Bessel beam, or the diffraction can be limited to some extent as the Bessel-Gauss beam. Another property that some nondiffracting beams have is the so-called self-healing \cite{broky2008self}. The reconstruction of the initial transverse distribution is attributed to the Bessel and Airy beam families. The Bessel and Airy beams, in addition to self-healing and nondiffraction, can be described with vortex-like structures in their phase or polarization. The obtained angular momentum, whether spin or orbital, can be transferred to nanoparticles for manipulation or trapping.

Furthermore, another degree of freedom for constructing optical beams is their polarization manipulation. Polarization can have a significant effect on highly focused optical fields \cite{dorn2003sharper} and their propagation dynamics, resulting in interesting phenomena such as singular polarization beams \cite{cardano2013generation} or spin-orbit interactions of the beam \cite{bliokh2015spin}. 

Moreover, there have been a lot of investigations into topological quasiparticles, called skyrmions. Firstly, a model for describing baryons as topological solitons was proposed by Tony Skyrme \cite{skyrme1961non}. Later, the idea of stable field topological configurations gained attention in different contexts, such as liquid crystals \cite{fukuda2011quasi}, thin magnetic films \cite{kiselev2011chiral}, spintronics \cite{wiesendanger2016nanoscale}, and others. Recently, the exploration of topological quasiparticles emerged in optics \cite{parmee2022optical}. The quasiparticles can be observed in the Stokes field domain \cite{shen2021generation}, electric field \cite{gutierrez2021optical}, spin of light \cite{du2019deep}, and any other derivative form of an electromagnetic wave. The main difficulties arise from the fact that free-space optics are usually treated as transverse fields, and the observation of the longitudinal components can be achieved only for strongly focused or evanescent fields. However, optical quasiparticles were investigated for optical lattices \cite{marco2024propagation, mcwilliam2022topological}, light pulses \cite{shen2021supertoroidal, vo2024closed}, photonic hopfions \cite{droop2023transverse, shen2023topological, wan2022scalar}. 

The study of optical lattices containing topological quasiparticles beginning is rooted in phase singularities, a good example being superpositions of nondiffracting Bessel beams \cite {orlov2002propagation, orlov2004propagation}. Periodic skyrmionic structures were observed within optics \cite{marco2024periodic}, even maintaining propagation-invariant properties \cite{marco2024propagation}. Poincare sphere analysis was successful in identifying it \cite{wang2024orientation}. These types of lattices were also observed within Airy beams with a particle-like polarization singularity in the center of the parabolic trajectory of an Airy beam \cite{bervskys2023accelerating}.

All these complex topologies can be efficiently created using metasurfaces, which can be realized using different concepts. Here, we select a beam shaping method based on birefringent nanogratings inscribed in the bulk volume of the glass \cite{Shimotsuma2003,Mirza2016}. Direct laser inscription of nanogratings, known as type~II modification \cite{Hnatovsky2011}, controls the retardance and orientation of the fast axis throughout the aperture area of the element. These metadurfaces created from layers of nanogratings are called geometric phase elements (GPE) \cite{Hasman2003,Drevinskas2016,Sakakura2020}. Alternatively, they are known as Pancharatnam-Berry phase elements because they are described by expressing a phase change that occurs due to a change in polarization state after light propagation through birefringent media \cite{Cohen2019}. 

GPEs are useful for many purposes, recognizing them with a high optical damage threshold, close to the material of which they are made, reaching $2$~J/cm$^2$ at $1030$~nm wavelength and $212$~fs pulse length \cite{lidt}. This way we, together with industrial partner \textit{Workshop of Photonics}, have created numerous photonic elements with specifically optimized phase patterns that were inscribed on a one-inch-sized fused silica substrate for various high-power ultra-fast applications \cite{vslevas2024optical, nacius2024polarization, nacius2025asymmetric}. 

In this research, we theoretically introduce radially and azimuthally polarized Airy-like beams. In these types of beam, the line-like axial polarization singularity interacts with the parabolically shaped propagation trajectory of an Airy beam. We investigate the spatial spectra and electric field distributions, present the analytical theory behind the modeling, and give a generalized description of radially and azimuthally polarized beams when the solution for the scalar wave equation is known. Analogically, the Fourier spectra description is derived for the nonuniform radial and azimuthal polarization beams when the scalar spectra are known. Second, we transfer the design to the geometrical phase elements, which we inscribe using a femtosecond laser system. We experimentally construct such high-power ultrafast beams using these geometrical phase elements and compare their Stokes field distributions for experimental validation of theoretical calculations. The Poincare sphere is depicted and investigated. Third, we explore the topological structure of the constructed beam, theoretically investigate the electric field domain topology, and experimentally and analytically compare the topological field distributions of the Stokes vector field.

\section{Theoretical background} \label{Theory}
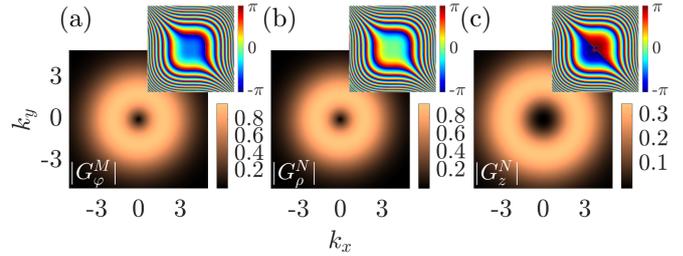
\begin{figure}
\includegraphics[scale=0.17]{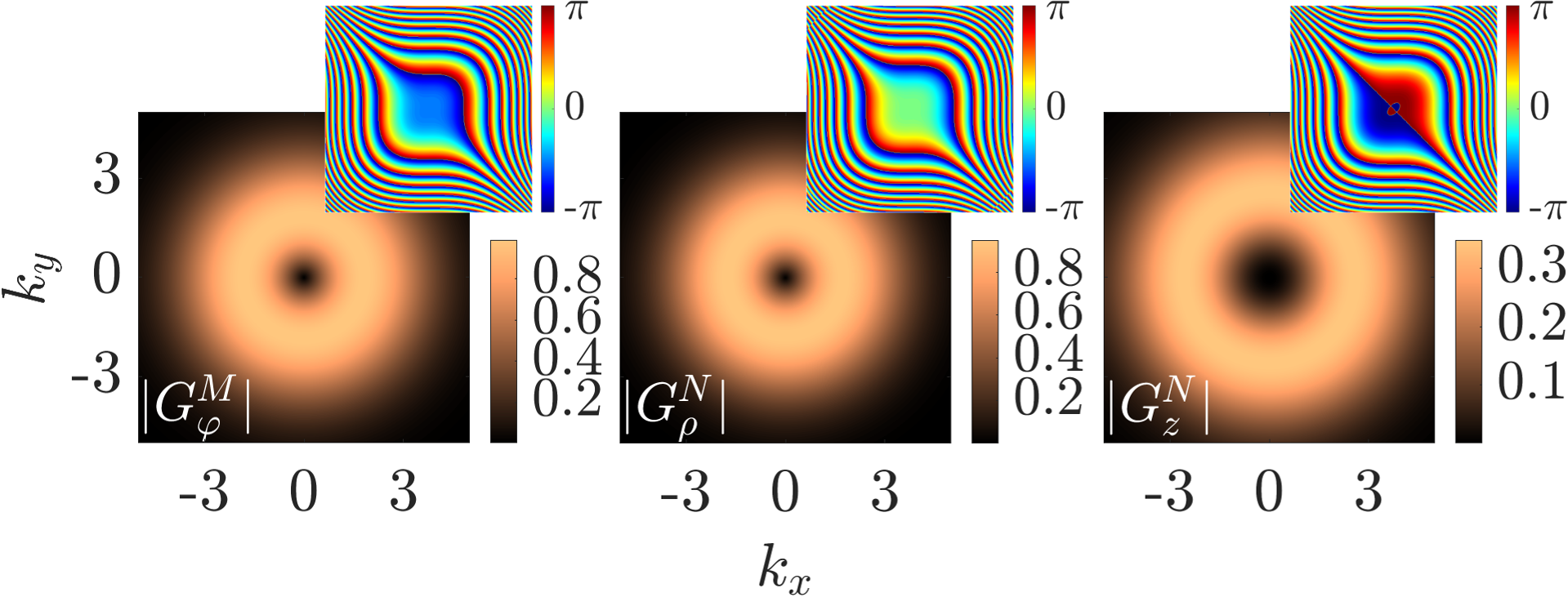}%
\begin{picture}(0,0)
\put(-228,85){(a)}
\put(-152,85){(b)}
\put(-78,85){(c)}
\end{picture}
\caption{An angular spectrum representation of Airy vector beams. An angular spectrum amplitude of an azimuthally polarized beam $\textbf{M}$ (a) and a radially polarized beam $\textbf{N}$ (b, c). The wave number is $k = 2 \pi$, the decay factors $a_x = a_y = 0.15$, and the normalization constants $x_0=y_0=1$.  The insets represent phase distributions.}
\label{fig:ang_spctr}
\end{figure}
The solution for the paraxial scalar (1+1)D Airy equation is given by \cite{siviloglou2007accelerating} 
\begin{equation}
\begin{aligned}
u_t(t, z) =& \text{Ai} \left[s_t-(\xi_t / 2)^{2}+\mathrm{i}  a_t \xi_t\right] \exp \left[a_t \left( s_t-\xi_t^2/2 \right)\right.\\
+&\left. \mathrm{i}  \xi_t / 2 \left(a_t^{2}+s_t-\xi_t^{2} / 6\right) \right],
\label{Eq:Airy_scalar}
\end{aligned}
\end{equation}
where $\xi_t=z/kt_0^2$ is a normalized longitudinal propagation coordinate $z$, $s_t=t/t_0$ is a normalized transverse coordinate $t$ ($x$ or $y$), $k=2\pi /\lambda$ is the wave vector in vacuum and $a_t$ is the decay factor (aperture factor) of the beam. The decay factor allows for the representation of experimentally realizable Airy beams, as the non-apertured Airy beam has infinite power. The three-dimensional solution of the scalar wave equation is the product of two two-dimensional equations described in the $x$ and $y$ directions and a plane wave as $U(\textbf{r})=u_x(x,z) u_y(y,z) \exp (\mathrm{i}  k_z z)$. Note that the time dependence of the field is omitted here. The spatial spectrum of Eq. \ref{Eq:Airy_scalar} is given by $g_t = \exp[{\left(a_t+\mathrm{i}  k_t t_0\right)^3}/3]$, where $k_t$ denotes the component $x$ or $y$ of the wave vector. Consequently, the spatial spectrum of the three-dimensional Airy beam is $g(k_x,k_y)=g_x(k_x) g_y(k_y)$.
Two orthogonal vector solutions, namely \textbf{M} and \textbf{N}, can be constructed from a scalar solution $U(\textbf{r})$, see \cite{morse1954methods}, given by
\begin{equation}\label{Eq:LMN}
	\textbf{M}(\textbf{r}) = \nabla \times [\textbf{e}_{z}\ U(\textbf{r})],\quad
	\textbf{N}(\textbf{r}) = \dfrac{1}{k} \nabla\times \textbf{M}(\textbf{r}),
\end{equation}
where $\textbf{e}_{z}=(0,0,1)$ and $\nabla$ is a nabla operator. There are alternative choices for the vector $\textbf{e}_{z}$, see \cite{morse1954methods}, in our case, we construct vector beams \textbf{M} and \textbf{N}, with the constant vector pointing in the $z$ direction. This choice is expected to result in the vector beam \textbf{M} having azimuthal polarization, and vector beam \textbf{N} with radial polarization in the transverse plane perpendicular to the propagation axis \cite{orlov2014vectorial}. The general expression for the vector spectra of both beams is
\begin{equation}\label{Eq:Spectra_LMN}
\textbf{G}^{M}(k_{x}, k_{y})=\mathrm{i} g \mathbf{k} \times \textbf{e}_{z}, \quad \textbf{G}^{N}(k_{x}, k_{y})=\frac{1}{k} g(\mathbf{k} \times \textbf{e}_{z}) \times \mathbf{k},
\end{equation}
where $\mathbf{k}=(k_x,k_y,k_z)$.
The spatial spectrum for the azimuthally polarized vector Airy beam (the transverse electric mode $\mathbf{M}$) is given by 
\begin{equation}\label{Eq:M_spectra}
\textbf{G}^{M}(k_{x}, k_{y})=\mathrm{i}  g \left( \mathbf{e}_{x} k_{y} - \mathbf{e}_{y} k_{x} \right) = -\mathrm{i}  g k_{\rho}\mathbf{e}_{\phi}.
\end{equation}
Here, ($\rho$, $\phi$) are the polar coordinates and $k_{\rho}$ is the transverse component of the wave vector.
Similarly, the radially polarized Airy beam (transverse magnetic mode $\mathbf{N}$) has a spatial spectrum given by 
\begin{align}\label{Eq:N_spectra}
\textbf{G}^{N}(k_{x}, k_{y})&=\frac{g}{k} \left[ \mathbf{e}_{x} k_{x} k_{z} + \mathbf{e}_{y} k_{y} k_{z} - \mathbf{e}_{z} \left(k_{x}^2 + k_{y}^2\right) \right] \nonumber \\
 & =\frac{g}{k} \left[ \mathbf{e}_{\rho} k_{\rho} k_{z}  - \mathbf{e}_{z} k_{\rho}^2  \right] 
\end{align}
It is worth noting that vector fields constructed using the Equation \ref{Eq:LMN} have their spatial spectra expressed by the Equations \ref{Eq:M_spectra} and \ref{Eq:N_spectra} also in a general case \cite{stratton2007electromagnetic}. 
\begin{figure}
\centering
\includegraphics[scale=0.4]{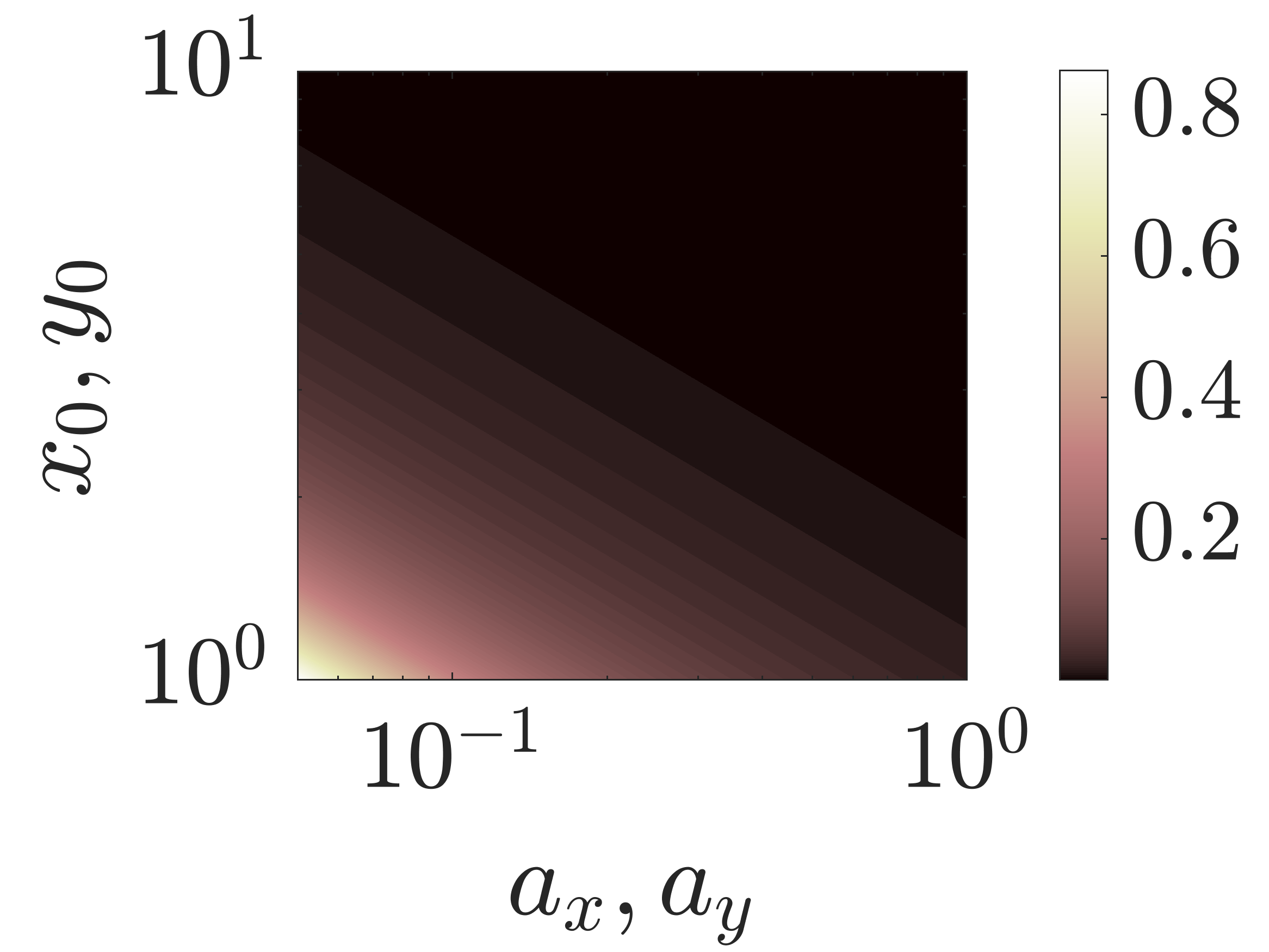}%
\begin{picture}(0,0)
\put(-80,90){\textcolor{white}{\LARGE $\frac{|\textbf{G}^{N}_z|^2}{|\textbf{G}^{N}_\rho|^2}$}}
\end{picture}
\caption{Power distribution among the $\textbf{e}_z$ and $\textbf{e}_\rho$ components of an angular spectrum of Airy-like vector beam $\textbf{G}^{N}$ for the decay factor when $a_x=a_y$ and scaling factor $x_0=y_0$ values. The wave number $k = 2 \pi$.}
\label{fig:paraxial_limit}
\end{figure}
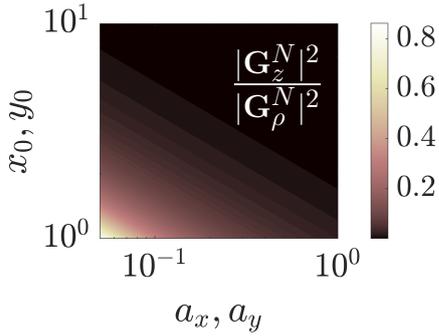

Furthermore, in the paraxial limit $k_z \approx k$ and $k_x^2, k_y^2 \approx 0$, the longitudinal spectral component of the radially polarized field $\textbf{G}^{N}(k_{x}, k_{y})$ decreases and disappears, and the transverse components of both fields $\textbf{G}^{N}(k_{x}, k_{y})$ and $\textbf{G}^{M}(k_{x}, k_{y})$ become similar.

The analysis of the spatial spectra is crucial for the experimental realization of these beams. Usually, one of the methods used for generating optical beams, whether scalar or vector, relies on the manipulation of phase and amplitude of the spatial spectra, which is performed in one of the focal planes of the Fourier lens, and at the second focal plane the desired beam configuration is realized.

The spatial spectrum of the vector beam $\textbf{G}^{M}$ has a single component in the polar coordinate system and is proportional to $\textbf{e}_\phi$. This is an indication of the fact that the $\textbf{G}^{M}$-type beam is azimuthally polarized. In Figure \ref{fig:ang_spctr}a, the amplitude and phase of the spatial spectra $\textbf{G}^{M}(k_{x}, k_{y})$ are presented. The amplitude of the spectra resembles a doughnut shape, and the phase sustains the scalar Airy beam's cubic profile. In comparison, the scalar Airy beam has a Gaussian shape of the amplitude distribution with the same phase profile but with an additional phase shift of $\pi$.

Next, the spectrum $\textbf{G}^{N}(k_{x}, k_{y})$ described primarily in the polar coordinate basis vector $\textbf{e}_\rho$ has a component in the direction of propagation, this arises from the fact that both vector fields \textbf{M} and \textbf{N} are solenoidal. Furthermore, the spectrum $\textbf{G}^{N}(k_{x}, k_{y})$ has the same doughnut shape and cubic phase-like distributions for its components $\textbf{e}_\rho$ and $\textbf{e}_z$.
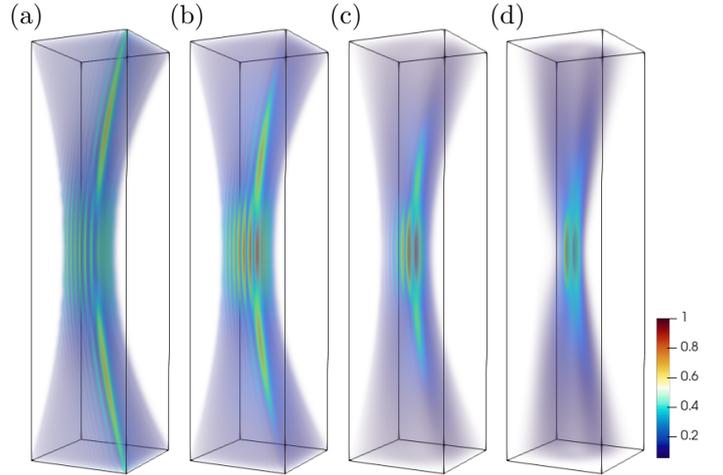
\begin{figure}
\centering
\includegraphics[scale=0.31, trim={465 0 460 0}, clip]{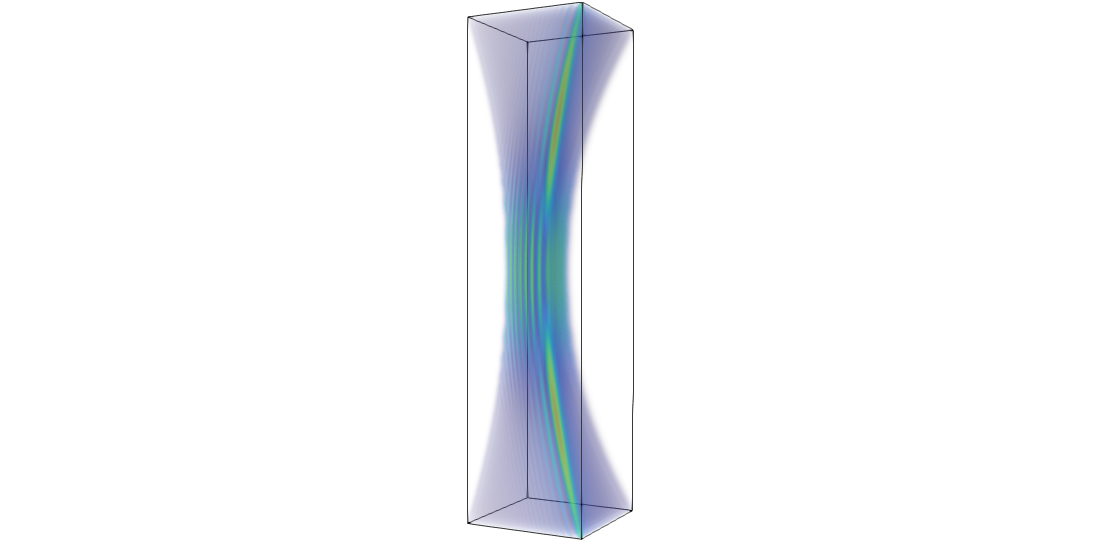}
\includegraphics[scale=0.31, trim={465 0 460 0}, clip]{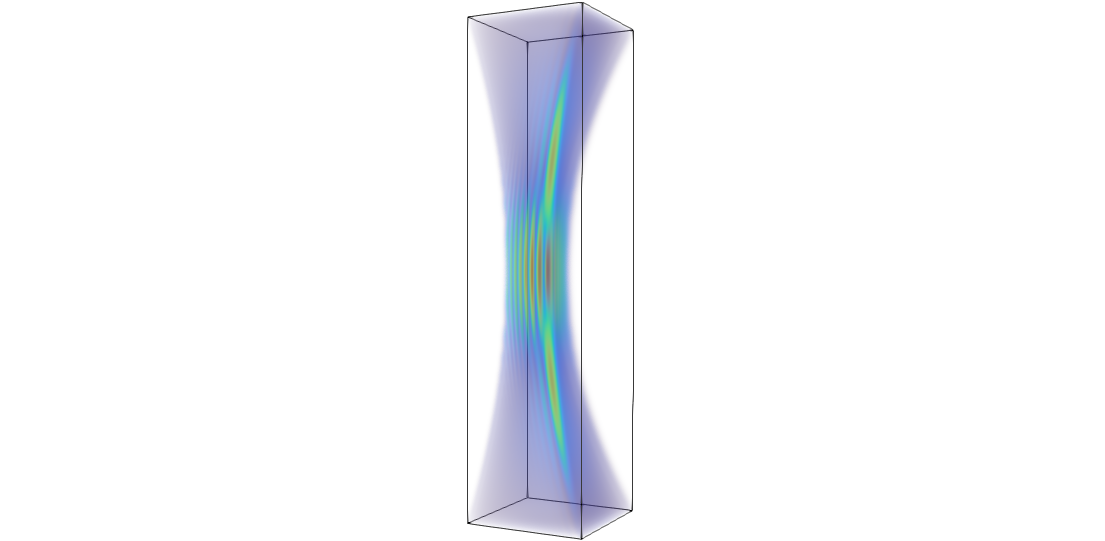}
\includegraphics[scale=0.31, trim={465 0 460 0}, clip]{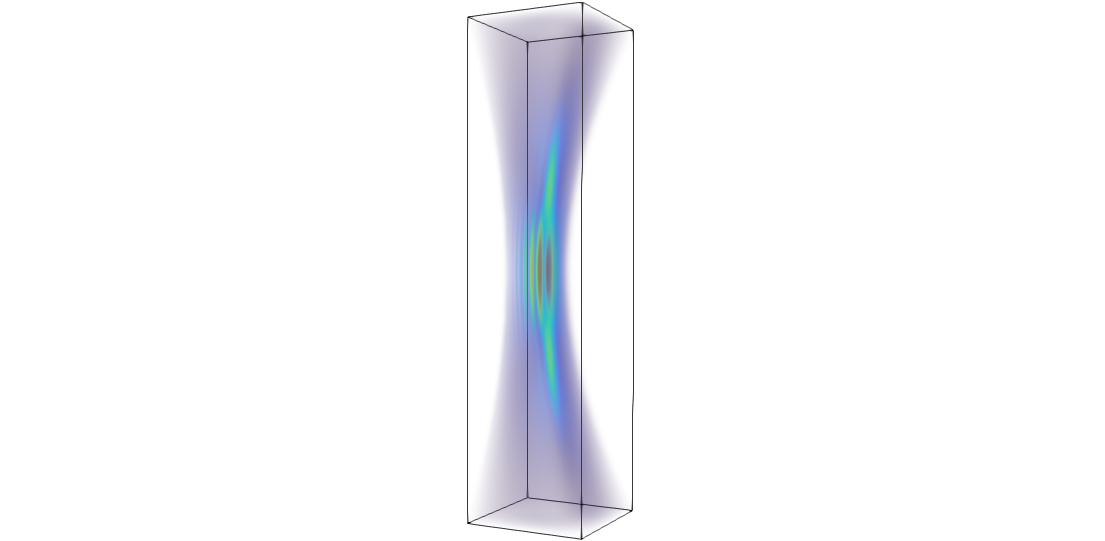}
\includegraphics[scale=0.31, trim={465 0 460 0}, clip]{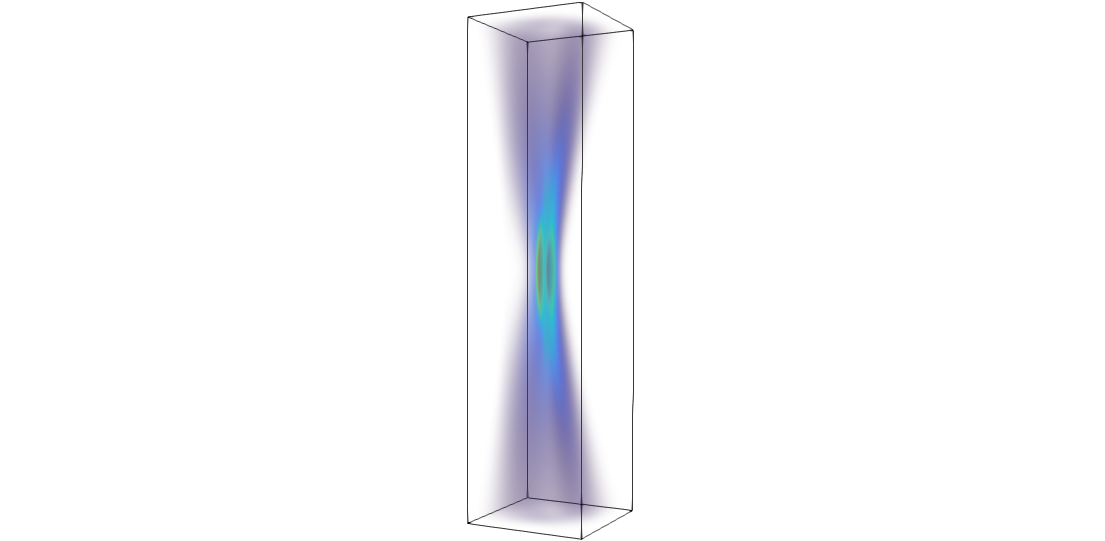}
\begin{picture}(0,0)
\put(-245,170){(a)}
\put(-185,170){(b)}
\put(-125,170){(c)}
\put(-65,170){(d)}
\put(-35,5){\includegraphics[scale=0.3, trim={1000 0 0 0}, clip]{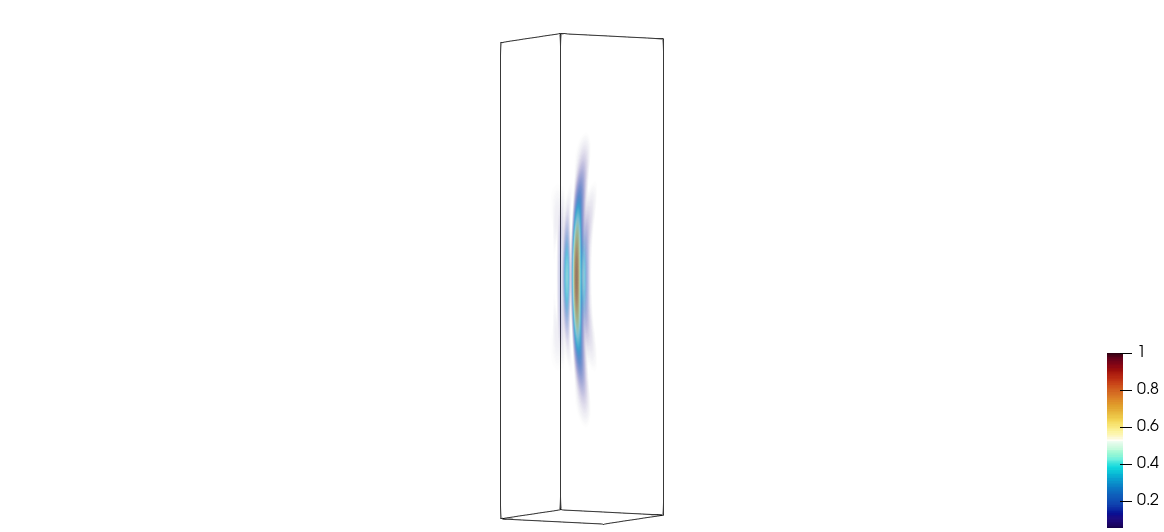}}
\end{picture}
\caption{Intensity distribution of Airy-like vector beams. Vector beam $\textbf{M}$ (a-d) spatial distributions of the intensity with decay factors $a = 0.05$ (a), $a = 0.1$ (b), $a = 0.2$ (c) and $a = 0.5$ (d), wave number $k = 2 \pi$ and scaling factors $x_0=y_0=1$. The distribution of the vector beam $\textbf{N}$ is very similar.}
\label{fig:MN_3D_profile}
\end{figure}

Further, investigating \textbf{N} type beam's spectra, the decay parameter and the choice of the scaling factor affect the power distribution between the components $\textbf{G}^{N}_z$ and $\textbf{G}^{N}_\rho$, see Figure \ref{fig:paraxial_limit}. This is important when evaluating the paraxiality of the beam. One can notice that at low values of the decay parameter and the scaling factor, the spatial spectra of the generated beam has a large spatial component in the direction of propagation. In contrast, at higher values of these parameters, the component in the direction of propagation decreases rapidly.

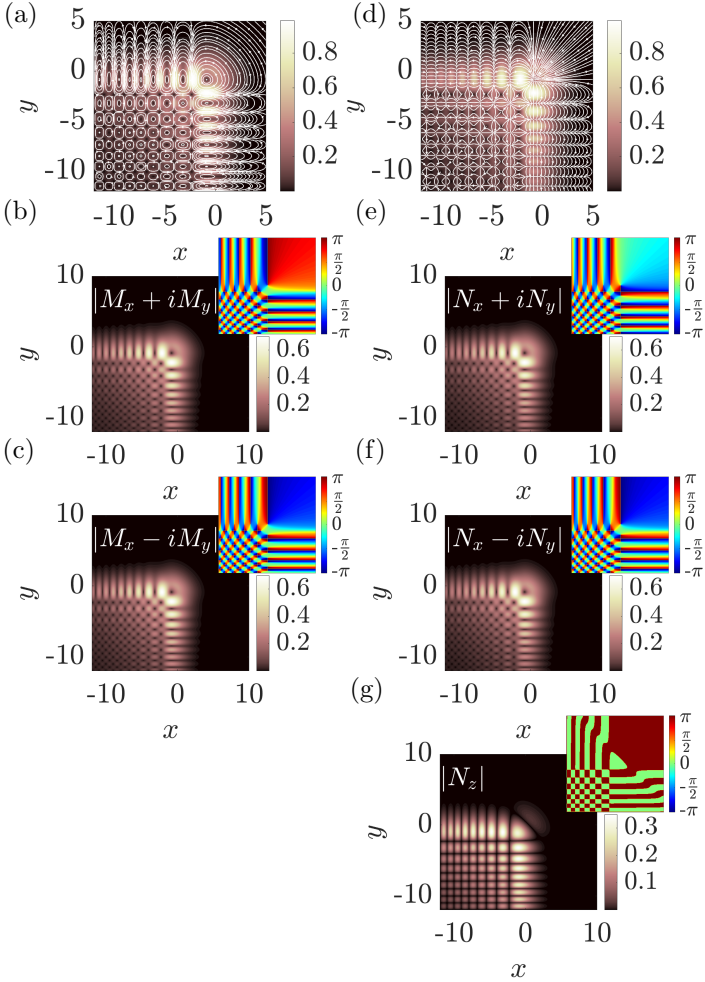
\begin{figure}[t!]
\includegraphics[scale=0.32]{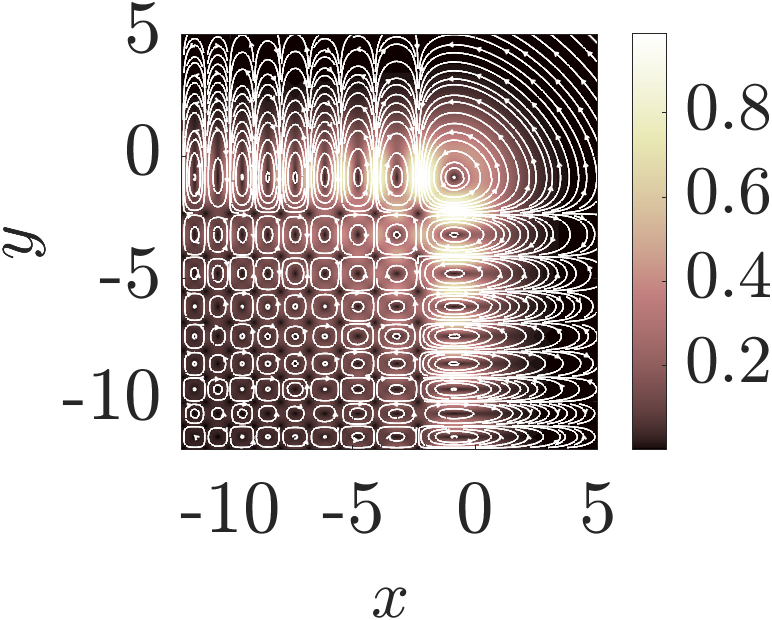}%
\begin{picture}(0,0)
\put(-125,90){(a)}
\end{picture}
\includegraphics[scale=0.32]{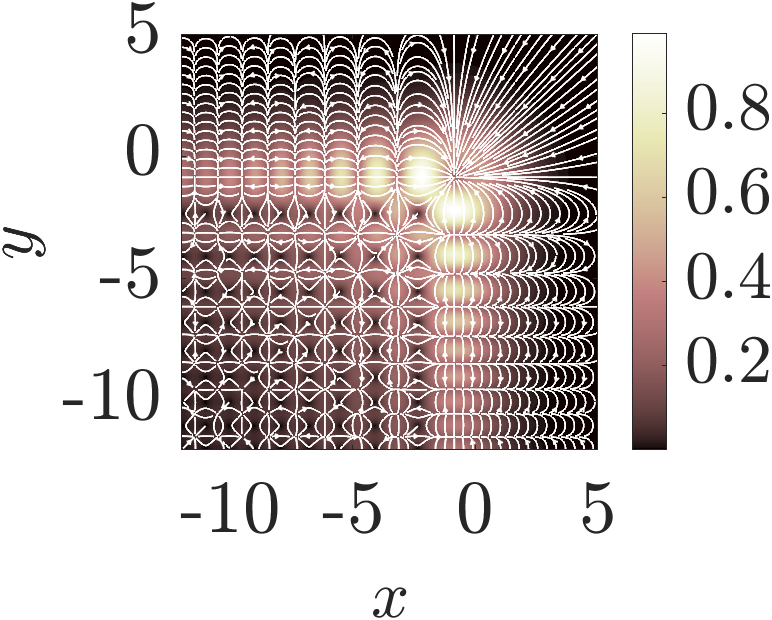}%
\begin{picture}(0,0)
\put(-117,90){(d)}
\end{picture}

\includegraphics[scale=0.3]{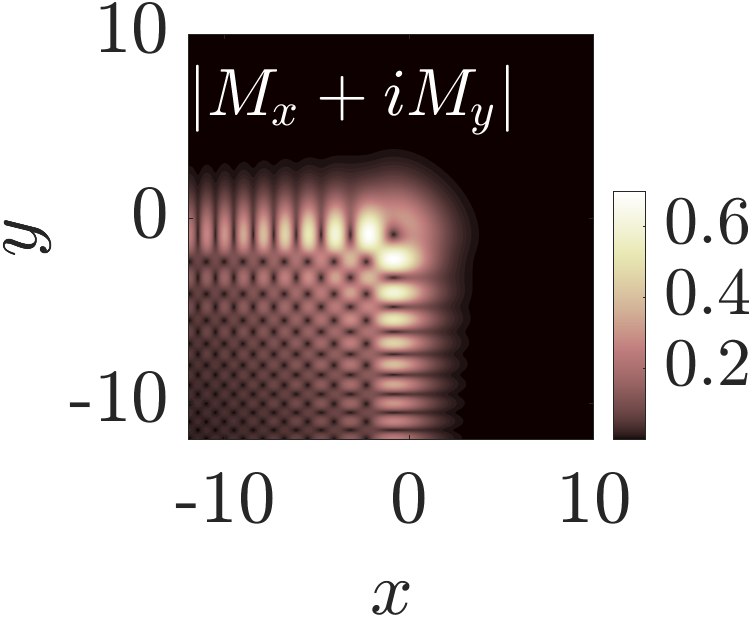}%
\begin{picture}(0,0)
\put(-34,61){\includegraphics[scale=0.14]{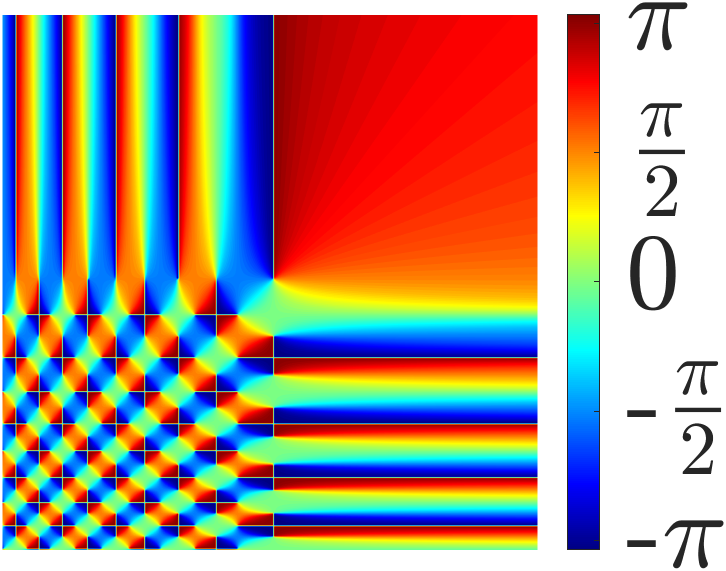}}
\put(-115,106){(b)}
\end{picture}\ \ \ \ \ \ \
\includegraphics[scale=0.3]{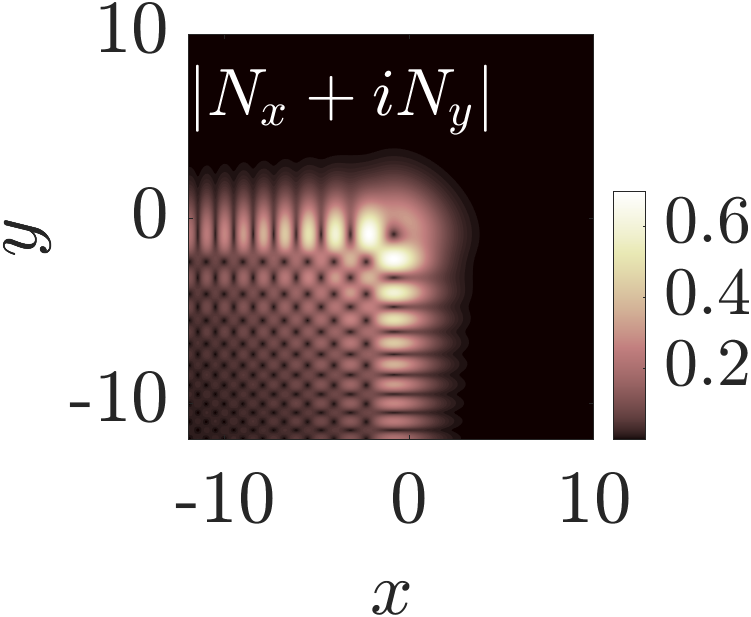}%
\begin{picture}(0,0)
\put(-34,61){\includegraphics[scale=0.14]{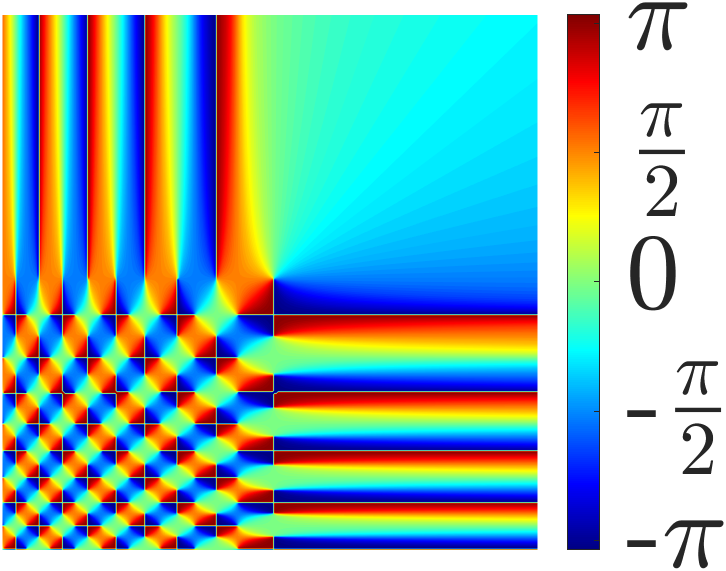}}
\put(-115,106){(e)}
\end{picture}

\includegraphics[scale=0.3]{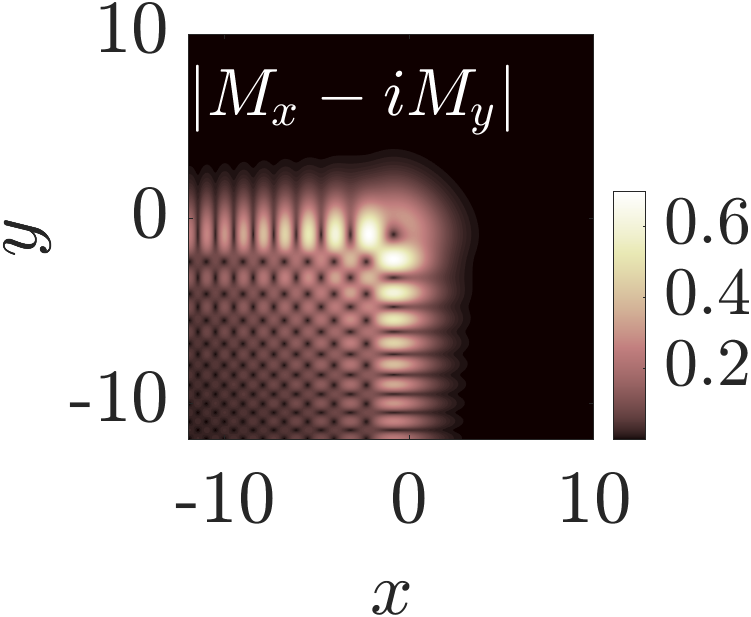}%
\begin{picture}(0,0)
\put(-34,61){\includegraphics[scale=0.14]{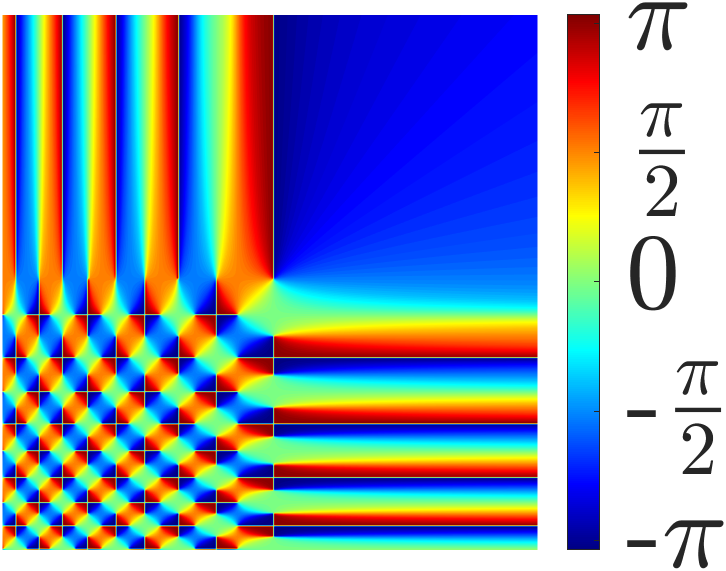}}
\put(-115,106){(c)}
\end{picture}\ \ \ \ \ \ \
\includegraphics[scale=0.3]{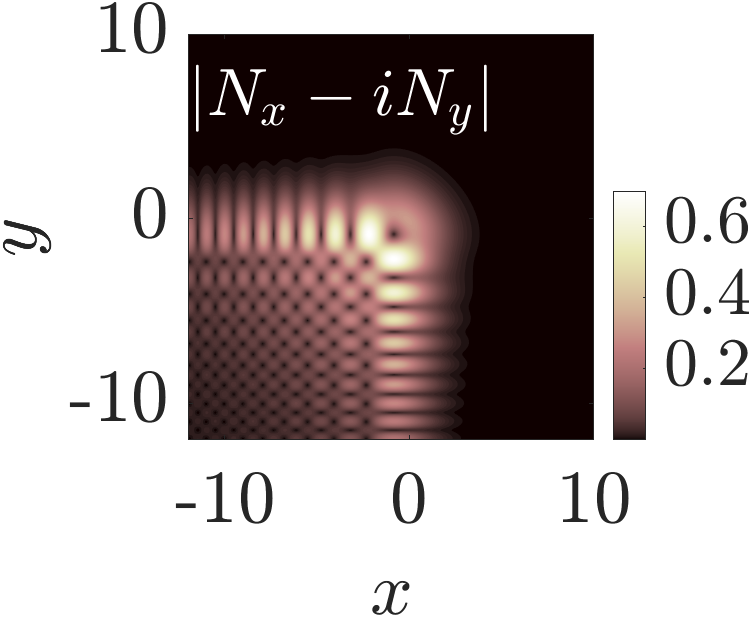}%
\begin{picture}(0,0)
\put(-34,61){\includegraphics[scale=0.14]{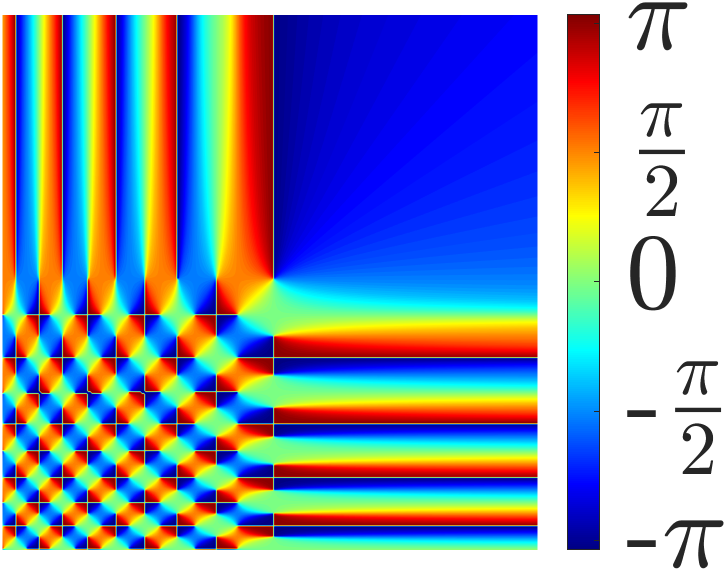}}
\put(-115,106){(f)}
\end{picture}

\qquad \qquad \qquad \qquad \qquad \qquad \quad
\includegraphics[scale=0.3]{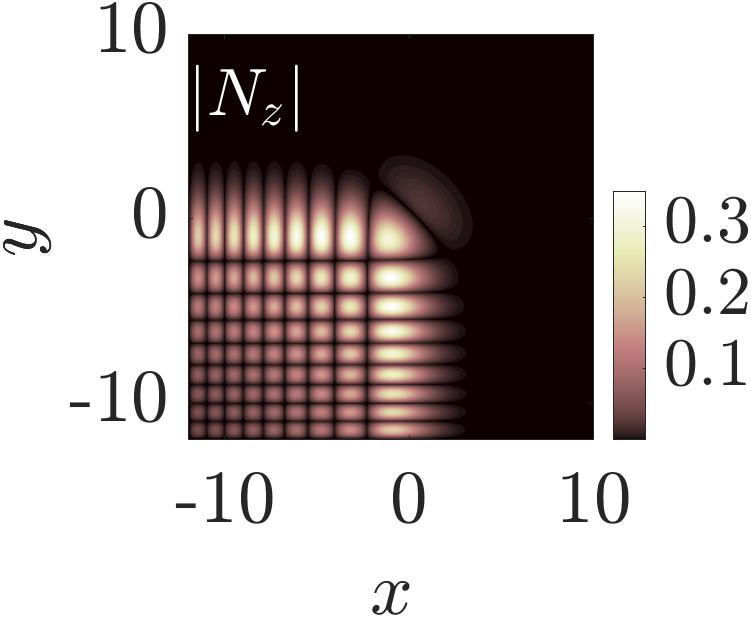}%
\begin{picture}(0,0)
\put(-34,61){\includegraphics[scale=0.14]{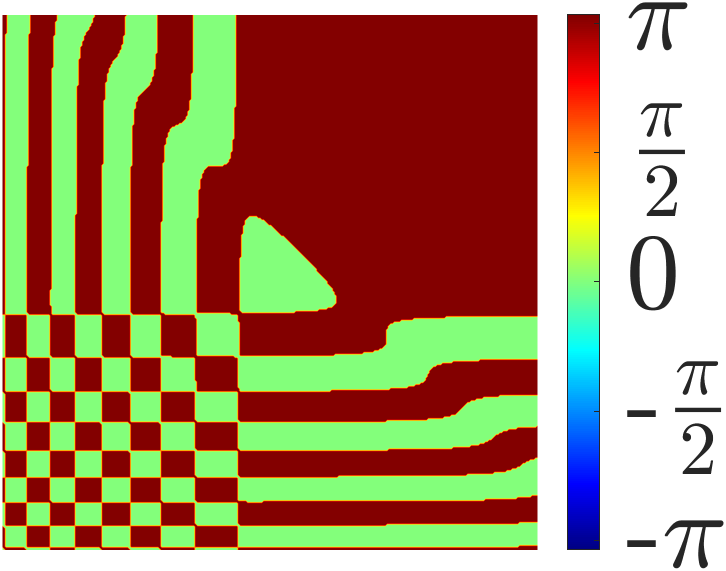}}
\put(-115,106){(g)}
\end{picture}

\caption{(a-c) The absolute value of the amplitude distributions of $\mathbf{M}$ and (d-g) of the $\mathbf{N}$ beam. (a,d) The total intensity at $z=0$, white arrows represent the flow of the vector field. The absolute value of the amplitude distribution of the components (b, e) $\mathbf{e}_+$, (c, f) $\mathbf{e}_-$, and (g) of the component $\mathbf{e}_z$ of the beams. The decay factors $a_x=a_y=0.15$, the normalization distances $x_0=y_0=1$, and the wavenumber $k=2\pi$. Insets: phase distributions of the given components. }
\label{fig:MN_xy}
\end{figure}

Analytical expressions for the Airy TE (\textbf{M}) and TM (\textbf{N}) vectors can be obtained by performing an inverse Fourier transform of equations (\ref{Eq:M_spectra}, \ref{Eq:N_spectra}) or substituting Eq. \ref{Eq:Airy_scalar} into Eq. \ref{Eq:LMN}. The resulting expressions for both vector fields are lengthy, so for simplicity reasons, they are expressed in simplified form as
\begin{equation}\label{Eq:M_beam}
\textbf{M}(\textbf{r})=\left( \mathbf{e}_x u_x u_y^y - \mathbf{e}_y u_x^x u_y \right).
\end{equation}
Analogically, the transverse magnetic vector of the Airy beam is given by 
\begin{equation}\label{Eq:N_beam}
\begin{aligned}
\mathbf{N}(\mathbf{r})&=\frac{1}{k}\left[\begin{array}{l}
(\mathrm{i} ku_y + u_y^z) u_x^x + u_x^{xz} u_y \\
(\mathrm{i} ku_x + u_x^z) u_y^y + u_x u_y^{yz} \\
- u_x u_y^{yy} - u_x^{xx} u_y
\end{array}\right]. \\
\end{aligned}
\end{equation}
Here, the subscript indicates the functions $u_x$ and $u_y$ of Eq. \ref{Eq:Airy_scalar} and the superscript indicates partial derivatives with respect to $x$, $y$ or $z$.

The field intensity distributions, which are equal for both $\textbf{M}$ and $\textbf{N}$ type beams, are shown in Figure \ref{fig:MN_3D_profile}. The parameters chosen for this simulation are the decay factor $\{a_x,a_y\} = \{0.05, 0.1, 0.2, 0.5\}$, the scaling factor $\{x_0, y_0\} = 1$, and the wavenumber $k=2\pi$. It can be seen that, at the lower values of the decay factor, the Airy beam has a more profound intensity distribution similar to that of the conventional scalar Airy beam compared to higher values of the decay factor. Another distinct feature of the azimuthal and radial vector Airy beams compared to a traditional scalar is that, in the focal section of the beam, the intensity distribution has two peaks and a polarization singularity between them. This can be explained because in the spectra of both beams, the doughnut-shaped spectral intensity leaves out the middle part responsible for the beam's intensity at the focus, producing a radial or azimuthal polarization intensity minimum of the beam. When the decay factor is larger, both beams are concentrated in the focal region, leaving out higher spatial frequencies, which leads to the beams losing their curved-shaped intensity distribution.
\begin{figure}
\includegraphics[scale=0.33]{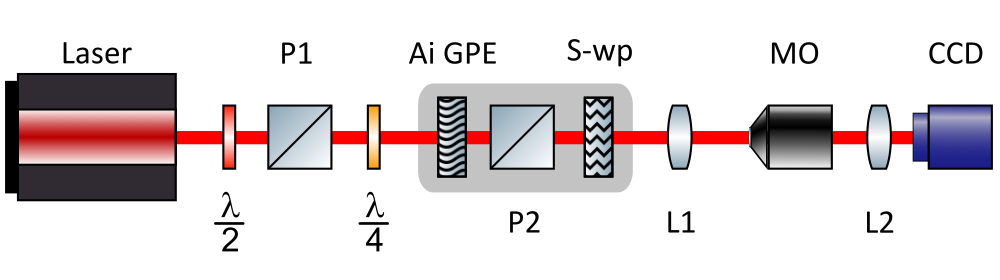}%
\caption{
The optical schematic for radially-like and azimuthally-like polarized Airy beam generation. $\lambda/2$ and $\lambda/4$ represent half and quarter-waveplates, P1 and P2 - polarizers, Ai GPE - Airy geometrical phase element, S-wp - S-waveplate, MO - magnifying objective, CCD - charge-coupled device camera.
}
\label{fig:OptSch}
\end{figure}

\begin{figure*}\centering
\includegraphics[scale=0.40, trim={0 0 0 0}, clip]{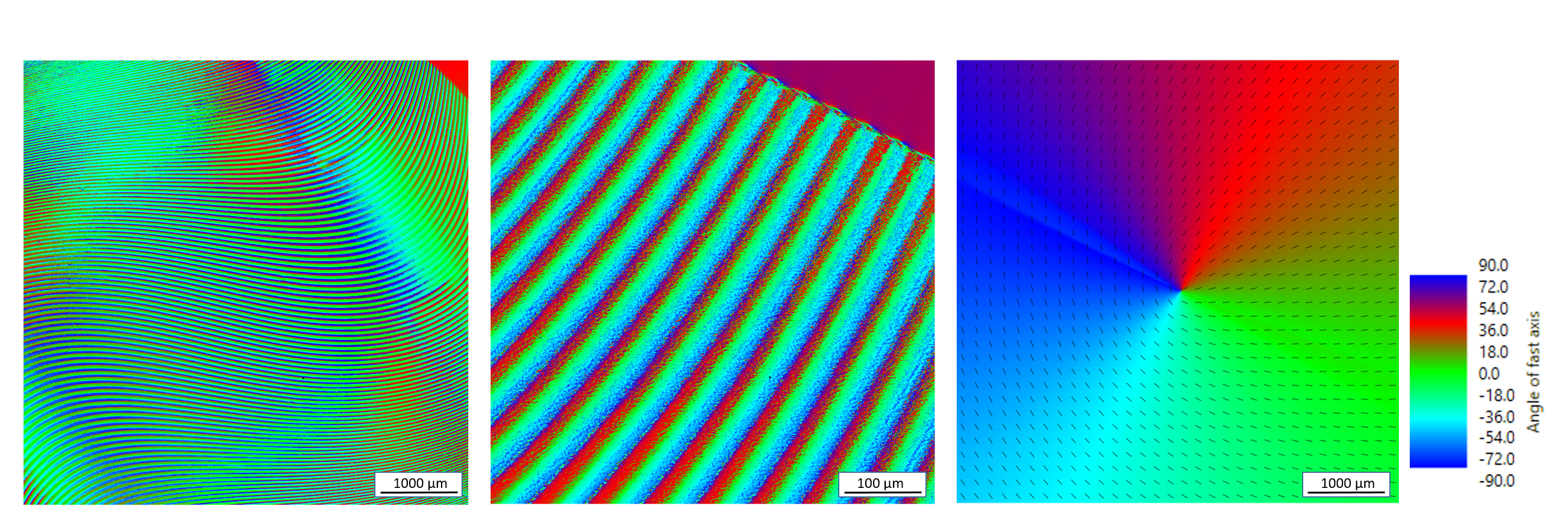}%
\begin{picture}(0,0)
\put(-500,160){(a)}
\put(-350,160){(b)}
\put(-200,160){(c)}
\end{picture}
\caption{Experimentally obtained fast axis angle distributions of the femtosecond laser inscribed geometrical phase elements. (a) Ai GPE element, (b) Ai GPE element magnified $\times\ 10$ compared to the (a), and (c) S-wp element, the lines here denote the fast axis angle.}
\label{fig:Measurments_azimuth}
\end{figure*}
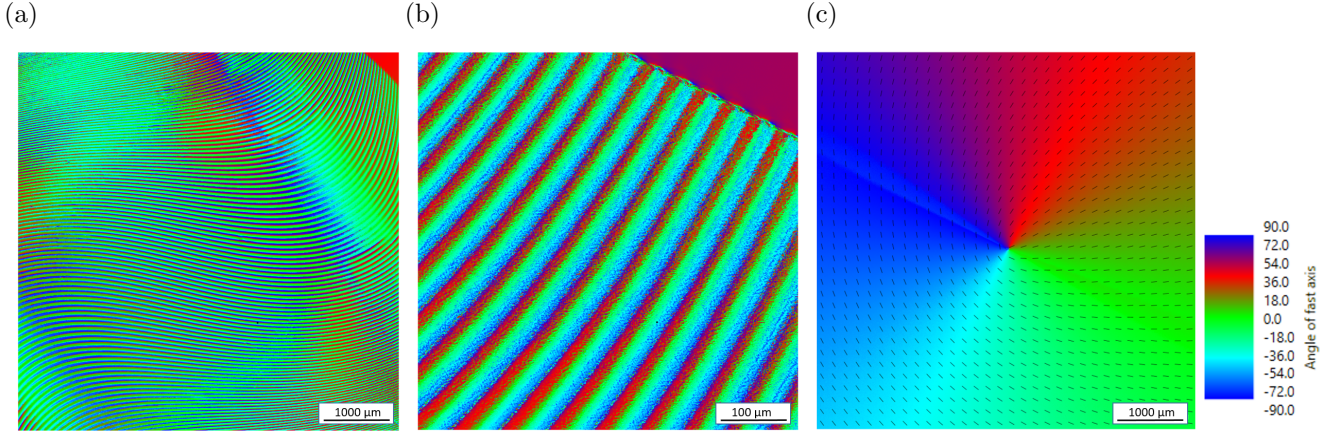

Another important feature of such beams is their spatial intensity distribution for different field components. The absolute value of the amplitude of the beam type $\mathbf{M}$ and $\mathbf{N}$ in the focal plane is shown in Figure \ref{fig:MN_xy}, with the decay factor $\{a_x,a_y\} = 0.15$, the scaling factor $\{x_0, y_0\} = 1$ and the wavenumber $k=2\pi$. In Figure \ref{fig:MN_xy} (a), we observe characteristics of the azimuthally polarized beam, a closed contour electric field that streamlines in the TE mode, which encloses the vector singularities of the beam. The vector field $\mathbf{M}$ in the focal plane is purely linear but nonuniformly polarized. The field's amplitude absolute value distribution resembles a scalar Airy beam with minima splitting the main lobe where the singularity is located. The vector field of type $\mathbf{M}$ in the basis of the circular polarization component is shown in Figures \ref{fig:MN_xy} (b,c). It can be seen that although both components have equal field distributions in the absolute value of the amplitude, they are antiphase to each other and $arg(M_+)+arg(M_-)=0$, where $M_+=M_x+iM_y$ and $M_+=M_x-iM_y$.

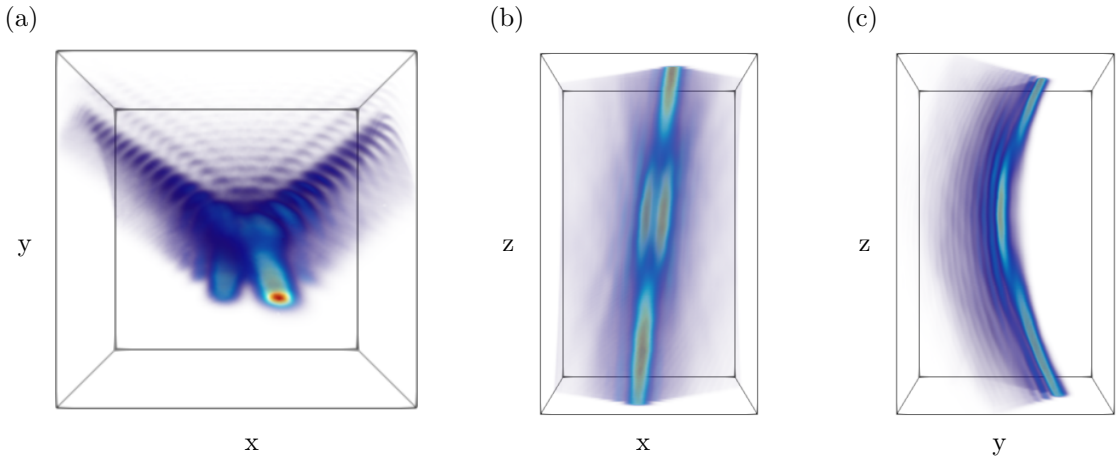
\begin{figure*}[t!]\centering
\includegraphics[scale=0.45, trim={0 0 0 0}, clip]{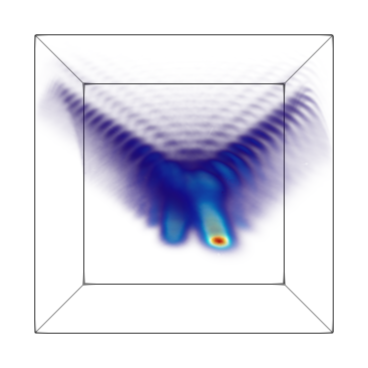}%
\begin{picture}(0,0)
\put(-170,160){(a)}
\put(-165,75){y}
\put(-80,0){x}
\end{picture}\hspace*{3pt}
\includegraphics[scale=0.3, trim={0 0 0 0}, clip]{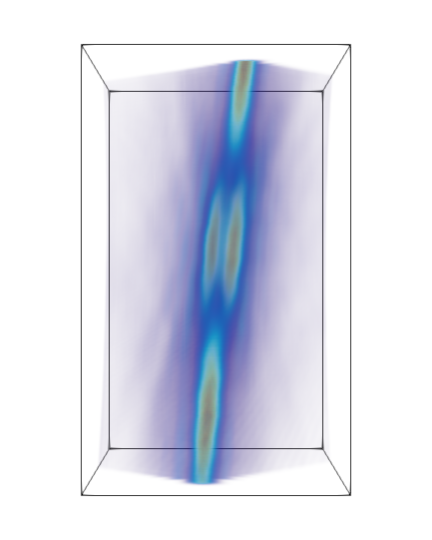}%
\begin{picture}(0,0)
\put(-125,160){(b)}
\put(-70,0){x}
\put(-120,75){z}
\end{picture}
\includegraphics[scale=0.3, trim={0 0 0 0}, clip]{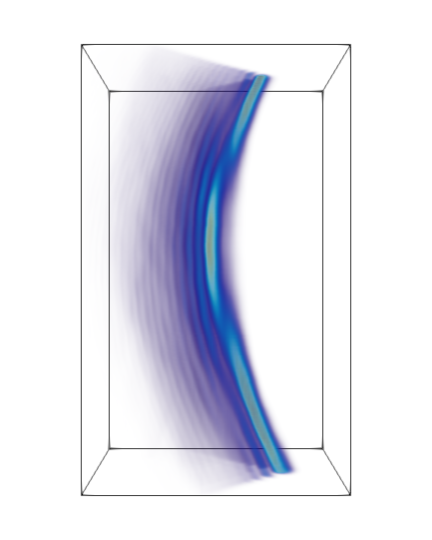}%
\begin{picture}(0,0)
\put(-125,160){(c)}
\put(-70,0){y}
\put(-120,75){z}
\end{picture}

\caption{ Intensity distribution of experimentally realized high-power azimuthally polarized vector Airy beam. A Fourier lens of focal distance $f=75\ mm$ was used to generate the beam, the $z$ coordinate spans  in the range of $z=(0\ mm, 49\ mm)$, the range of $x$ coordinate $x=(0\ mm,0.6\ mm)$, and the $z$ coordinate changes in $y=(0\ mm,0.8\ mm)$.}
\label{fig:Measurments_int}
\end{figure*}

The vector beam $\mathbf{N}$, shown in Figure \ref{fig:MN_xy} (d), is at each point in the focal plane linearly polarized but nonuniformly oriented resembles a radially polarized beam, known from the literature \cite{dorn2003sharper}. The circular components depicted in Figure \ref{fig:MN_xy} (e,f) have the same distribution of the absolute value of the amplitude as in the previous case, and the phases are shifted by $\pi$, so $arg(N_+e^{i\pi})+arg(N_-)=0$. We see that in the given example in Figure \ref{fig:MN_xy} (d) the flow of the electric field streamlines converges to locations of polarization singularities, at these points the electric field is orientated perpendicularly to the focal plane and has component orientated along the propagation direction; see Figure \ref{fig:MN_xy} (g). At some of these points, we observe the electric field oriented both outwards and inward perpendicular to the focal plane, and the direction can be deduced from the phase of the $N_z$ component. Although both beams $\mathbf{M}$ and $\mathbf{N}$ have similar spatial distributions, differences in electric field orientations are caused by phase differences between the components $M_+$, $M_-$ and $N_+$, $N_-$ and by the presence of the $z$ component of the beam of type $\mathbf{N}$.
\begin{figure}[b!]
\includegraphics[scale=0.31, trim={0 80 0 0}, clip]{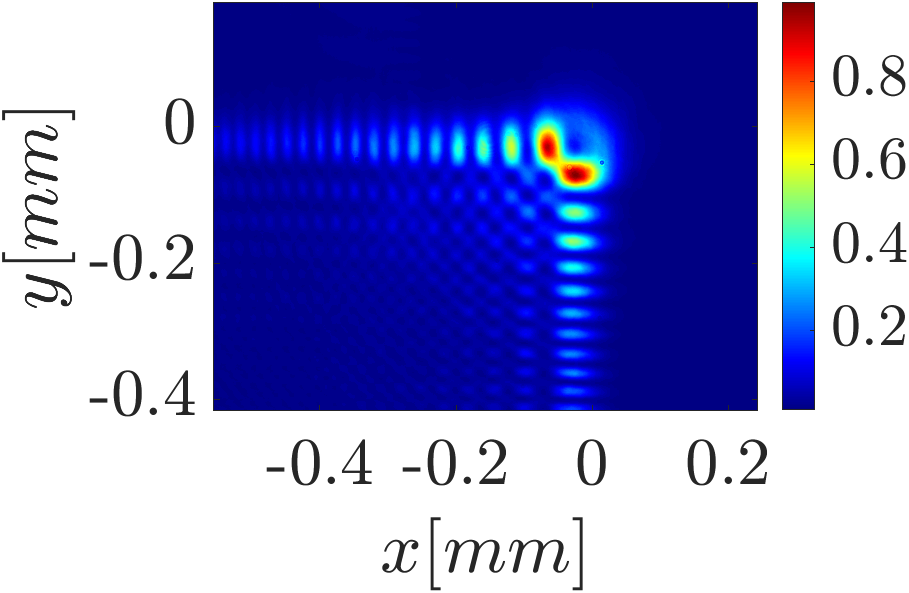}%
\begin{picture}(0,0)
\put(-95,10){\textcolor{white}{$S_0$}}
\end{picture}\hspace*{3pt}
\includegraphics[scale=0.31, trim={100 80 0 0}, clip]{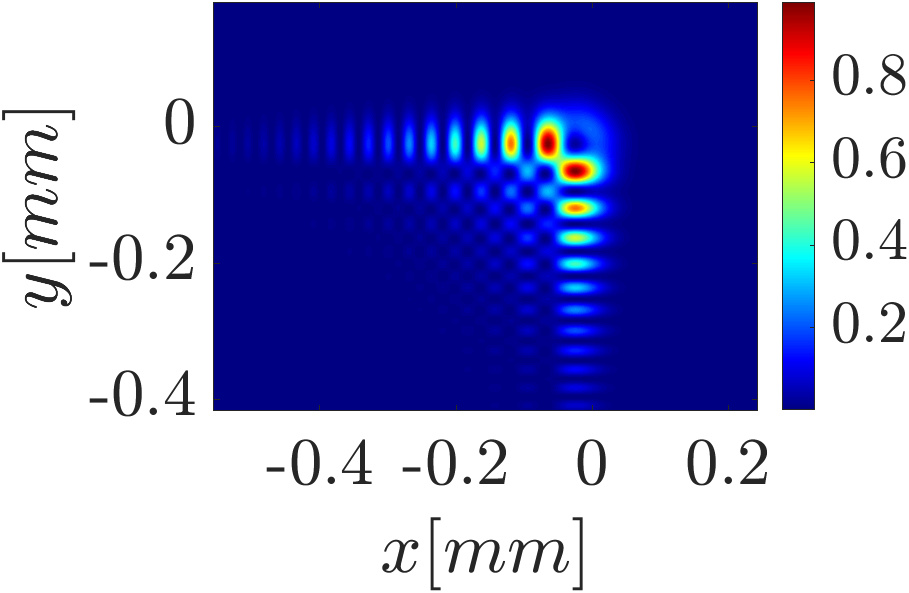}%
\begin{picture}(0,0)
\put(-95,10){\textcolor{white}{$S_0$}}
\end{picture}

\includegraphics[scale=0.31, trim={0 80 0 0}, clip]{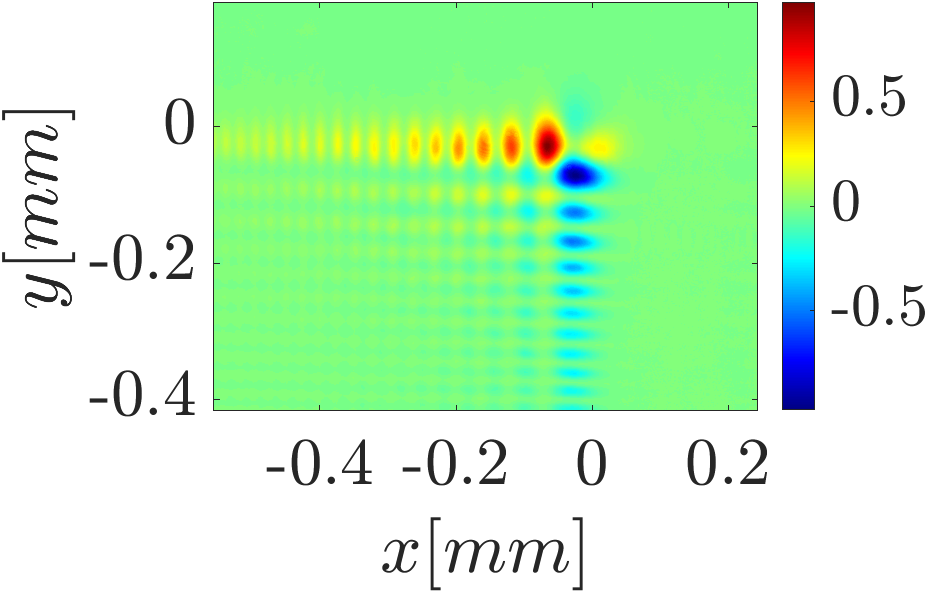}%
\begin{picture}(0,0)
\put(-95,10){$S_1$}
\end{picture}\hspace*{0.3pt}
\includegraphics[scale=0.31, trim={100 80 0 0}, clip]{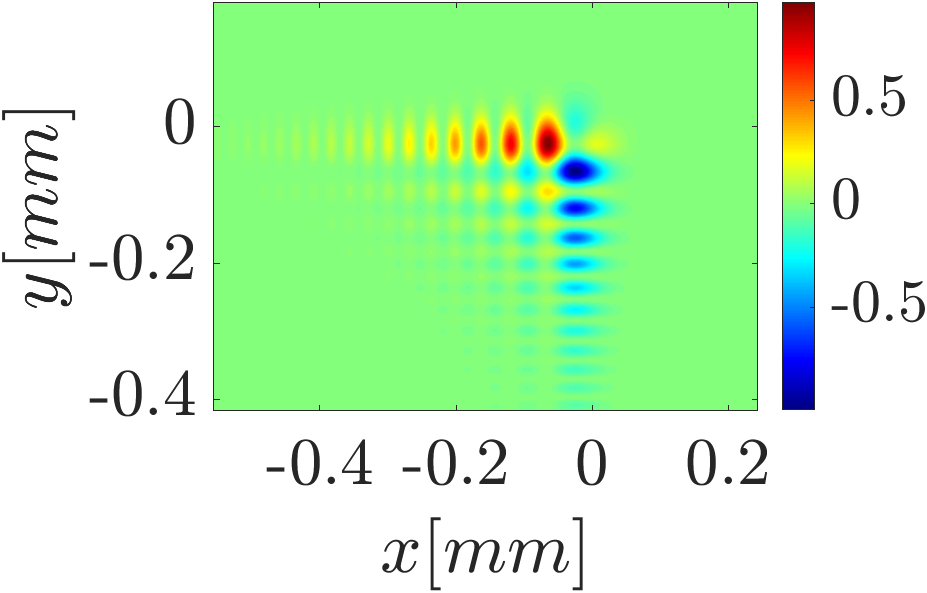}%
\begin{picture}(0,0)
\put(-95,10){$S_1$}
\end{picture}

\includegraphics[scale=0.31, trim={0 80 0 0}, clip]{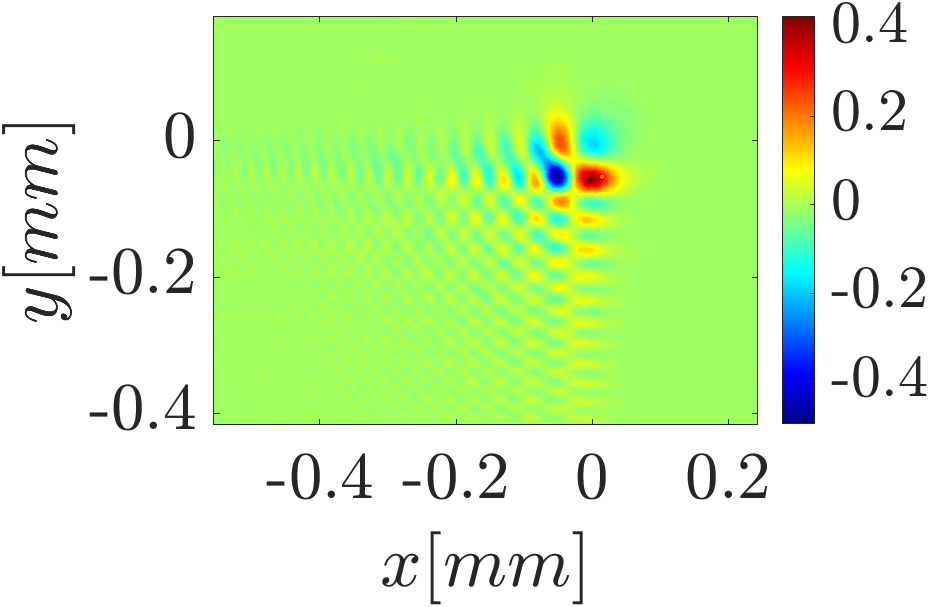}%
\begin{picture}(0,0)
\put(-95,10){$S_2$}
\end{picture}
\includegraphics[scale=0.31, trim={100 80 0 0}, clip]{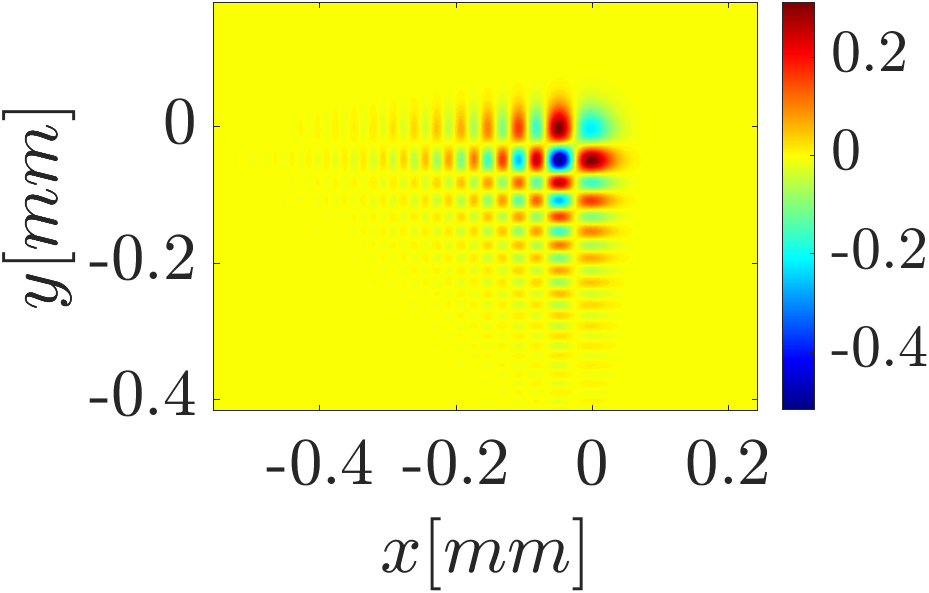}%
\begin{picture}(0,0)
\put(-95,10){$S_2$}
\end{picture}

\includegraphics[scale=0.31, trim={0 0 0 0}, clip]{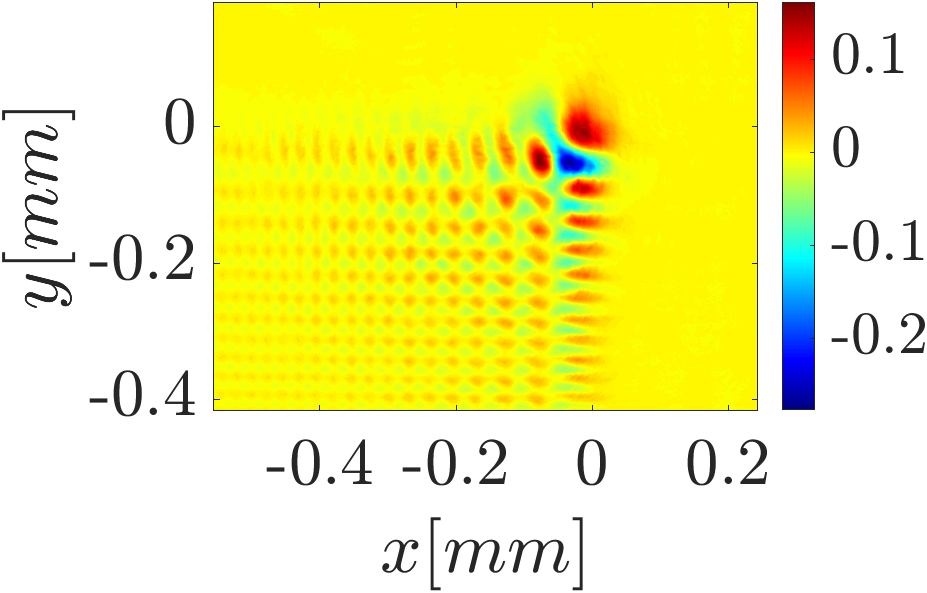}%
\begin{picture}(0,0)
\put(-95,35){$S_3$}
\end{picture}\hspace*{0.3pt}
\includegraphics[scale=0.31, trim={100 0 0 0}, clip]{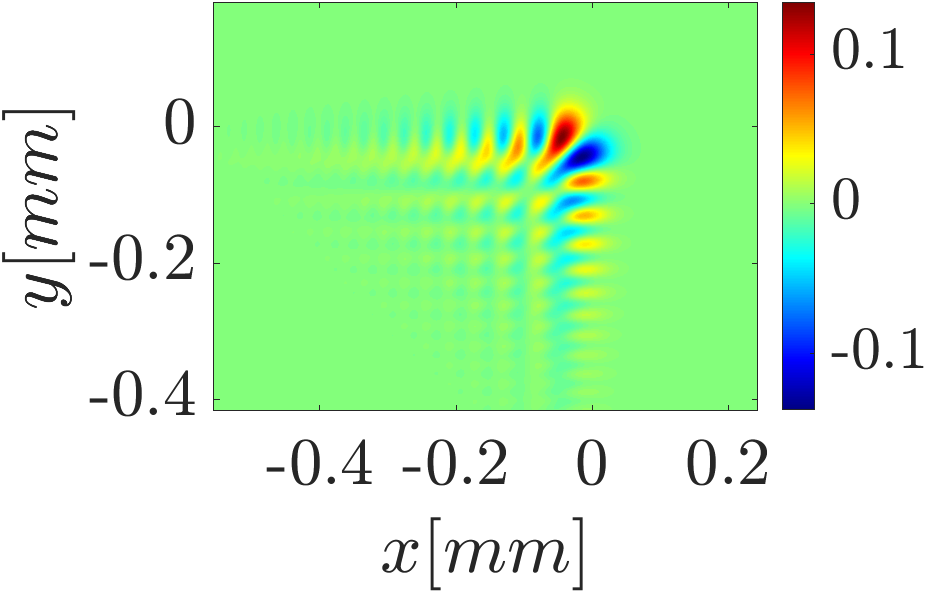}%
\begin{picture}(0,0)
\put(-95,35){$S_3$}
\end{picture}

\caption{ Stokes vector distributions for the $\mathbf{M}$ type beam. Left column: experimental results, right column numerical simulation. The parameters of the simulation are: the decay factor $a_x=a_y=0.15$, the normalization distances $x_0=y_0=30\ \mu m$, distance from the focus $z=-2\ mm$, and the wavelength $\lambda=1028\ nm$. }
\label{fig:Stokes_M}
\end{figure}
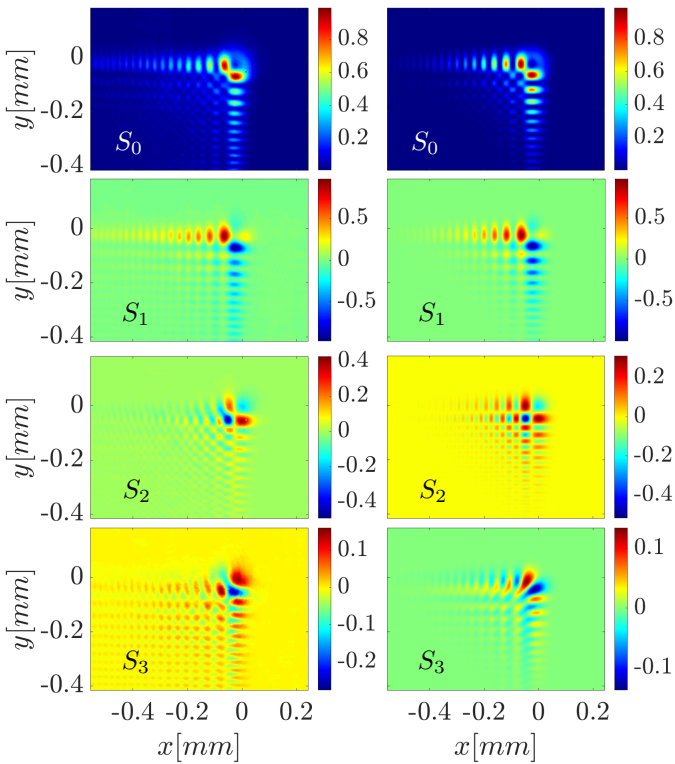
We continue the discussion by recalling the Jones matrix formalism for the experimental verification of our simulations. Jones calculus is a comfortable notation in which polarized fields are described by Jones vectors, and linear optical elements by Jones matrices. An element that converts linearly polarized light to azimuthally or radially nonuniformly polarized beams by manipulating the phases of the incoming wave is called the S-waveplate element. The Jones matrix for the S-waveplate element \cite{beresna2011polarization} is described by the
\begin{equation}
\begin{aligned}
J&=\left[\begin{array}{l}
\cos{\theta} \quad \sin{\theta} \\
\sin{\theta} \quad -\cos{\theta}
\end{array}\right]. \\
\end{aligned}
\end{equation}
where $\theta$ is the azimuthal angle.
Interestingly, an incoming linearly polarized beam $E^x_{in} = g\textbf{e}_x$ or $E^y_{in} = ig\textbf{e}_y$, when interacting with such an element, produces transverse electric or transverse magnetic modes.
\begin{equation}
\begin{aligned}
JE^y_{in} =ig\left[\begin{array}{l}
\sin{\theta} \\
-\cos{\theta}
\end{array}\right]=\frac{1}{k}\textbf{G}^{M}(k_{x}, k_{y}).
\end{aligned}
\label{eq:Spectra_M}
\end{equation}

\begin{equation}
\begin{aligned}
JE^x_{in} =g\left[\begin{array}{l}
\cos{\theta} \\
\sin{\theta}
\end{array}\right]\approx \textbf{G}^{N}(k_{x}, k_{y}).
\end{aligned}
\label{eq:Spectra_N}
\end{equation}
In general, the S-waveplate element acts as an element that transforms any linearly polarized scalar beam spectra into two vectorial spectral wave solutions $\textbf{M}$ given by Eq. \ref{eq:Spectra_M} and $\textbf{N}$ by Eq. \ref{eq:Spectra_N}, when vector beams are constructed from a constant vector $\textbf{e}_z$.

\section{Experimental verification of the concept}
\begin{figure}[t!]
\includegraphics[scale=0.31, trim={0 80 0 0}, clip]{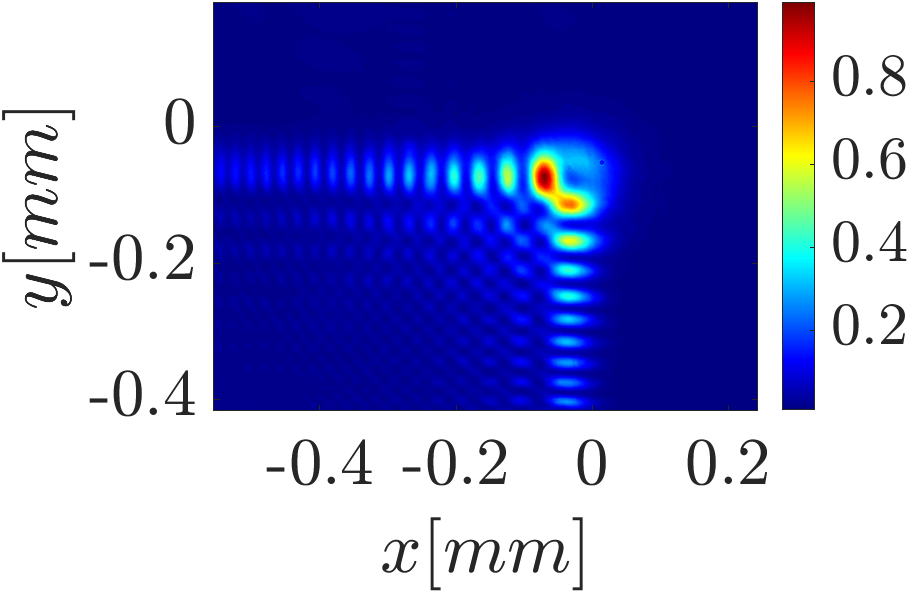}%
\begin{picture}(0,0)
\put(-95,10){\textcolor{white}{$S_0$}}
\end{picture}\hspace*{3pt}
\includegraphics[scale=0.31, trim={100 80 0 0}, clip]{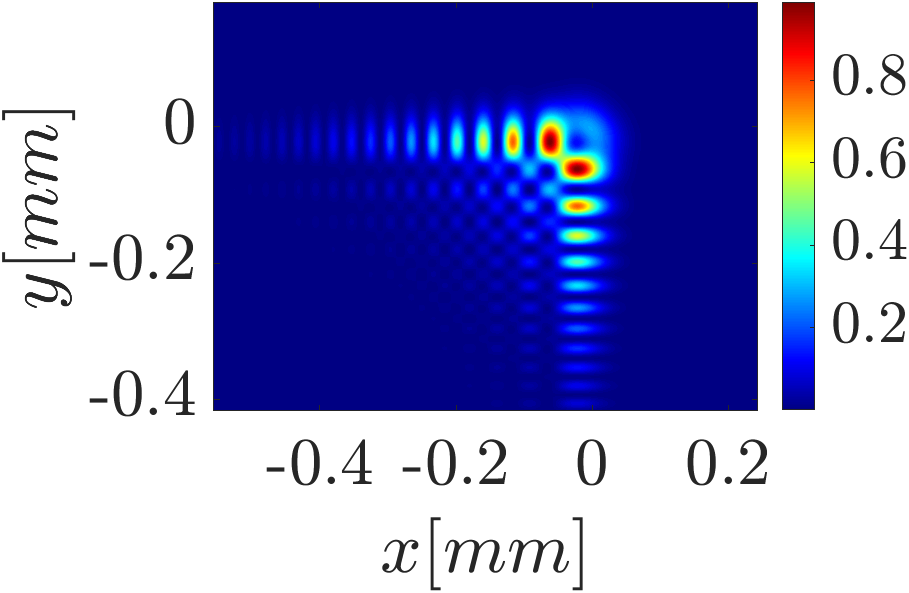}%
\begin{picture}(0,0)
\put(-95,10){\textcolor{white}{$S_0$}}
\end{picture}

\includegraphics[scale=0.31, trim={0 80 0 0}, clip]{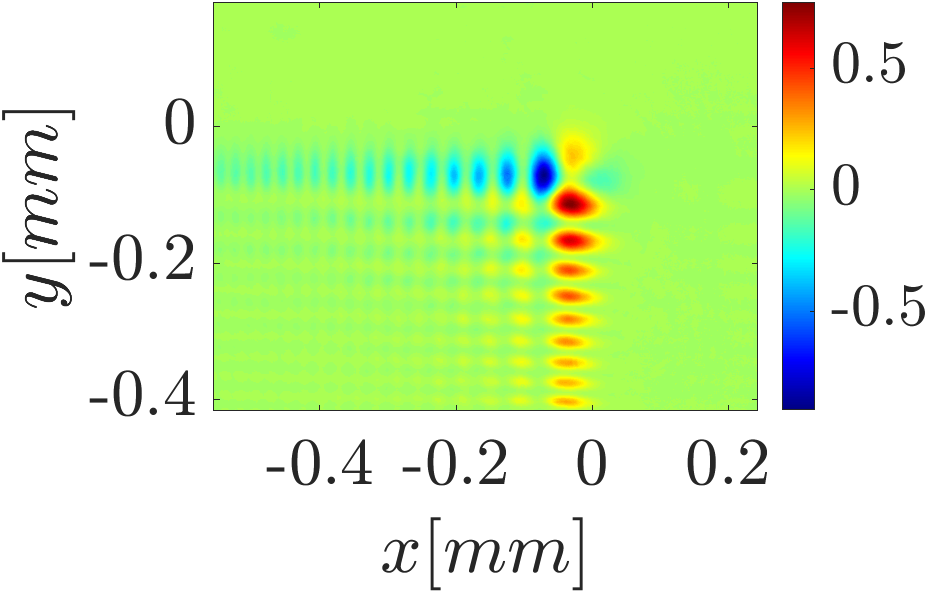}%
\begin{picture}(0,0)
\put(-95,10){$S_1$}
\end{picture}\hspace*{0.3pt}
\includegraphics[scale=0.31, trim={100 80 0 0}, clip]{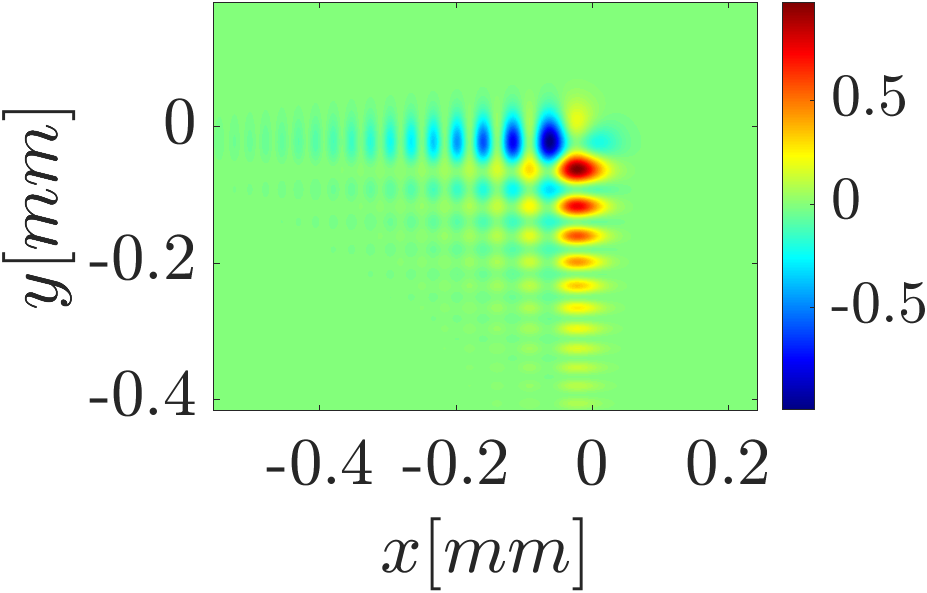}%
\begin{picture}(0,0)
\put(-95,10){$S_1$}
\end{picture}

\includegraphics[scale=0.31, trim={0 80 0 0}, clip]{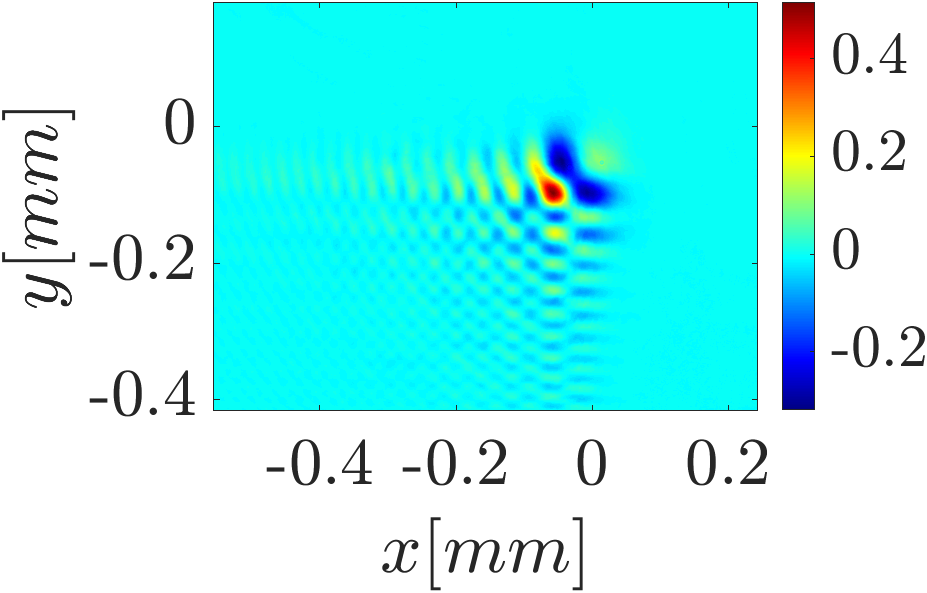}%
\begin{picture}(0,0)
\put(-95,10){$S_2$}
\end{picture}
\includegraphics[scale=0.31, trim={100 80 0 0}, clip]{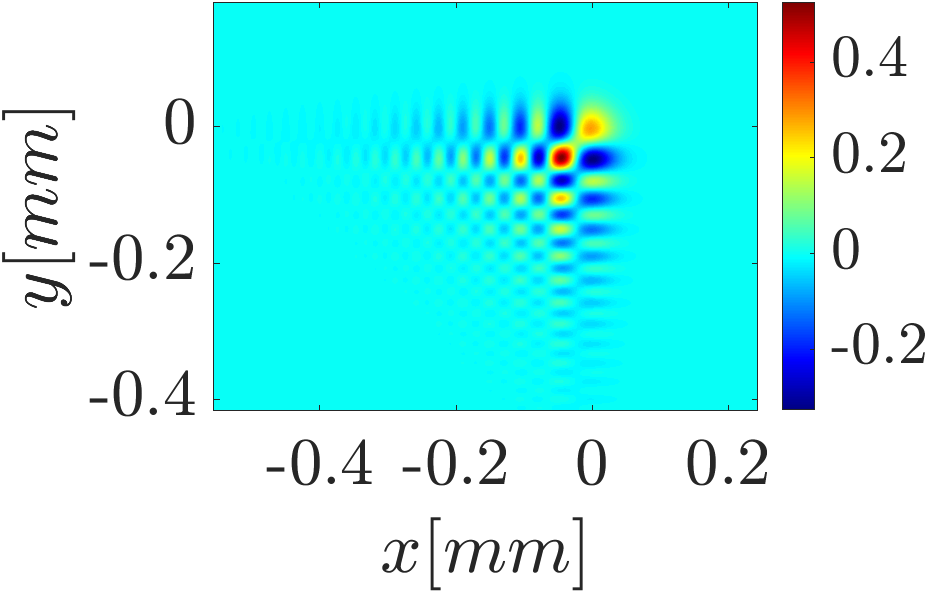}%
\begin{picture}(0,0)
\put(-95,10){$S_2$}
\end{picture}

\includegraphics[scale=0.31, trim={0 0 0 0}, clip]{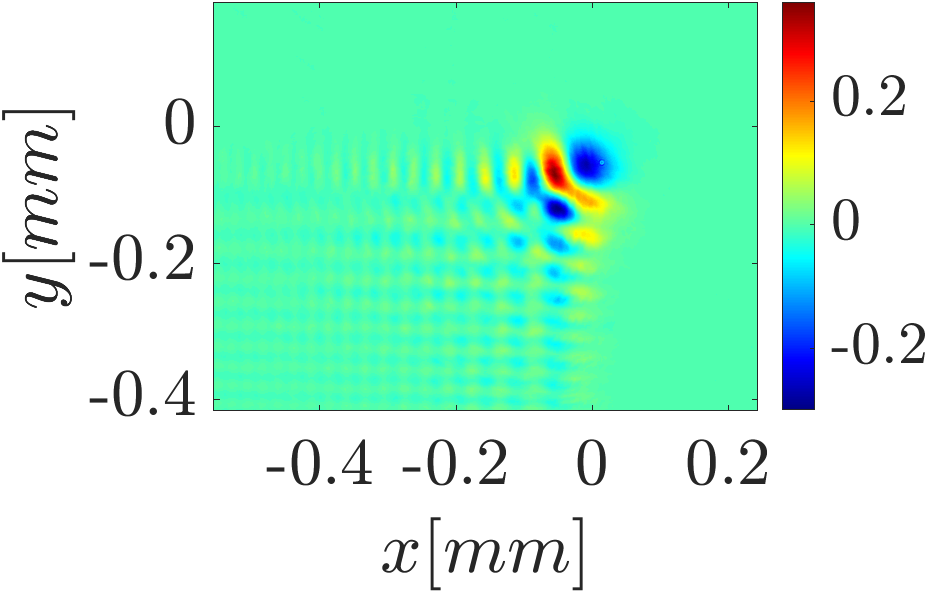}%
\begin{picture}(0,0)
\put(-95,35){$S_3$}
\end{picture}\hspace*{0.3pt}
\includegraphics[scale=0.31, trim={100 0 0 0}, clip]{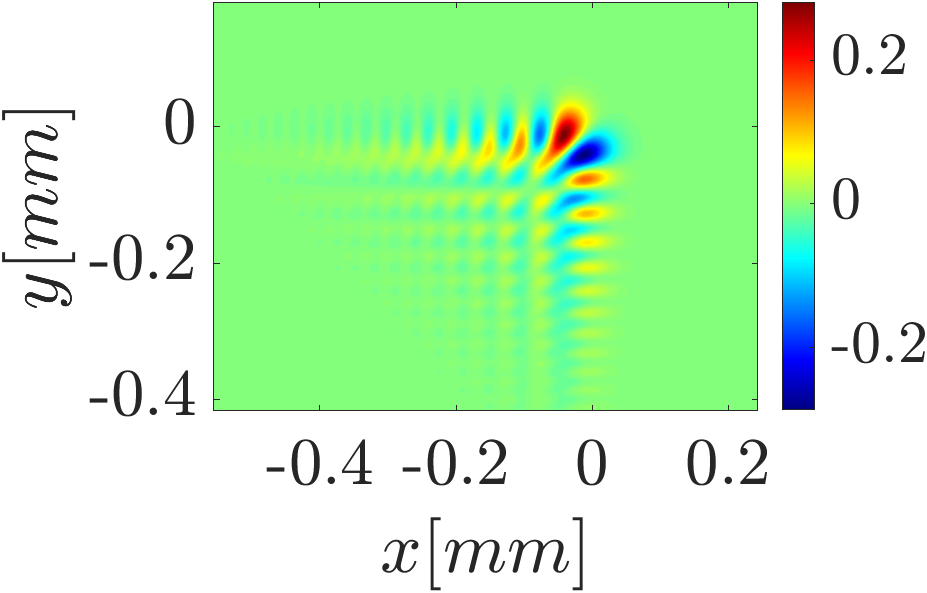}%
\begin{picture}(0,0)
\put(-95,35){$S_3$}
\end{picture}

\caption{ Stokes vector intensity distributions for the $\mathbf{N}$ type beam. Left column: experimental results, right column numerical simulation. The simulation parameters are the decay factor $a_x=a_y=0.15$, the normalization distances $x_0=y_0=30\ \mu m$, the distance from the focus $z=-4\ mm$, and the wavelength $\lambda=1028\ nm$. }
\label{fig:Stokes_N}
\end{figure}
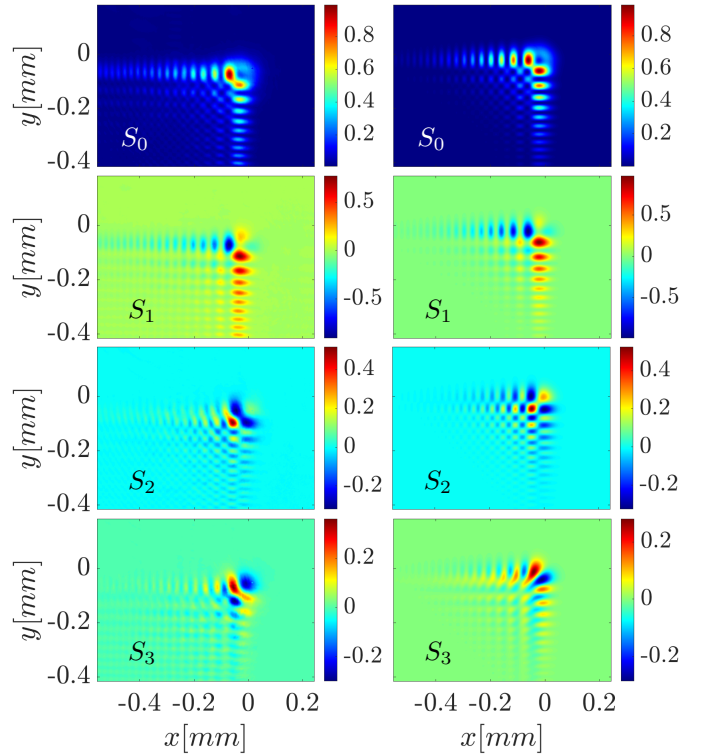
For the validation of the concept described in the previous Section and the creation of high-power ultrafast non-homogeneously polarized Airy beams, an experiment was conducted, High average power ultra-short pulsed laser ("Pharos", Light Conversion) was used. This laser generates pulses as short as $250\ fs$ with the ability to control the duration up to $20\ ps$, while the maximum pulse energy can be $600\ \mu J$ and $20\ W$ power. In this setup, the laser generates a $4.2\ mm$ width (at $1/e^2$ intensity) Gaussian beam of $M^2 < 1.3$. The power of the laser was controlled using a motorized external attenuator, consisting of a rotating half-waveplate and a pair of Brewster polarizers. The optical schematic is presented in Figure \ref{fig:OptSch}. Firstly, a linearly polarized Gaussian beam is emitted at wavelength $\lambda=1028\ nm$. The half-waveplate and polarizer (P1), together, work as a power attenuator, and the quarter-waveplate is used to prepare a circularly polarized Gaussian beam for the Airy geometrical phase element (Ai GPE), see Fig. \ref{fig:Measurments_azimuth}(a,b),  to produce an Airy phase distribution. After the Gaussian beam propagates through an Ai GPE element, the beam's polarization remains circular with additional cubic phase distribution. An appearance of the difference in the phase is rooted in the phenomena that were discovered by S. Pancharatnam and M. Berry \cite{pancharatnam1956generalized,berry1987adiabatic}. When two states of polarization are achieved from the initial polarization state a geodesic triangle can be drawn on the surface of the Poincare sphere and half of the encompassed solid angle of the sphere is the so-called geometrical angle. The concept was further investigated not only in the time domain but also for transversely inhomogeneous space-invariant beams \cite{bomzon2001pancharatnam}.

To produce an inhomogeneously polarized Airy beam, a Fourier spectra of the beam type $\textbf{M}$ or $\textbf{N}$ must first be obtained. For this purpose, we have a P2 polarizer that converts circularly polarized light to linearly without disturbing the intensity distribution.

Further, depending on the desired vector beam type, the S-wp (S-waveplate, see Fig. \ref{fig:Measurments_azimuth}(c)) element's orientation must be set. For the beam of type $\textbf{M}$, the S-wp element must be rotated by $90^{\circ}$ degrees with respect to the polarizer P2, and for the beam of type $\textbf{N}$ the polarization axis must align with the S-wp element.

In general, the gray-marked area in Figure \ref{fig:OptSch} is the physical representation of Eqs. \ref{eq:Spectra_M} and \ref{eq:Spectra_N}, because each element can be represented in the Jones matrix formalism and the propagation through multiple elements in the mathematical sense is equal to matrix multiplication. Furthermore, to obtain the nonhomogeneously polarized Airy beam, a prepared spectrum is Fourier transformed by the lens (L1), so the element Ai GPE is placed in one focal spot and the magnifying objective in another. After the Fourier spectra are transformed and the vector beam is magnified, it is collected and observed with the CCD camera, see Figure \ref{fig:Measurments_int}.

To ensure that the experimentally observed vector fields are the same as described by Equations \ref{Eq:M_beam} and \ref{Eq:N_beam}, the Stokes parameters were measured. The measurement setup is also similar to the one shown in Figure \ref{fig:OptSch}, but additionally contains a rotating quarter-waveplate and fixed linear polarizer after the L2 collecting lens, see \cite{schaefer2007measuring}. The Stokes parameters of the electromagnetic beams, see \cite{collett2005field}, are described as
\begin{equation}\label{eq:StokesParam}
\begin{aligned}
    S_0 &= E_x E_x^* + E_y E_y^*, \\
    S_1 &= E_x E_x^* - E_y E_y^*, \\
    S_2 &= E_x E_y^* + E_y E_x^*, \\
    S_3 &= i(E_x E_y^* - E_y E_x^*),
\end{aligned}
\end{equation}
where $\mathbf{S}(x,y)=\{S_0,S_1,S_2,S_3\}$ are Stokes parameters, $E_x$ and $E_y$ are electric field components in the $x$ and $y$ direction, the asterisk indicates complex conjugation and $i$ is the imaginary number.

The Stokes vector for analytically calculated and experimentally measured data for the vector beam $\textbf{M}$ is given in Figure \ref{fig:Stokes_M}, the left column of the picture corresponds to the measured data, and the right to the numerical simulations. Numerical calculations for the Stokes parameters of the vector beam $\textbf{M}$ were obtained using Equations \ref{Eq:M_beam} and \ref{eq:StokesParam} and the best agreement with the measurements corresponds to the decay parameter $\{a_x,a_y\}=0.15$, the scaling constant $\{x_0,y_0\}=30\ \mu m$ and the propagation coordinate $z=-2\ mm$ with wavelength $\lambda=1028\ nm$. It can be seen that the parameter $S_0$ that corresponds to the total intensity of the beam agrees very well. The two main lobes of the $\textbf{M}$-type beam are visible, as well as the more peculiar inner structure. Next, from the component $S_1$, which shows the polarized light in the direction of $\textbf{e}_x$ and $\textbf{e}_y$, one can notice that half of the beam has orientation in one direction and another half in the other, which agrees well both the in experiment and analytical calculations. The parameter $S_2$ corresponds to the linearly polarized but rotated by $45^{\circ}$ degrees to the $\textbf{e}_x$ and $\textbf{e}_y$ components. The experimentally obtained components sustain the main features of the analytically calculated ones but also exhibit some intensity redistribution on one "arm" of the beam. Comparison of the calculated and measured Stokes parameters $S_3$, see Figure \ref{fig:Stokes_M}, reveals that the analyzed vector beams at the focal plane are described only by non-uniformly linearly polarized light, without any circular polarization. In case there is a shift in the direction of propagation, when the propagation distance $z \neq 0$, the circularly polarized components are observed. Quantitatively, the intensity of the $S_3$ parameter agrees within the experiment with the numerically simulated one, but slight rotation of the central part is observed, the reasons for that will be discussed later.

Next, we investigate the Stokes parameters of the $\textbf{N}$ type beam which were obtained in the same way as for the $\textbf{M}$ type beam given in Figure \ref{fig:Stokes_N}. The experimentally obtained data is best matched with the analytically calculated one using Equations \ref{Eq:N_beam} and \ref{eq:StokesParam} with the decay parameter $\{a_x,a_y\}=0.15$, the scaling constant $\{x_0,y_0\}=30\ \mu m$ and the propagation coordinate $z=-4\ mm$ with wavelength $\lambda=1028\ nm$. The experimentally obtained Stokes parameters $S_0$ and $S_1$ show good agreement with the numerically predicted ones. Two lobes in the distribution of the parameter $S_2$ in the measurement merge, but in the calculations, a clear gap is seen, although the main lobe is observed both experimentally and numerically. The parameter $S_3$ seems to agree less with a small rotation of the center observed when compared to the numerically evaluated one.
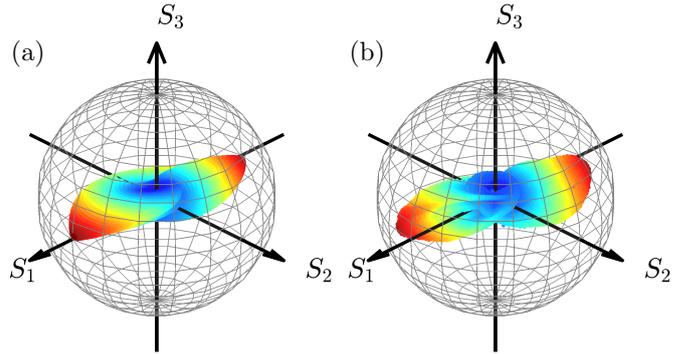
\begin{figure}
\includegraphics[scale=0.65, trim={0 0 0 0}, clip]{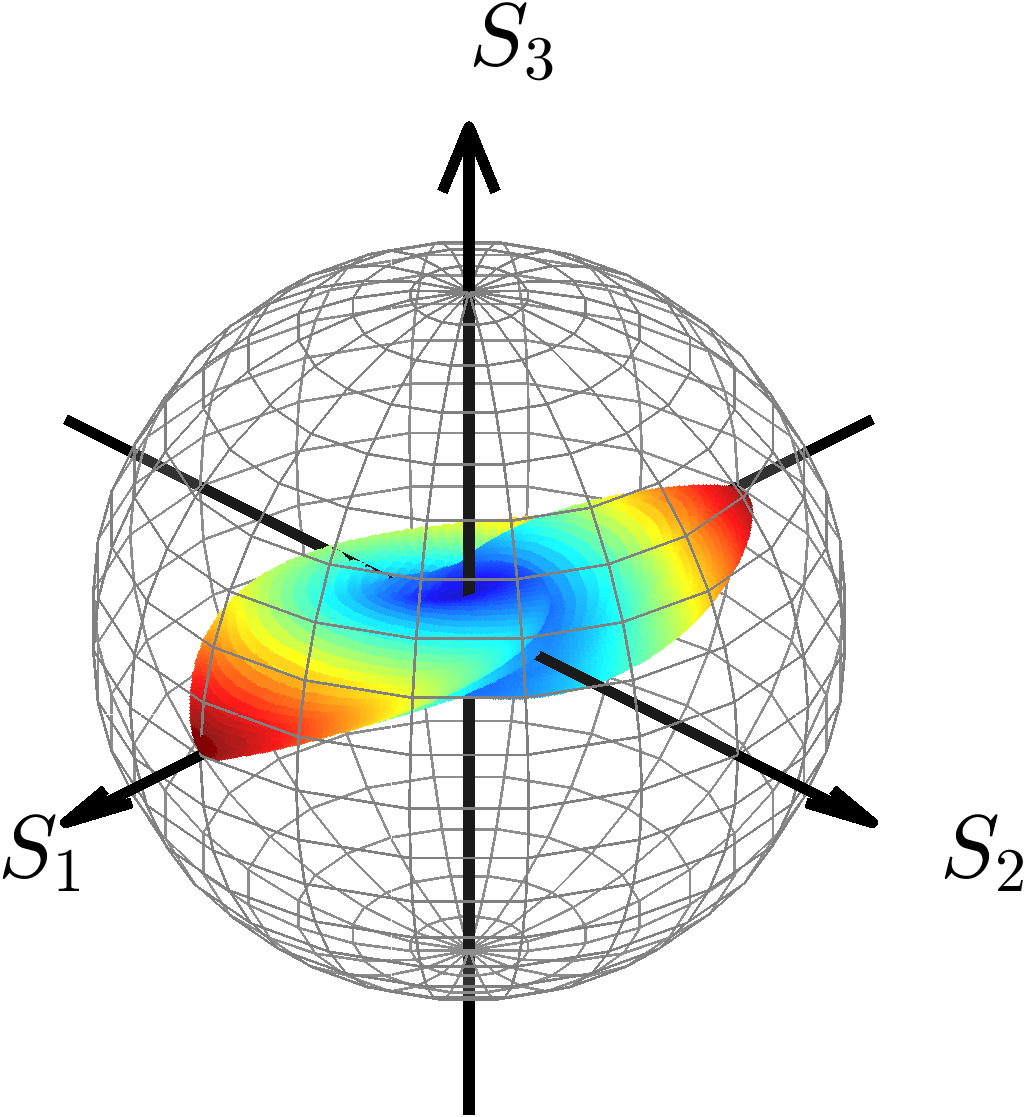}%
\begin{picture}(0,0)
\put(-120,110){(a)}
\end{picture}\hspace*{3pt}
\includegraphics[scale=0.65, trim={0 0 0 0}, clip]{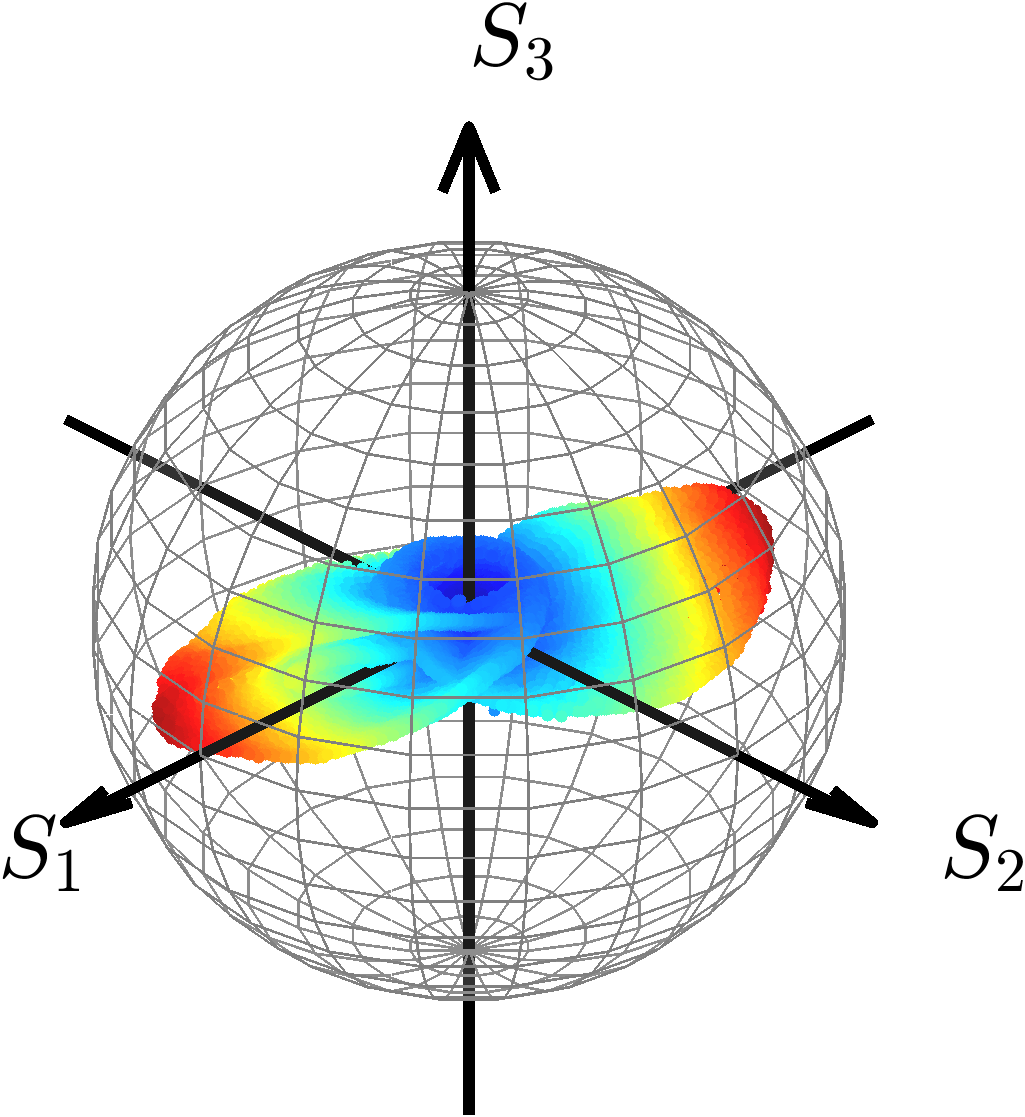}%
\begin{picture}(0,0)
\put(-120,110){(b)}
\end{picture}

\caption{ Poincare-like sphere representation of the azimuthally polarized Airy beam. (a) Analytical simulation, where the decay factor $a_x=a_y=0.15$, the normalization distances $x_0=y_0=30\ \mu m$, the distance from the focus $z=-2\ mm$, and the wavelength $\lambda=1028\ nm$, and (b) experimental results. The color bar and radius of the simulation sample points are given by the Stokes vector $S_0$. }
\label{fig:Poincare_M}
\end{figure}

\begin{figure}[t!]
\includegraphics[scale=0.65, trim={0 0 0 0}, clip]{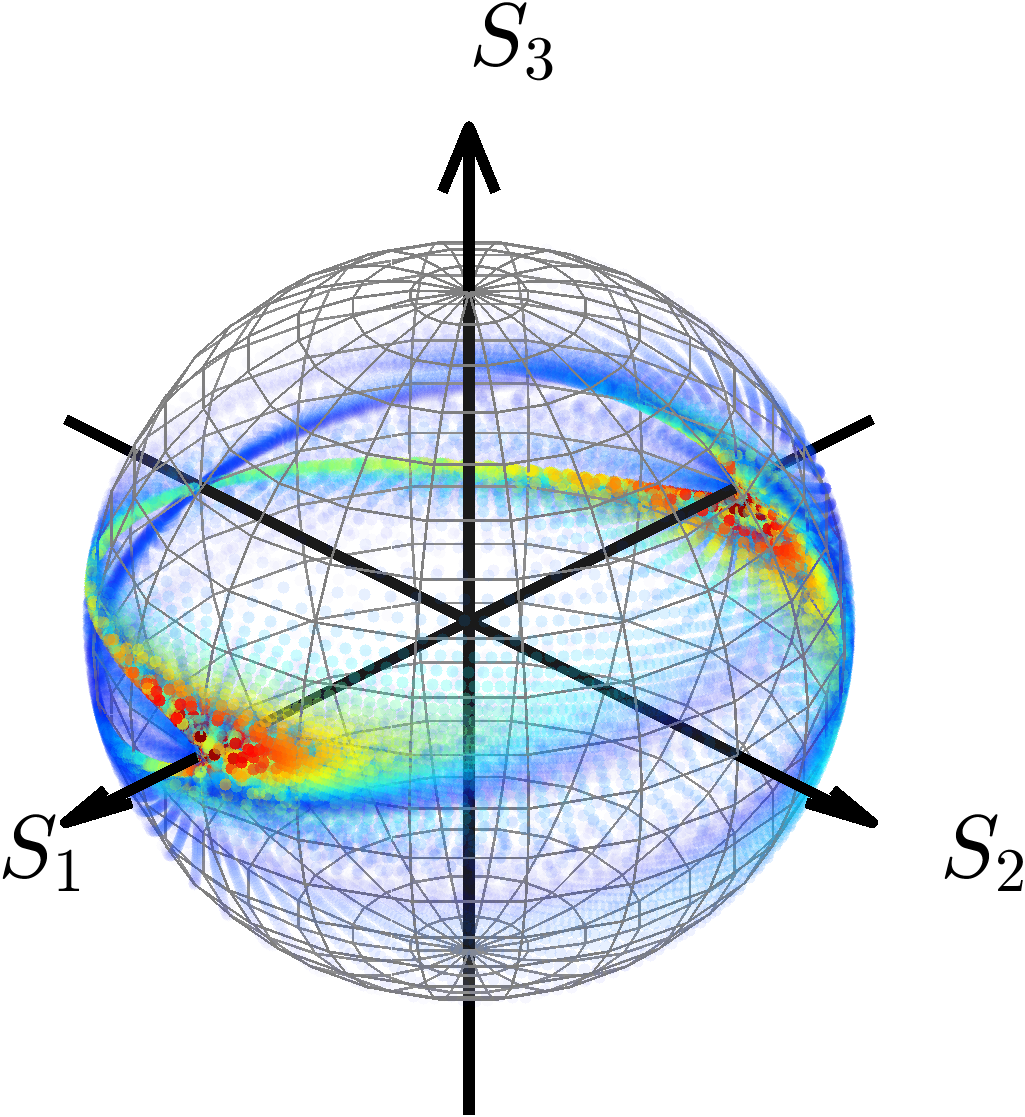}%
\begin{picture}(0,0)
\put(-120,100){(a)}
\end{picture}\hspace*{3pt}
\includegraphics[scale=0.65, trim={0 0 0 0}, clip]{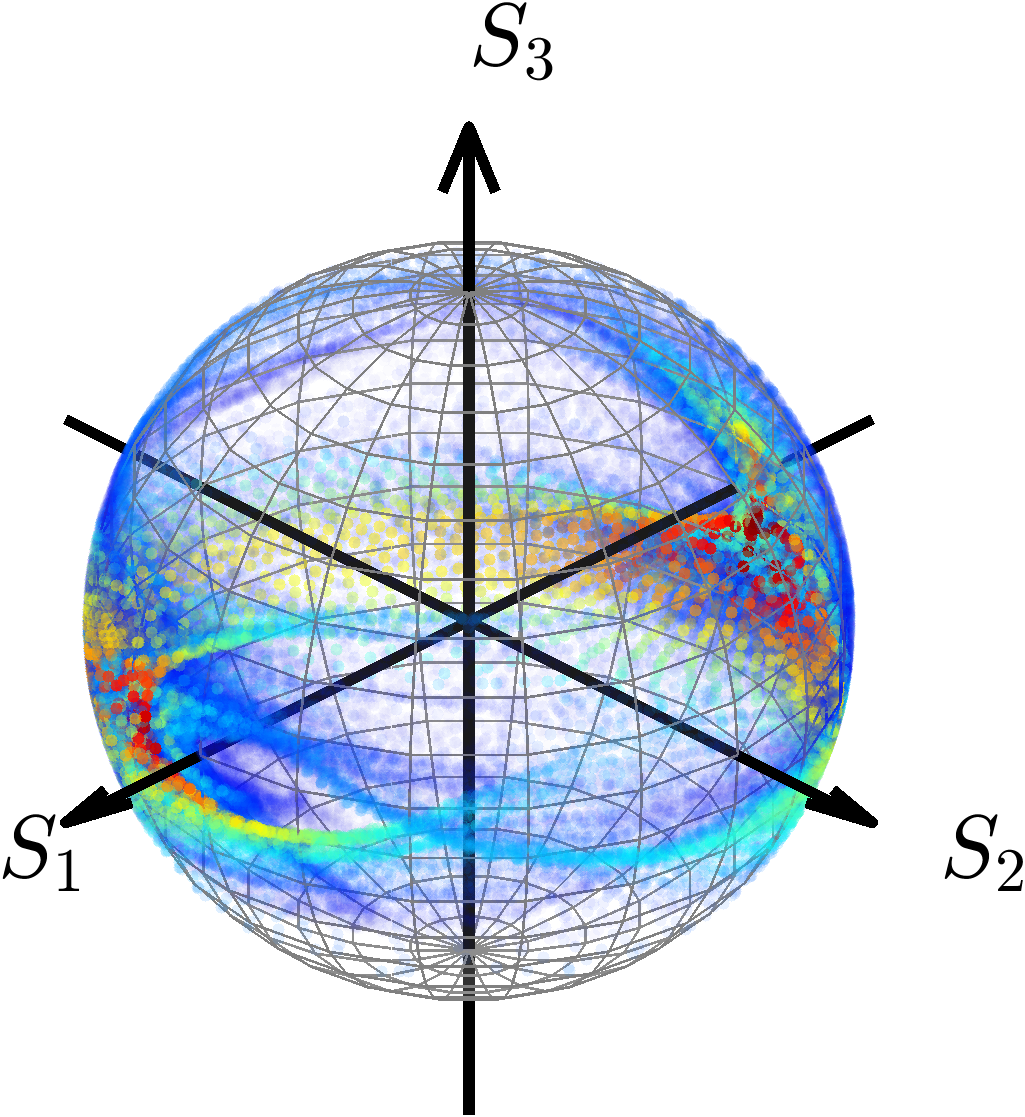}%
\begin{picture}(0,0)
\put(-120,110){(b)}
\end{picture}
\caption{Conventional Poincare sphere representation of the azimuthally polarized Airy beam. (a) Simulation, where the decay factor $a_x=a_y=0.15$, the normalization distances $x_0=y_0=30\ \mu m$, the distance from the focus $z=-2\ mm$ and the wavelength $\lambda=1028\ nm$, and (b) experimental results. The colorbar is given by the Stokes vector $S_0$ and the sample points are placed at $R=1$. }
\label{fig:Poincare_M2}
\end{figure}
\begin{figure}[t!]
\includegraphics[scale=0.65, trim={0 0 0 0}, clip]{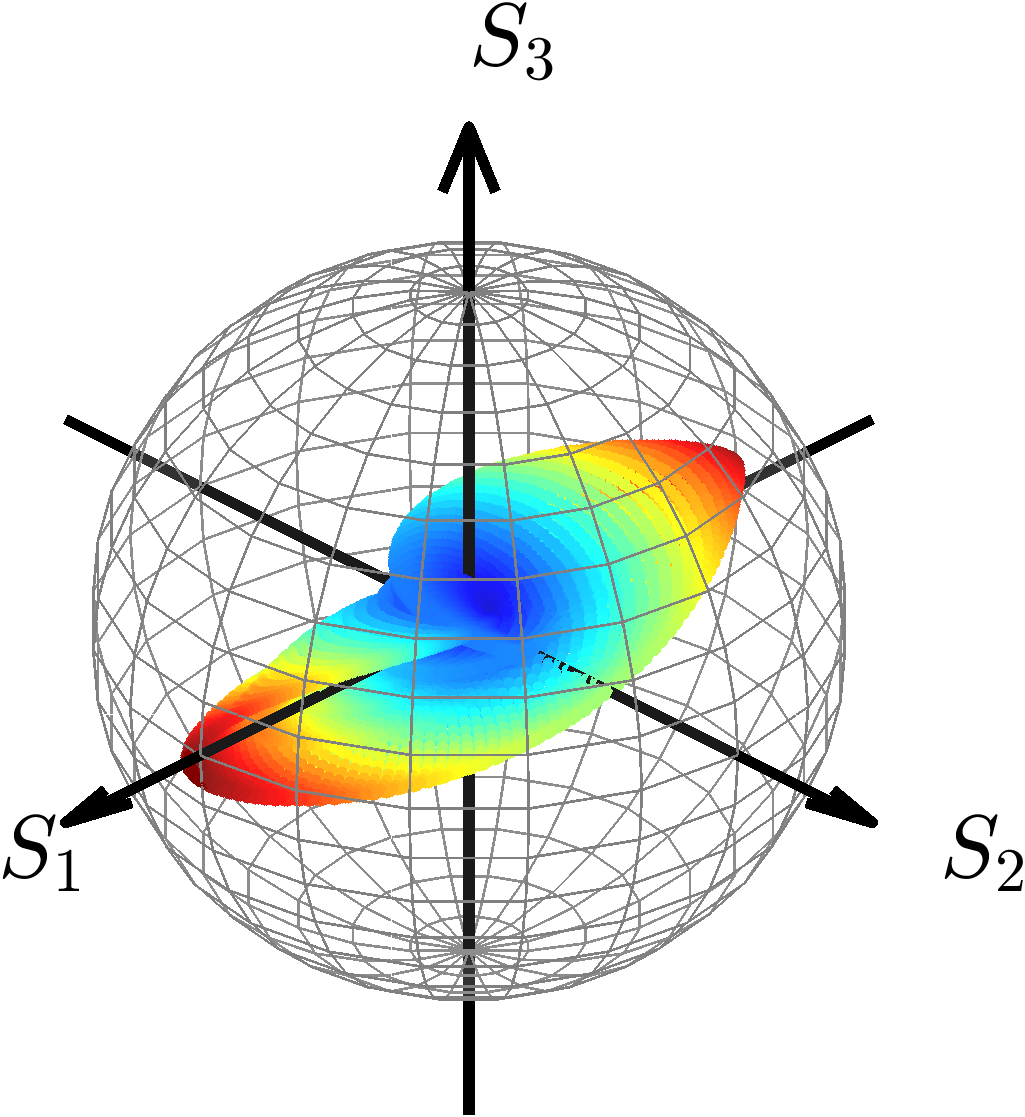}%
\begin{picture}(0,0)
\put(-120,110){(a)}
\end{picture}\hspace*{3pt}
\includegraphics[scale=0.65, trim={0 0 0 0}, clip]{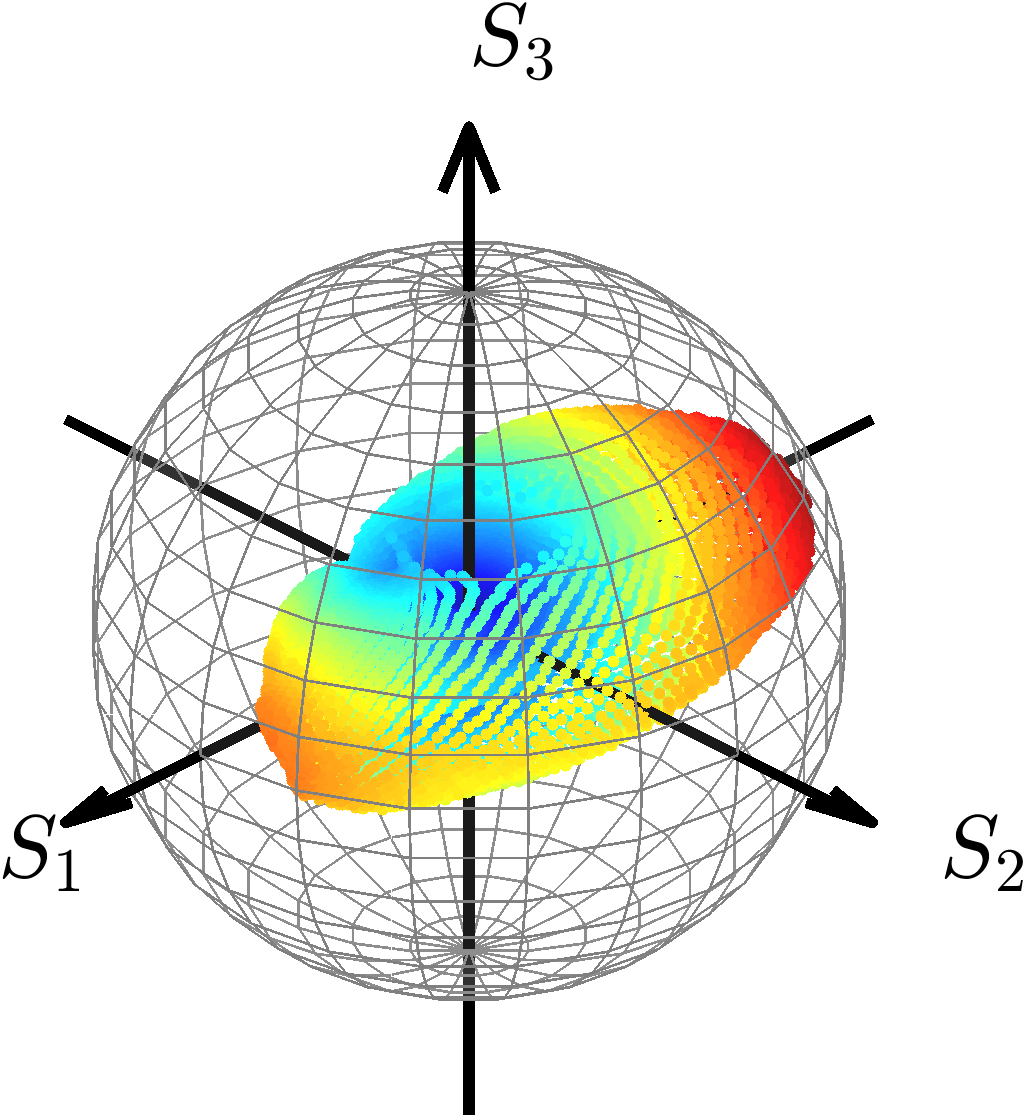}%
\begin{picture}(0,0)
\put(-120,110){(b)}
\end{picture}
\caption{ Poincare-like sphere representation of the radially polarized Airy beam. (a) Simulation, where the decay factor $a_x=a_y=0.15$, the normalization distances $x_0=y_0=30\ \mu m$, the distance from the focus $z=-4\ mm$ and the wavelength $\lambda=1028\ nm$, (b) experimental results. The color bar and radius of the simulation sample points are given by the Stokes vector $S_0$. }
\label{fig:Poincare_N}
\end{figure}

\begin{figure}
\includegraphics[scale=0.65, trim={0 0 0 0}, clip]{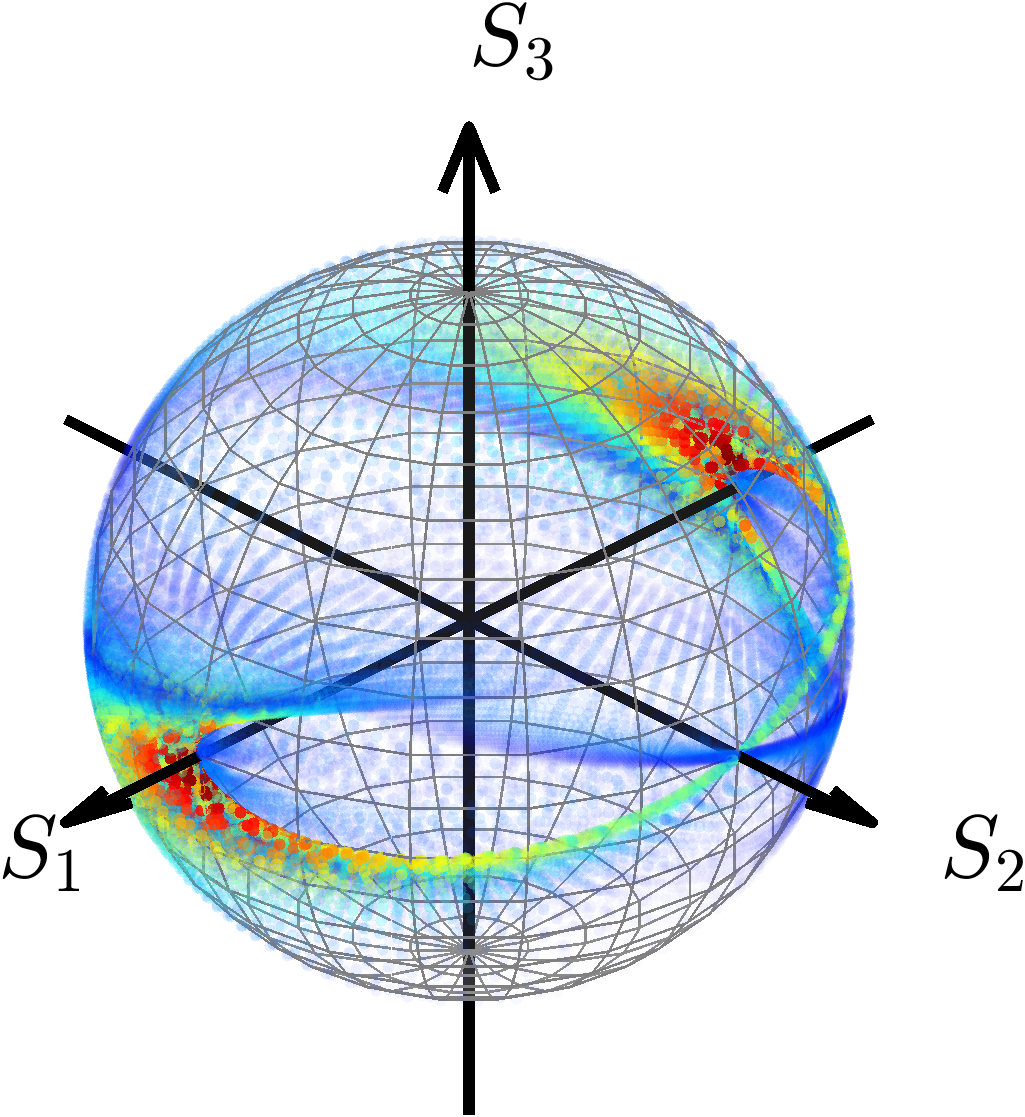}%
\begin{picture}(0,0)
\put(-120,110){(a)}
\end{picture}\hspace*{3pt}
\includegraphics[scale=0.65, trim={0 0 0 0}, clip]{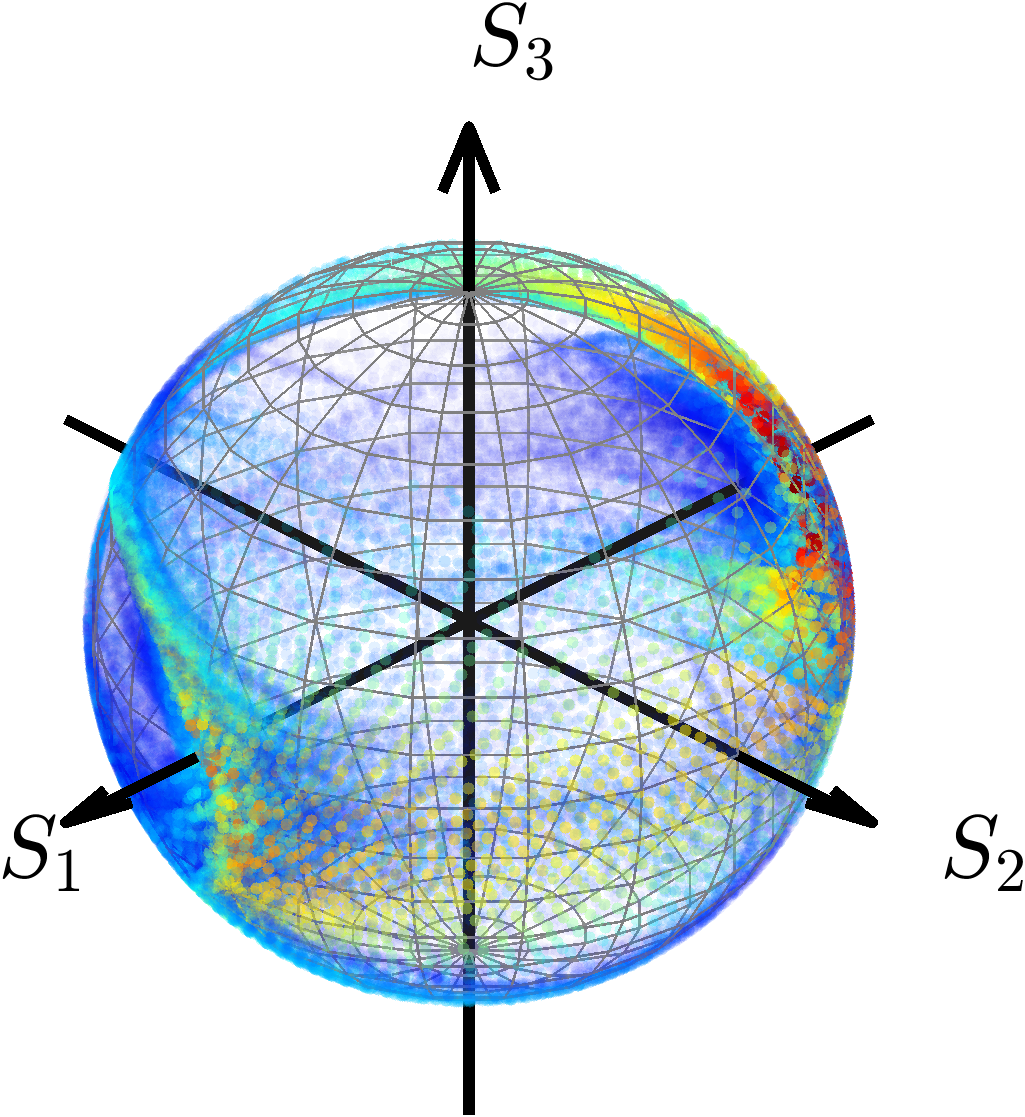}%
\begin{picture}(0,0)
\put(-120,110){(b)}
\end{picture}
\caption{Conventional Poincare sphere representation of the radially polarized Airy beam. (a) Simulation, where the decay factor $a_x=a_y=0.15$, the normalization distances $x_0=y_0=30\ \mu m$, the distance from the focus $z=-4\ mm$ and the wavelength $\lambda=1028\ nm$, (b) experimental results. The simulation colorbar is given by the Stokes vector $S_0$ and sample points are placed at $R=1$. }
\label{fig:Poincare_N2}

\end{figure}
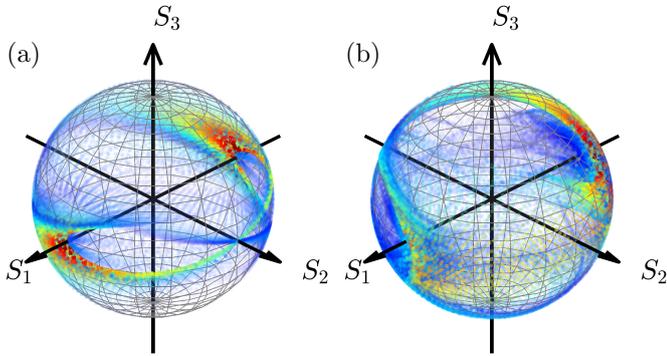
A few things were noticed by a detailed comparison of the experimentally obtained and analytically calculated Stokes parameters for radially and azimuthally polarized Airy-like vector beams. Firstly, the intensity profile, corresponding to the Stokes parameter $S_0$, is similar for both beams and resembles the conventional Airy scalar beam, with the main lobe divided into two equal-intensity lobes. The intensity distribution of the parameter $S_1$ for the $\textbf{M}$ and $\textbf{N}$ type beams differs only in that it is reflected around the diagonal $x=y$. The parameter $S_2$, when comparing two beams, is antisymmetric in the value of the amplitude. They differ by a small amount due to the different propagation distances: the azimuthally polarized beam $\textbf{M}$ was measured and calculated at $z=-2\ mm$ and the radially polarized beam $\textbf{N}$ at $z=-4\ mm$. This confirms that in the vicinity of the focal point, the parameter $S_2$ is sensitive to the exact location of the measurement. The most noticeable mismatch between the analytically calculated and experimentally obtained data appears in the distribution of the Stokes parameter $S_3$. Although they seem to be the same, there are small differences. The difference could be explained by two operations, the inversion in the amplitude of the parameter $S_3$ and the rotation of the two main lobes of this component in opposite directions.

Multiple factors can be attributed to possible experimental imperfections, which can be accounted for by the inaccuracies in the Stokes parameters. Let us start with the fact that the AiGPE element in addition to the Airy phase mask is superimposed with a blaze-grating mask. It has been realized that the refracted Airy beam has a cleaner intensity distribution compared to one without a blaze grating. The blaze-grated Airy beam does not interfere with the incoming Gaussian beam at the element's output. On the downside, it makes it more difficult to align the refracted Airy beam with the rest of the experimental setup, as it requires additional precision in placing elements and introduces an angle between the refracted Airy beam and the optical axis. 

The second type of inaccuracy might be caused by the polarizer P2, which may not be fully aligned with the $x$ or $y$ axis, depending on the type of vector beam produced, and does not produce a perfect polarization of the incoming beam. Consequently, even very small misalignment angles in the construction of the Airy beam might bring an additional circular polarization component after the P2 polarizer. The misalignment of the S-wp element with the incoming Airy beam means that the output is a coherent addition of the azimuthally and radially polarized vector beams. Moreover, the Stokes parameter measurement setup includes a rotating quarter-waveplate and a fixed polarizer that must also be perfectly aligned with the resulting beam. The last type of error might be due to the uncertainty in the positive or negative direction of the reference focal plane at $z=0$. This possibly results in the fact that although the Stokes parameters $S_0$, $S_1$, $S_2$ remain unchanged, the $S_3$ parameter is slightly rotated and inverted.
\begin{figure}
\includegraphics[scale=0.37, trim={70 0 105 0}, clip]{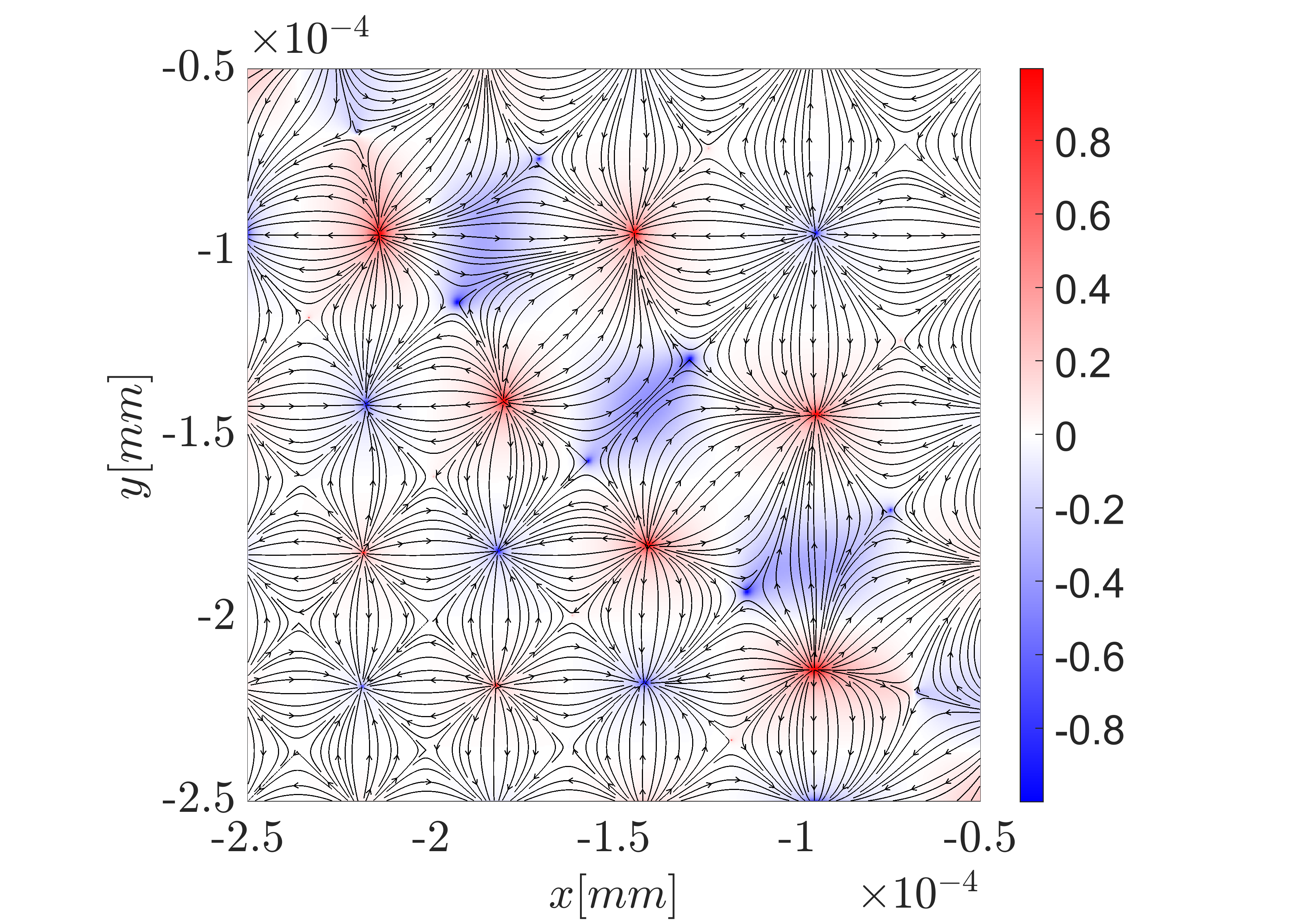}%
\begin{picture}(0,0)
\put(-120,90){}
\end{picture}
\caption{ Distribution of the normalized imaginary part of the radially polarized Airy beam, when $n_x = \text{Im}\{N_x\}/|\textbf{N}|$, $n_y = \text{Im}\{N_y\}/|\textbf{N}|$ and $n_z = \text{Im}\{N_z\}/|\textbf{N}|$. The decay factor $a_x=a_y=0.15$, the normalization distances $x_0=y_0=30\ \mu m$, distance from the focus $z=-2\ mm$, and the wavelength $\lambda=1028\ nm$ were used in modeling of the radially polarized beam. The colorbar indicates the value of the $n_z$ component and arrows depict the direction of the transverse components of the \textbf{n} field. }
\label{fig:Im{N}}
\end{figure}
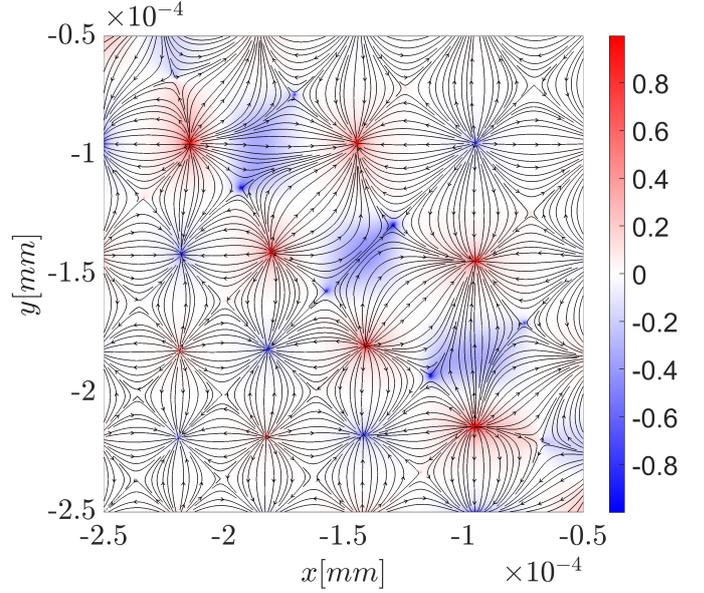

\begin{figure}
\includegraphics[scale=0.28, trim={30 0 50 0}, clip]{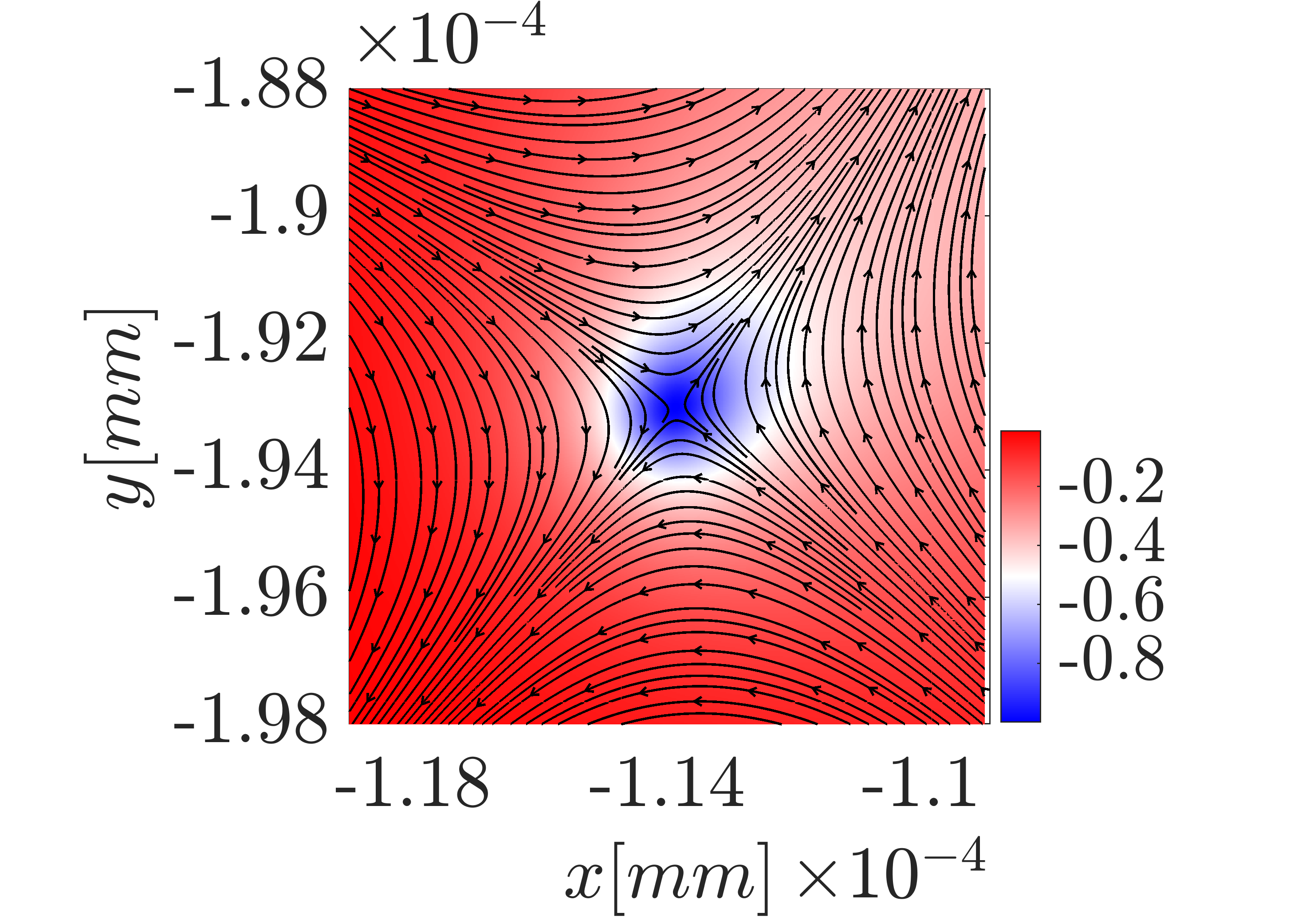}%
\begin{picture}(0,0)
\put(-40,65){\frame{\includegraphics[scale=0.16, trim={100 50 80 25}, clip]{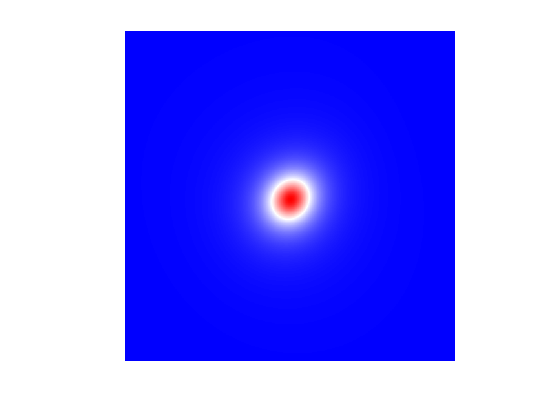}}}
\put(-115,106){(a)}
\end{picture}\includegraphics[scale=0.28, trim={30 0 50 0}, clip]{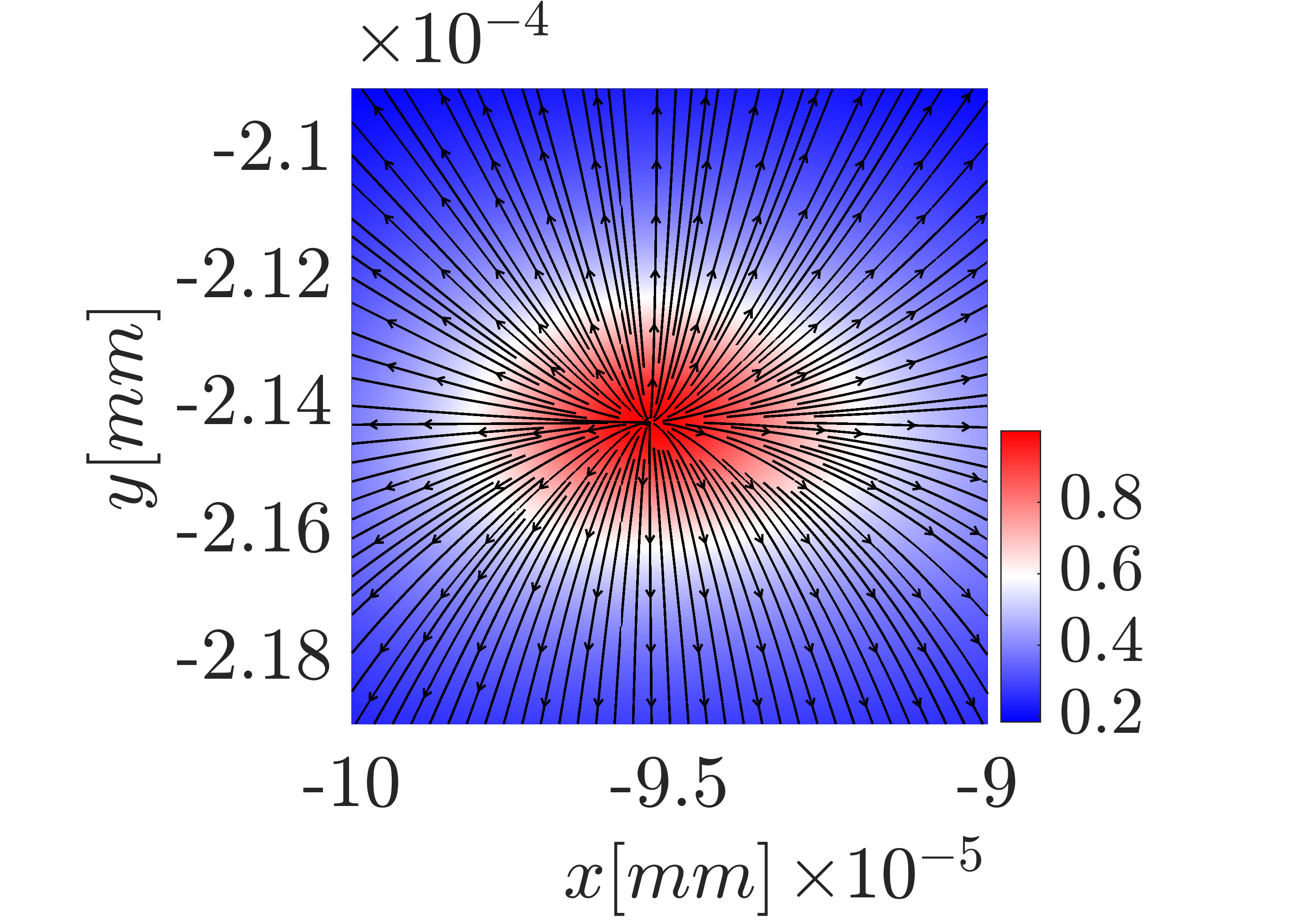}%
\begin{picture}(0,0)
\put(-40,65){\frame{\includegraphics[scale=0.16, trim={100 50 80 25}, clip]{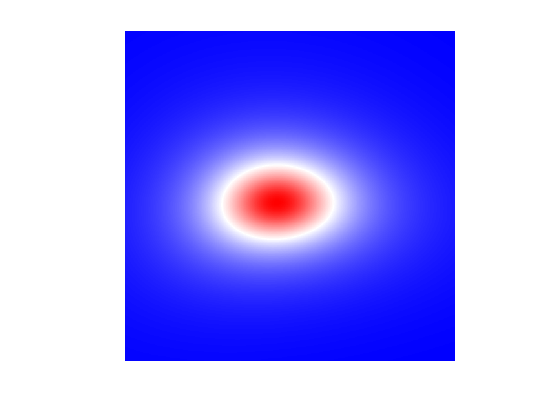}}}
\put(-115,106){(b)}
\end{picture}
\caption{ Distribution of the normalized imaginary part of the radially polarized Airy beam, when $n_x = \text{Im}\{N_x\}/|\textbf{N}|$, $n_y = \text{Im}\{N_y\}/|\textbf{N}|$ and $n_z = \text{Im}\{N_z\}/|\textbf{N}|$. The decay factor $a_x=a_y=0.15$, the normalization distances $x_0=y_0=30\ \mu m$, the distance from the focus $z=-2\ mm$, and the wavelength $\lambda=1028\ nm$ were used in the modeling of the \textbf{N} type beam. Insets: representation of the calculated skyrmionic density of the field \textbf{n}, blue corresponds to the zero value and red to the max value of a skyrmionic density. Arrows indicate the direction of the transverse components of the \textbf{n} field. }
\label{fig:Im{N}_zoomed}
\end{figure}

We continue our discussion by examining the Stokes parameter distribution on the Poincare-like sphere for both azimuthally and radially polarized Airy beams. Each measurement point of the Stokes vector $\mathbf{S}(x,y)$ is mapped to the Poincare-like sphere with $\Psi$ being orientation angle and $\chi$ ellipticity, of the polarization ellipse, see \cite{collett2005field}, as
\begin{equation}
\begin{aligned}
    \Psi &= \frac{1}{2} \tan^{-1} \left(\frac{S_2}{S_1}\right), \\
    \chi &= \frac{1}{2} \sin^{-1} \left(\frac{S_3}{S_0}\right), \\
\end{aligned}
\end{equation}
where the angles $2\Psi$ and $2\chi$ also represent the latitude and longitude angles of the Poincare sphere. For both types of beams in addition to the conventional Poincare sphere (Figures \ref{fig:Poincare_M2} and \ref{fig:Poincare_N2}), where the radius of each measurement sample is normalized to the total intensity at that sample point, we introduce an alternative Poincare-like sphere where all sample points are normalized to the maximum intensity point of the whole beam and not to each measurement sample (Figures \ref{fig:Poincare_M} and \ref{fig:Poincare_N}). In our opinion, this addition gives us different insights into the internal structure of the beams. The color scheme of the Poincare plots coincides with the color scheme of the \ref{fig:Poincare_M} and \ref{fig:Poincare_N} parameter's $S_0$ color scheme (Figures \ref{fig:Stokes_M} and \ref{fig:Stokes_N}).

\begin{figure*}[t!]
\centering
\includegraphics[scale=0.3, trim={0 90 0 0}, clip]{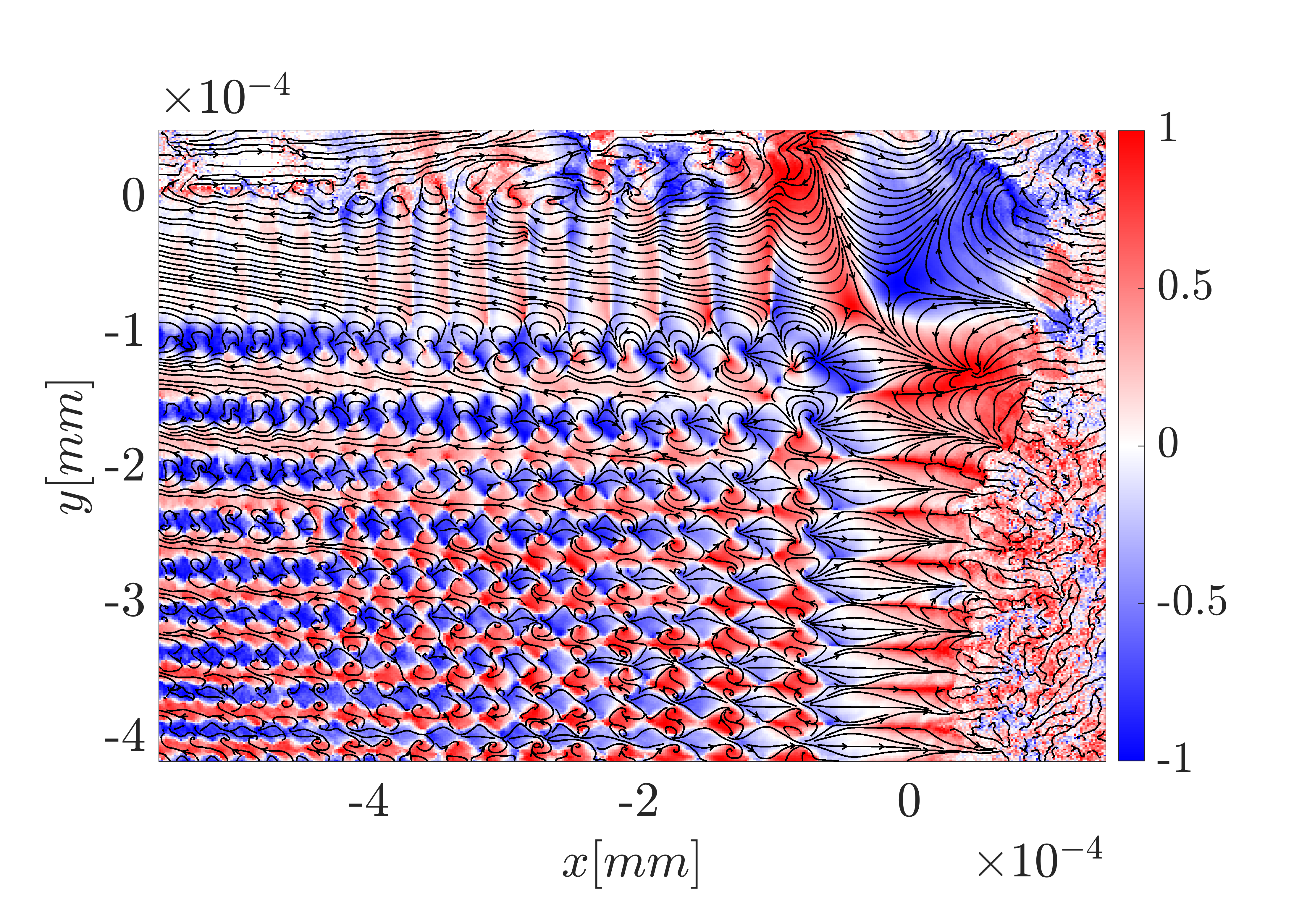}\includegraphics[scale=0.28, trim={90 90 0 50}, clip]{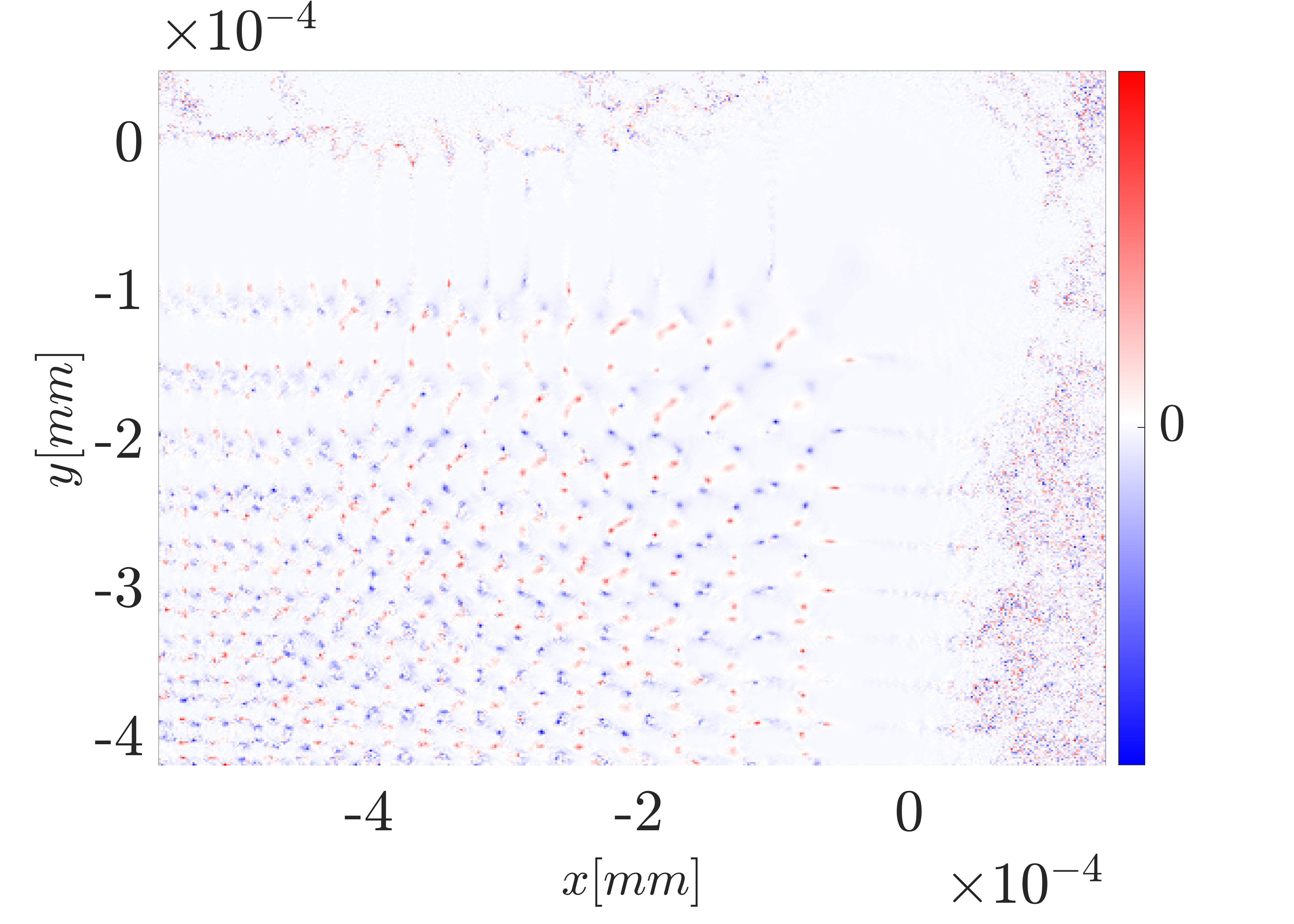}
\begin{picture}(0,0)
\put(-450,125){(a)}
\end{picture}
\includegraphics[scale=0.3, trim={0 0 0 0}, clip]{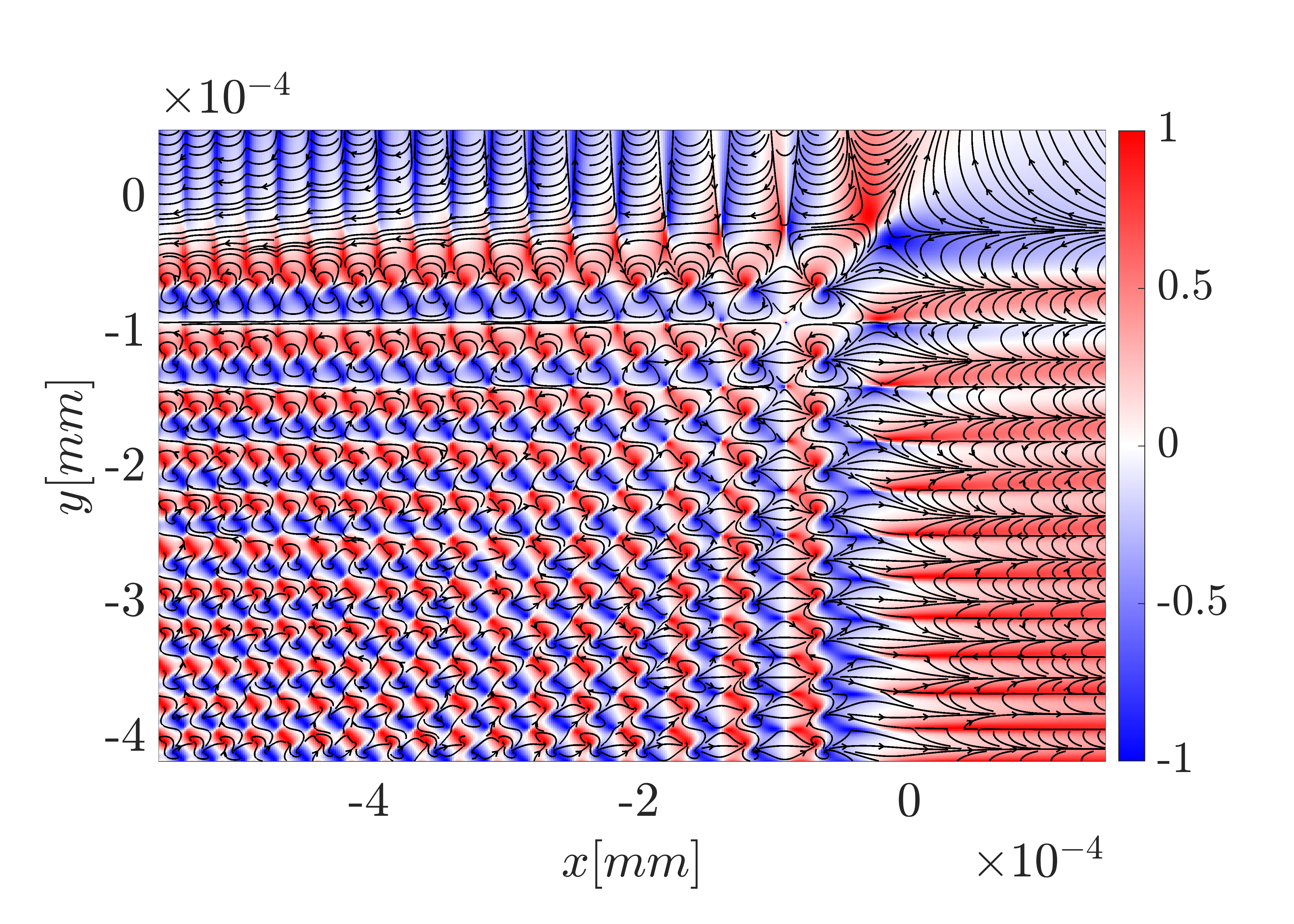}\includegraphics[scale=0.28, trim={90 0 0 50}, clip]{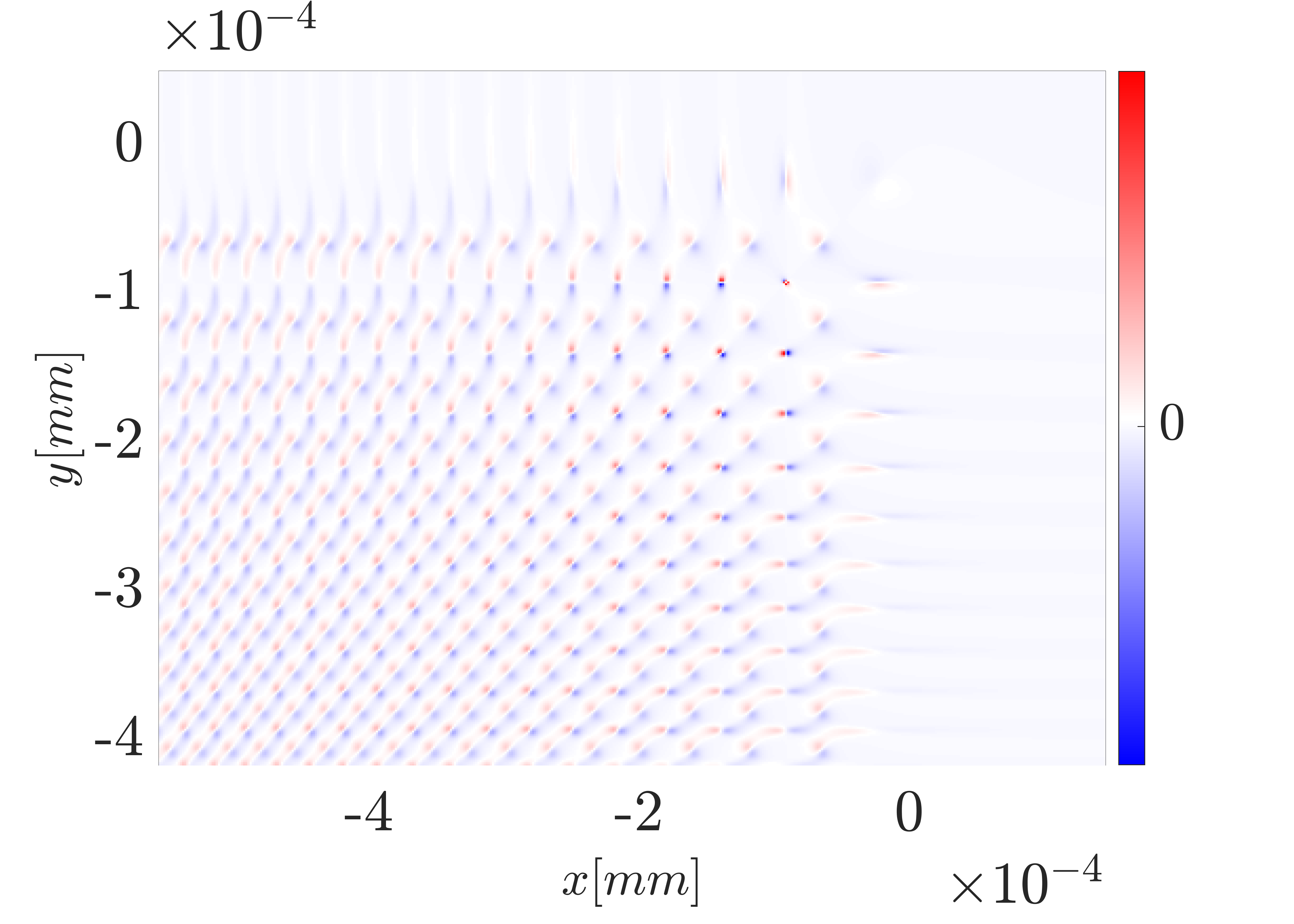}
\begin{picture}(0,0)
\put(-450,150){(b)}
\end{picture}

\caption{  (Left column) distribution of the normalized Stokes vector of the radially polarized Airy beam, when $n_x = S_1$, $n_y = S_2$ and $n_z = S_3$. a) experimental measurement of the beam with wavelength $\lambda=1028\ nm$. b) modeling, when the decay factor $a_x=a_y=0.15$, normalization distances $x_0=y_0=30\ \mu m$, distance from the focus $z=-2\ mm$, and wavelength $\lambda=1028\ nm$ were used. (Right column) the topological charge density representation of the calculated and measured field \textbf{n}. Arrows indicate the direction of the transverse components of the \textbf{n} field. }
\label{fig:Stokes_Skyrmion}
\end{figure*}
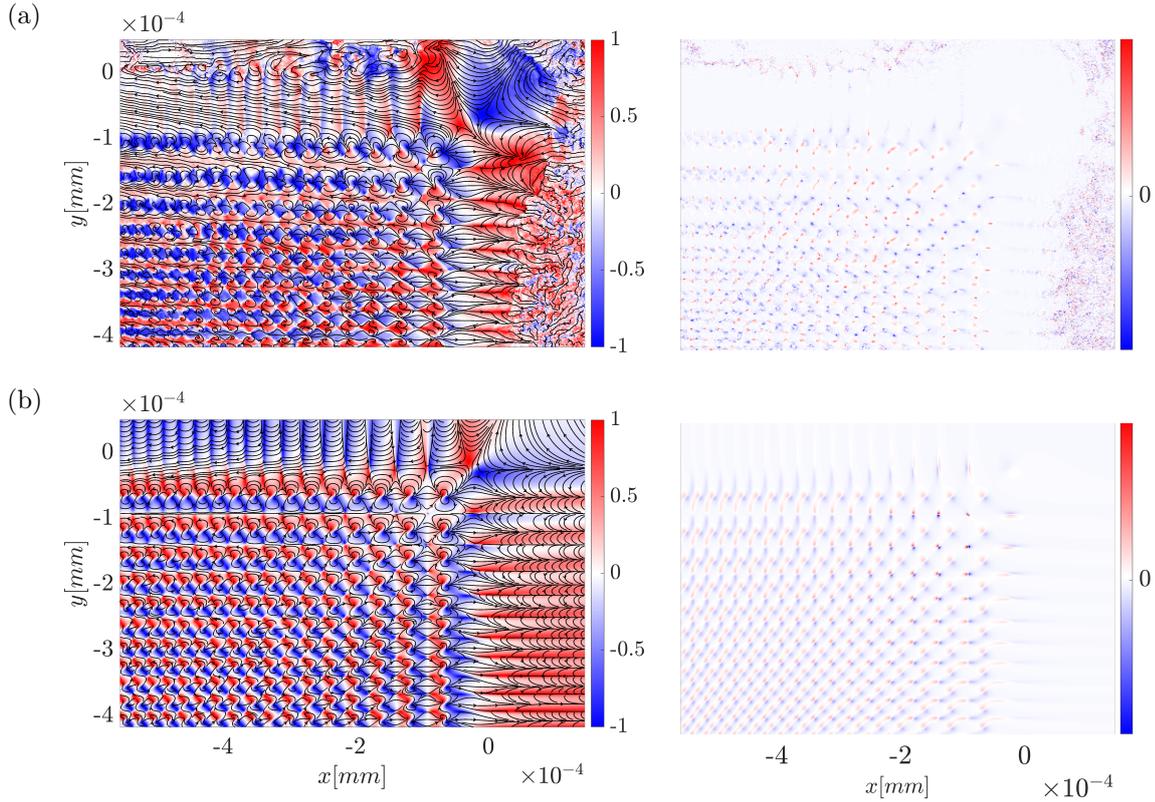

The Poincare sphere of azimuthally polarized Airy beams is depicted in Figures \ref{fig:Poincare_M} and \ref{fig:Poincare_M2}. Each figure contains two plots with theoretical calculations (left) and experimental measurements (right).  Figure \ref{fig:Poincare_M} is an unconventional Poincare-like sphere, where each point is not only color-coded by $S_0$ but is also located at the distance from the center, proportional to $S_0$. As a result, the beam is represented as a two-leaf structure oriented along the $S_1$ axis. One leaf is rotated $90^{\circ}$ degrees around the $S_1$ axis with respect to the other leaf. A similar structure is observed in the experimental measurement. 

The conventional color-coded Poincare sphere is shown in Figure \ref{fig:Poincare_M2}. This plot can be thought of as being obtained by projecting each point in Figure \ref{fig:Poincare_M} onto the surface of the sphere. This method gives us additional insights into the structure of the polarization of the azimuthally polarized Airy beam. As mentioned above, the state of polarization in the focus, when $z=0$, is a non-homogeneous linear polarization, so the Poincare sphere representation in that case resembles a line around the equator with the most intense points located around the axis $S_1$. Although Figure \ref{fig:Poincare_M2} presents some of the differences between the theoretical and experimental results, the main characteristics are maintained, and the observable differences are due to the reasons discussed above.

Figures \ref{fig:Poincare_N} and \ref{fig:Poincare_N2} show the two variants of the Poincare spheres for the radially polarized Airy beam. The logic of the beam representation is the same as it was for the azimuthally polarized beam. By comparing the theoretical and experimental results in Figure \ref{fig:Poincare_N} one can observe that experiment measurement points of the polarization structure are more dispersed than those predicted by the authors numerically and the high values are shifted towards the negative direction of $S_1$ axes rather than equally distributed along it. However, in general, an elongated beam structure that is extended in the $S_1$ direction is still clearly recognized in theoretical and experimental results. In Figure \ref{fig:Poincare_N2} the theory and the experiment show an acceptable resemblance. 

When comparing the distributions of azimuthally and radially polarized beams, it should be noted that the different locations along the longitudinal axis were used, this extends the Poincare sphere representation more towards the $S_3$ axis. Another feature that can be deduced from the figures \ref{fig:Poincare_M2} and \ref{fig:Poincare_N2} is that the points on the Poincare sphere for the azimuthally polarized Airy beam surround the $S_1$ axis from one side and for the radially polarized Airy beam from the opposite side.

\section{Topological structures in the nonhomogeneously polarized Airy beams}

Lastly, we investigate the topological structure of the azimuthally and radially polarized Airy-like beams. This investigation is usually carried out by mapping the points of a normalized 2D field to the unit sphere $S^2$. The normalized vector field \textbf{n} can be selected to be any form (linear or quadratic) of the electromagnetic field, in this work we look into the Stokes domain and the polarization domain of the fields. The topological density of a normalized field is given by 
\begin{equation}
\rho_s = \mathbf{n} \cdot\left(\frac{\partial \mathbf{n}}{\partial x} \times \frac{\partial \mathbf{n}}{\partial y}\right).
\label{eq:Skyrmionic_density}
\end{equation}
When this density is integrated over the region $d^2r$ the topological charge is calculated. It provides information on the field wrapping properties around the unit sphere, such as polarity, vorticity, and helicity \cite{gobel2021beyond}.

Let us begin a discussion by exploring the distribution of the normalized electric field for the radially polarized Airy beam depicted in Figure \ref{fig:Im{N}}. The plot portrayed shows the absolute value of the normalized electric field in the direction of propagation, and the streamlines field direction in the transverse plane. The parameters used for the calculation are the decay factor $a_x=a_y=0.15$, the normalization distances $x_0=y_0=30\ \mu m$, the distance from the focus $z=-2\ mm$, the wavelength $\lambda=1028\ nm$, and the normalized field is chosen to be $\textbf{n} = \text{Im}\{\textbf{N}\}/|\textbf{N}|$. 

In this configuration of the field, two types of different topological structures can be observed, see Figure \ref{fig:Im{N}_zoomed}. The first one is presented in Figure \ref{fig:Im{N}_zoomed} (a), it resembles a topological configuration of the antiskyrmion. In the top right corner of the plot, a calculated topological density is given as inset; note that not the value of the density per se is important, but that the density is positive and the topological-particle-like structure is observed. It can be seen that the field is directed toward the center from two opposite directions, the outgoing stream is perpendicular to the incoming stream, and in total four distinct directions are present. The second type of topological structure is observed at different locations of the beam, shown in Figure \ref{fig:Im{N}_zoomed} (b). This topological structure seems to be described by a Néel-type skyrmion. The field lines are radially emerging from the center of the structure. The inset showing the topological density has a positive density and resembles a deformed ellipse shape. 

The target for the investigation is a quadratic form of the electric field - the Stokes vector. For the sake of brevity, we only investigated the radially polarized Airy beam, as the azimuthally polarized Airy beam has a similar topological structure. We chose a normalized vector \textbf{n} to be $\{n_x, n_y, n_z\} = \{S_1, S_2, S_3\}/|\{S_1, S_2, S_3\}|$, where $S_1$, $S_2$ and $S_3$ are previously analyzed Stokes parameters. The emerging structure of the normalized Stokes topological field is shown in \ref{fig:Stokes_Skyrmion}. The blue and red color scheme depicts the magnitude of the normalized parameter $S_3$ and the streamlines direct the flow of the normalized transverse Stokes field. In Figure\ref{fig:Stokes_Skyrmion} (a) is an experimentally measured result, the left figure shows the topology and the right figure shows the topological density. It is noted that the topological density plot has round-shaped local structures. In Figure \ref{fig:Stokes_Skyrmion} (b) the computational modeling is given that produces similar results. In addition, a particle-like topological density is observed in both situations. The apparent difference is that the modeled and measured fields seem to have a rotation or skew when compared to each other. 

\begin{figure}
\includegraphics[scale=0.28, trim={30 0 50 0}, clip]{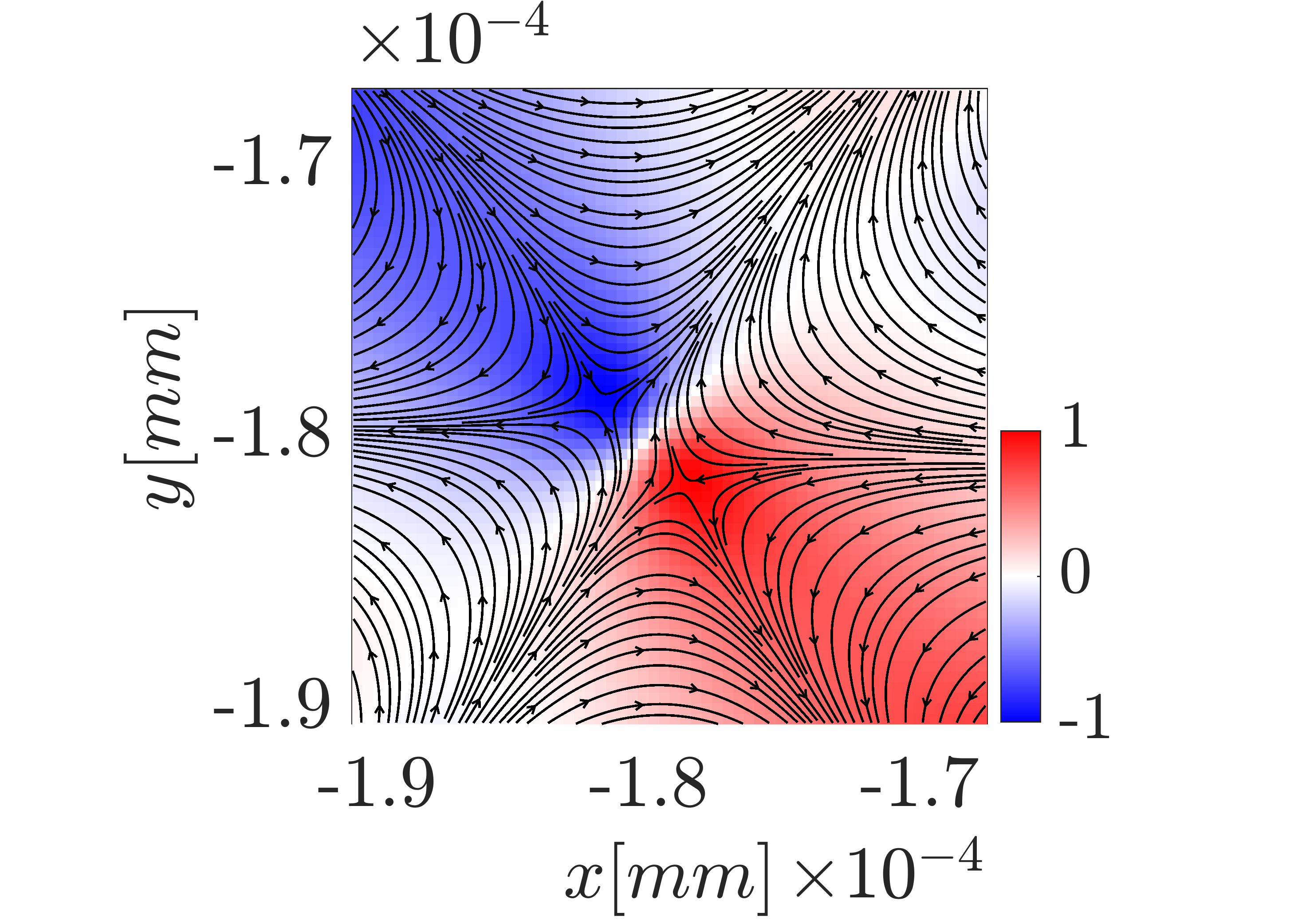}%
\begin{picture}(0,0)
\put(-40,65){\frame{\includegraphics[scale=0.17, trim={120 50 100 80}, clip]{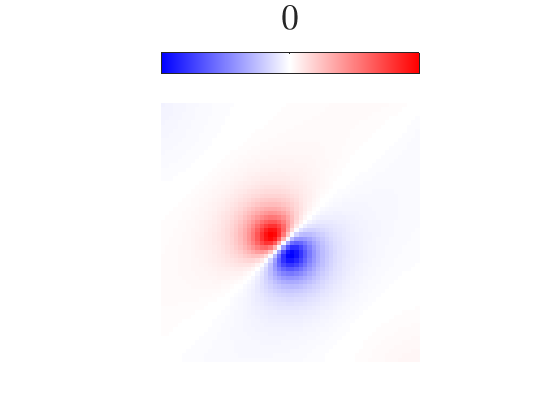}}}
\put(-40,98){\includegraphics[scale=0.17, trim={120 250 100 0}, clip]{N_Stokes_th_main_1_density.png}}
\put(-115,106){(a)}
\end{picture}\includegraphics[scale=0.28, trim={30 0 50 0}, clip]{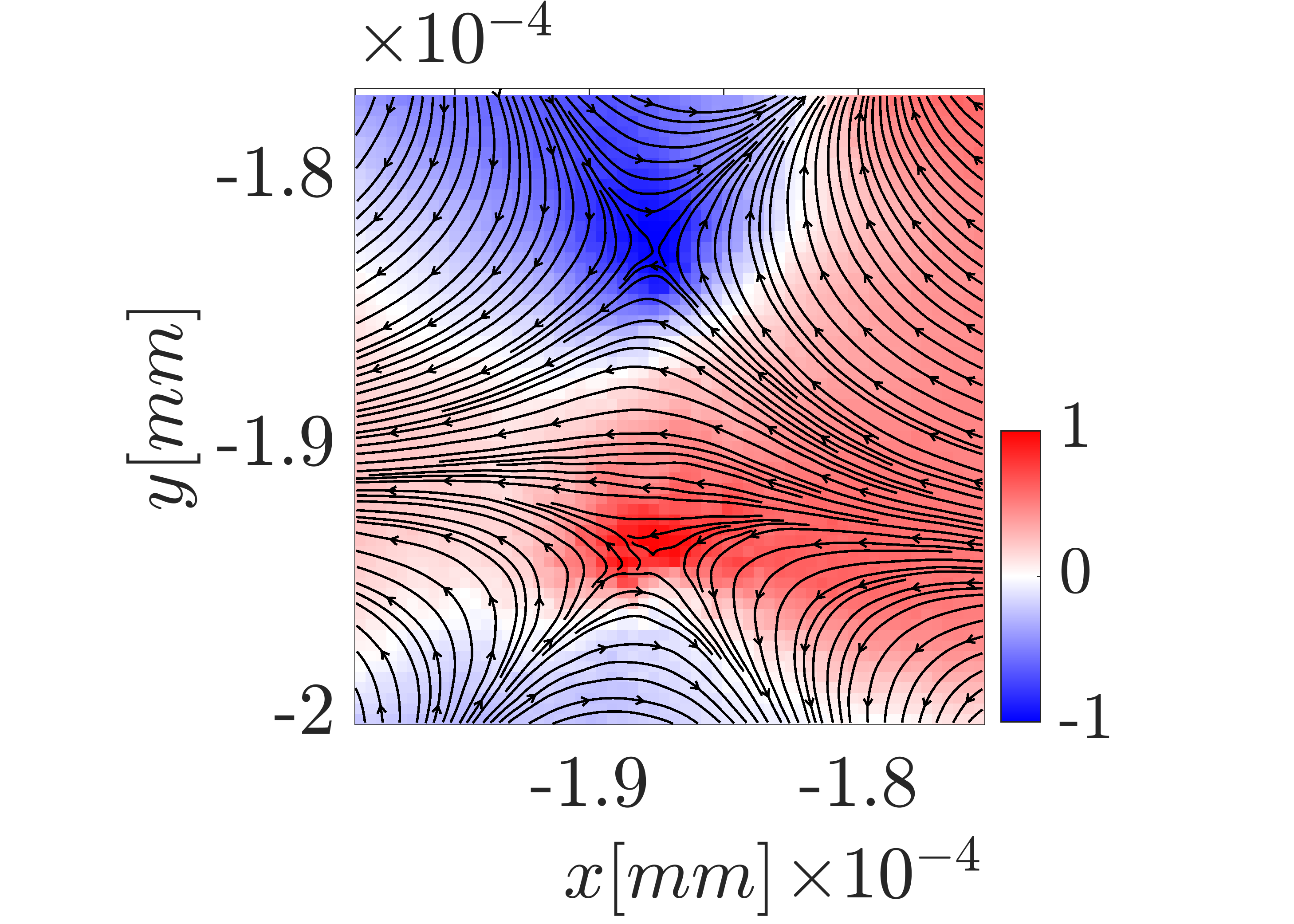}%
\begin{picture}(0,0)
\put(-40,65){\frame{\includegraphics[scale=0.17, trim={120 50 103 80}, clip]{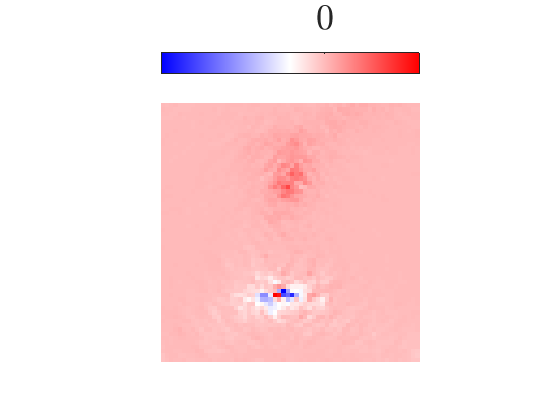}}}
\put(-40,98){\includegraphics[scale=0.17, trim={120 250 100 0}, clip]{N_Stokes_ex_main_1_density.png}}
\put(-115,106){(b)}
\end{picture}

\includegraphics[scale=0.28, trim={30 0 50 0}, clip]{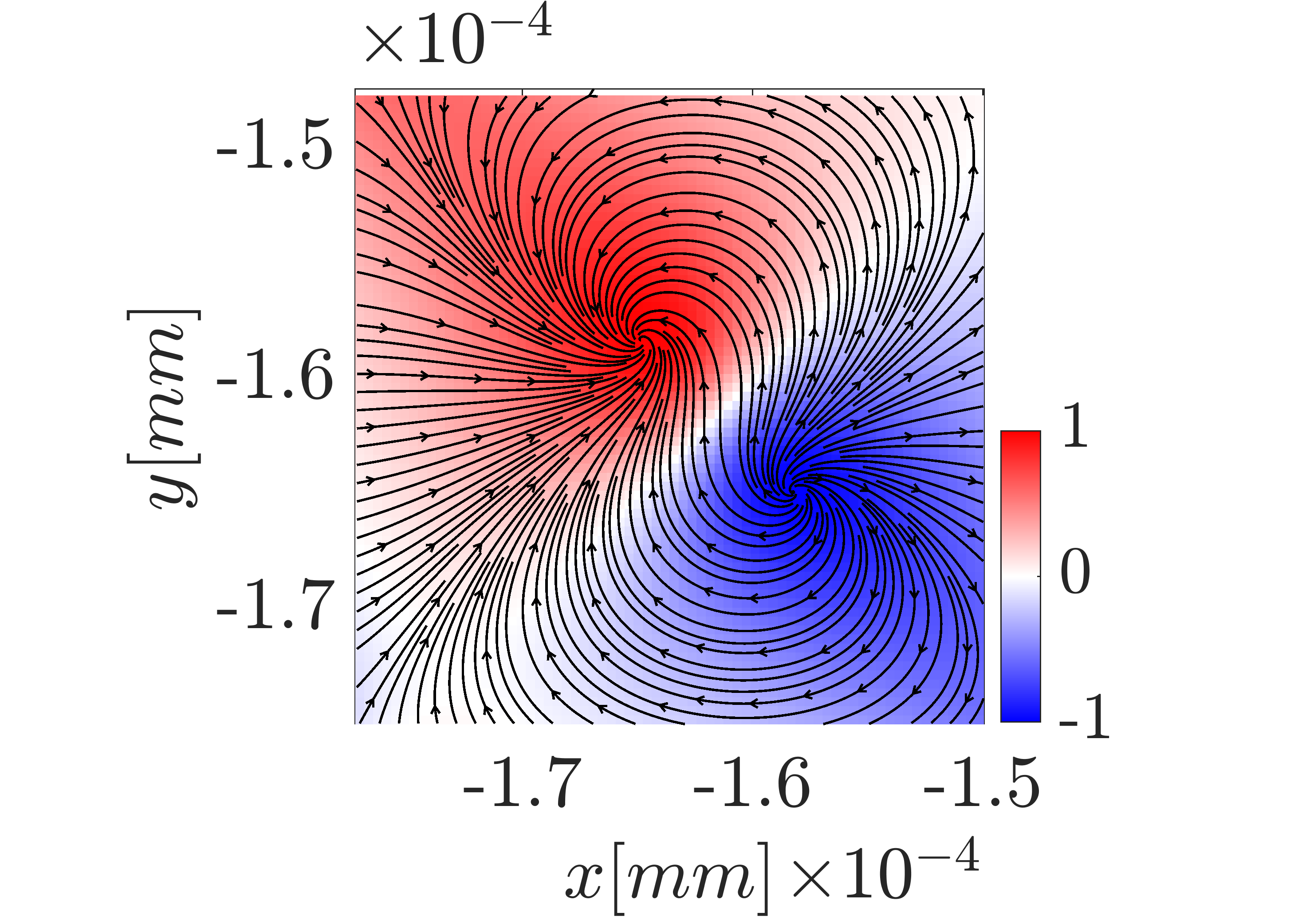}%
\begin{picture}(0,0)
\put(-40,65){\frame{\includegraphics[scale=0.17, trim={120 50 100 80}, clip]{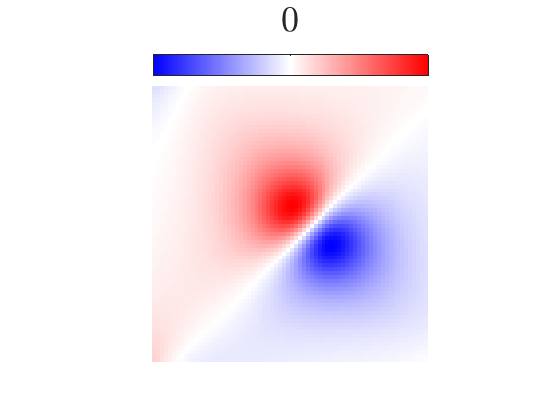}}}
\put(-40,98){\includegraphics[scale=0.17, trim={120 250 100 0}, clip]{N_Stokes_th_main_2_density.png}}
\put(-115,106){(c)}
\end{picture}\includegraphics[scale=0.28, trim={30 0 50 0}, clip]{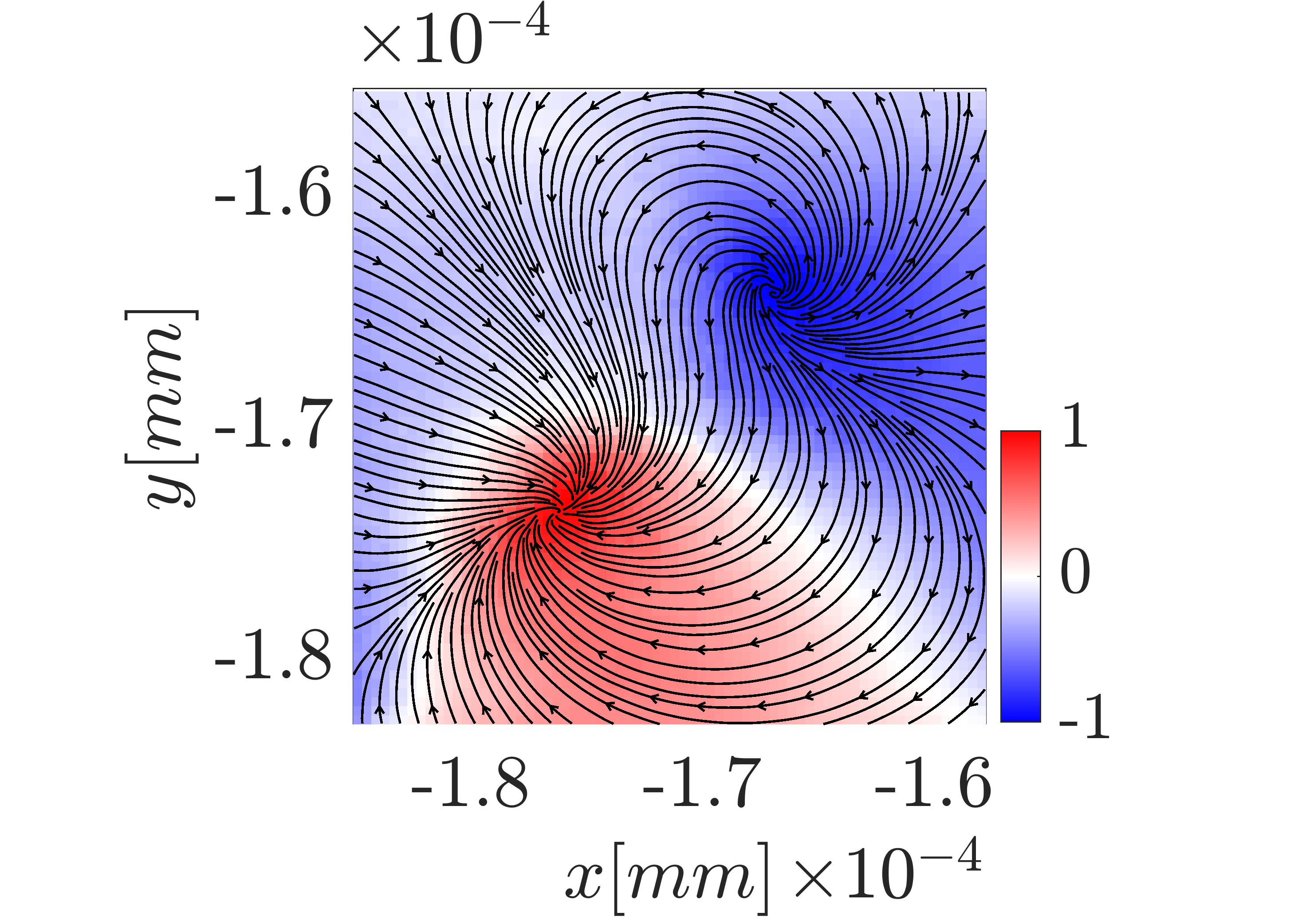}%
\begin{picture}(0,0)
\put(-40,65){\frame{\includegraphics[scale=0.17, trim={120 50 100 80}, clip]{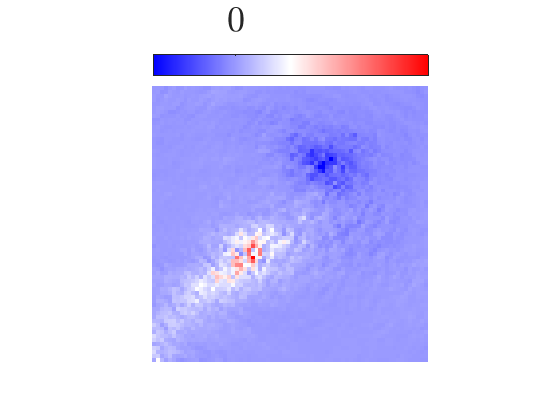}}}
\put(-40,98){\includegraphics[scale=0.17, trim={120 250 100 0}, clip]{N_Stokes_ex_main_2_density.png}}
\put(-115,106){(d)}
\end{picture}

\caption{ Distribution of the normalized Stokes vector of the radially polarized Airy beam, when $n_x = S_1$, $n_y = S_2$ and $n_z = S_3$. (a, c) modeling, when the decay factor $a_x=a_y=0.15$, normalization distances $x_0=y_0=30\ \mu m$, distance from the focus $z=-4\ mm$, and wavelength $\lambda=1028\ nm$ were used. (b, d) experimental measurement of the beam with wavelength $\lambda=1028\ nm$. Insets: representation of the calculated skyrmionic density of the field \textbf{n}. Arrows indicate the direction of the transverse components of the \textbf{n} field. }
\label{fig:N_Stokes_Skyrmion_1}
\end{figure}

As a closer investigation reveals, two types of topological field configurations. The first field configuration is zoomed in in Figure \ref{fig:N_Stokes_Skyrmion_1} (a, b) and depicts the results of the modeling (a) and experimental (b). When investigated closer, two antiskyrmion topological quasiparticles are observed, the insets are the topological density of the zoomed field.

The second topological structure detected in the field is shown in Figure \ref{fig:N_Stokes_Skyrmion_1} (c, d). It represents two opposite topological density quasiparticles. The field is swirling into and out of the center of the topological density field extremes.



\section{Conclusions}

In conclusion, we have realized high-power ultrafast Airy-like azimuthally and radially polarized vector beams. 

First, we have theoretically investigated the Fourier spectra and provided an analytical expression for azimuthally and radially polarized vector beam spatial spectra. Our analysis revealed that the beam spectrum of the azimuthally polarized Airy beam has a single angular component in the cylindrical coordinate system, and the beam spectrum of the radially polarized Airy beam has radial and longitudinal field components. Moreover, the decay factor not only apertures the beam but influences the whole beam structure, since when it is increased most of the beam energy is concentrated around the focal point, and on the contrary when the decay parameter is small, two distinct and elongated intensity lobes are present due to the interaction of the curved propagation trajectory with a line-like polarization singularity, located on the $z$-axis.

Secondly, we carried out an experiment and have verified experimentally the realization of high-power azimuthally and radially polarized Airy beams. In addition to the classical optical elements, the experiment contained two femtosecond laser inscribed optical metasurfaces with the manipulation of the geometrical phase: an Airy phase mask (Ai GPE) and an S-waveplate. These elements produce ultrafast vector Airy beams of high power. We showed that when a linearly polarized beam spectrum is propagated through the S-waveplate element, the element works as an operator transforming an incoming field into two vectorial spectral wave solutions, which could be described analytically as azimuthally and radially polarized beams. We measured Stokes parameters of these beams and represented them using two types of Poincare spheres, showing not only the constituent polarization components but also their intensities. 

Lastly, we continued by investigating the topological structure of the beam. When investigating the radially polarized Airy beam and its electric field, we have found two types of topological structures - antiskyrmion and skyrmion appearing in lattice formations. We have continued by investigating the quadratic form of the radially polarized Airy beam - its Stokes parameter domain - and did detect theoretically and experimentally complex topological lattice configurations and have encountered similar-looking biquasiparticles, anti-skyrmion pair, and a swirling topological structures appearing in lattices.

These findings not only advance our understanding of high-power ultrafast vector beams but also open up new avenues for their application in various fields. The ability to manipulate and control the polarization and topological structures of these beams can lead to significant advancements in optical communications, storage, laser micromachining, and medical imaging. Future research can build on these results to explore even more complex beam configurations and their interactions with different materials and within different environments. This work lays the foundation for innovative technologies that leverage the unique properties of high-power Airy-like azimuthally and radially polarized vector beams, paving the way for new breakthroughs in photonics and beyond. Future research may focus on further increasing the power and complexity of these beams as well as investigating their interactions with matter. By continuing to push the boundaries of ultrafast laser technology and advanced optical techniques, we can unlock the full potential of these intriguing light fields.

In summary, this work provides the community with a comprehensive investigation into the generation, propagation, and topological properties of high-power ultrafast vector Airy beams. Theoretical analysis and experimental results here are advancing our understanding of these complex light fields, paving the way for potential applications in various fields, including optical manipulation, nonlinear optics, and quantum optics.

\section*{Acknowledgments}
This research has received funding from the Research Council of Lithuania (LMTLT) via agreement No [S-MIP-23-71].


\end{document}